# Much Faster Algorithms for Matrix Scaling

Zeyuan Allen-Zhu
zeyuan@csail.mit.edu
Institute for Advanced Study

Yuanzhi Li
yuanzhil@cs.princeton.edu
Princeton University

Rafael Oliveira
rmo@cs.princeton.edu
Princeton University

Avi Wigderson
avi@ias.edu
Institute for Advanced Study

April 7, 2017

**Abstract**

We develop several efficient algorithms for the classical *Matrix Scaling* problem, which is used in many diverse areas, from preconditioning linear systems to approximation of the permanent. On an input $n \times n$ matrix $A$, this problem asks to find diagonal (scaling) matrices $X$ and $Y$ (if they exist), so that $XAY$ $\varepsilon$-approximates a doubly stochastic matrix, or more generally a matrix with prescribed row and column sums.

We address the general scaling problem as well as some important special cases. In particular, if $A$ has $m$ nonzero entries, and if there exist $X$ and $Y$ with polynomially large entries such that $XAY$ is doubly stochastic, then we can solve the problem in total complexity $\widetilde{O}(m+n^{4/3})$. This greatly improves on the best known previous results, which were either $\widetilde{O}(n^4)$ or $O(mn^{1/2}/\varepsilon)$.

Our algorithms are based on tailor-made first and second order techniques, combined with other recent advances in continuous optimization, which may be of independent interest for solving similar problems.

## 1 Introduction

The *matrix scaling* problem is natural and simple to describe. Given a non-negative real matrix $A$, can one *scale* its rows and columns (namely *multiply* each by a non-negative constant) to yield prescribed row *sums* and column *sums*. Note that the number of constraints is the same as the number of degrees of freedom; however, what makes it interesting (beyond the many applications that we detail below) is that the constraints are additive while the scalings are multiplicative.

Taking real non-negative entries and computing the row and column sums actually captures a much more general problem: one can allow $A$ to have complex entries and require the $\ell_p$ norms of rows and columns, after scaling, to equal prescribed values.[1]

For non-negative $d \times n$ matrix $A$, we say $A$ is an $(r,c)$-matrix if $r$ and $c$ are respectively the vectors of row and column sums of $A$. Given vectors $r$ and $c$, the problem of *matrix $(r,c)$-scaling* is to find *diagonal* matrices $X$ and $Y$ for which the matrix $XAY$ is an $(r,c)$-matrix.

When $d = n$ and $r = c = \mathbb{1} \in \mathbb{R}^n$ where $\mathbb{1}$ is the all-one vector, the matrix $(\mathbb{1},\mathbb{1})$-scaling problem becomes the *doubly stochastic scaling problem*.

While the above exact scaling problem is of interest, its asymptotic version is even more so, both from the algorithmic viewpoint and from the structural one, as it captures natural combinatorial problems. We say that $A$ is *asymptotically $(r,c)$-scalable* if the row and column sums can reach $r$ and $c$ asymptotically: that is, for every $\epsilon > 0$, there exist positive diagonal matrices $X, Y$ such that, letting $B = XAY$, we have $\|B\mathbb{1} - r\| \leq \varepsilon$ and $\|\mathbb{1}^\top B - c\| \leq \varepsilon$.[2]

---

[1]The simple reduction replaces any entry $\alpha$ in the matrix by $|\alpha|^p$.

The combinatorial essence of asymptotic scaling follows from a well-known characterization (see Proposition 2.2). A matrix $A$ is asymptotically $(\mathbb{1}, \mathbb{1})$-scalable if and only if the permanent of $A$ is positive, namely if the bipartite graph defined by the positive entries in $A$ has a perfect matching. A matrix $A$ is asymptotically $(r,c)$-scalable if and only if a natural flow on the same bipartite graph[3] has a solution. Duality (Hall's theorem and max-flow-min-cut theorem) gives simple certificates of non-scalability in terms of the patterns of 0's in the matrix $A$.

The main computational problem we study is: given a matrix $A$, vectors $r, c$ and $\varepsilon > 0$, determine if $A$ is $\varepsilon$-approximately $(r,c)$ scalable, and if so, find the scaling matrices $X, Y$.

Before diving into the history of matrix scaling, we explain one of its most basic applications, which also demonstrates its algorithmic importance.

**Preconditioning Linear Systems.** When solving a linear system $Az = b$, it is often desirable —for numerical stability and efficiency purposes— to have matrix $A$ be well-conditioned. When this is not the case, one tries to transform $A$ into a "better conditioned" matrix $A'$. Matrix scaling provides a natural and efficient reduction to do so. For instance, one would *hope* that a scaled matrix $A'$, in which e.g. all row and column $p$-norms are (say) 1, is better conditioned.[4] For this reason, we can use a matrix scaling algorithm to obtain diagonal matrices $X, Y$, and define $A' = XAY$. Now, the solution to $Az = b$ can be obtained by solving the (hopefully more numerically stable) linear system $A'z' = Xb$ and setting $z = Y^{-1}z'$. We stress here that $A'$ and $A$ have the same sparsity.

## 1.1 History and Prior Work

The matrix $(r,c)$-scaling problem is so natural and important that it was discovered independently by many different scientific communities, starting in 1937 with the work of Kruithof [20] in telephone forecasting, Deming and Stephan [10] in transportation science, Brown [8] in engineering, Stone [36] in economics, Wilkinson [37] in numerical analysis, and Friedlander [11] and Sinkhorn [34] in statistics. It has been applied in image reconstruction [14], operations research [30, 33], decision and control [5], theoretical computer science [22], and other scientific disciplines. For more references, we refer the reader to the survey [15], the paper [33] and references therein.

Perhaps the most famous algorithm for solving matrix $(r,c)$-scaling is the *RAS method* [34]. [5]The RAS method alternatively applies row and column normalization, where in a row (resp. column) normalization we multiply each $A_{i,j}$ by $r_i \cdot \left(\sum_j A_{ij}\right)^{-1}$ (resp. by $c_j \cdot \left(\sum_i A_{ij}\right)^{-1}$).

In the original paper, Sinkhorn [34] only proved the convergence of the RAS method when $A$ has only strictly positive entries and when $r = c = \mathbb{1}$. The best known complexity result for the RAS method is given by Kalantari *et al.* [16]. In particular, they showed that if the entries of $A$ are polynomially bounded, then the RAS method converges in $\widetilde{O}(n/\varepsilon^2)$ iterations[6] for $(\mathbb{1}, \mathbb{1})$-scaling, or in $\widetilde{O}(h^2/\varepsilon^2)$ iterations for $(r,c)$-scaling, where $r$ and $c$ are integral vectors and $h \stackrel{\text{def}}{=} \|r\|_1 = \|c\|_1$. Kalantari *et al.* also analyzed the RAS method when $A$ is strictly positive in all $n^2$ entries, or is

---
[2]The choice of norm in these expressions is not too important, and can be take to be $\ell_2$.
[3]Connect the source to the row vertex $i$ with capacity $r_i$, and the column vertex $j$ to the sink with capacity $c_j$.
[4]This indeed is widely use in practice (see [9, 29]), and indeed tends to numerically stabilize systems (see [7, 23, 29]), although no theoretical bounds are known (see [**?** ]).
[5]Also known as the Sinkhorn process, discovered by Sinkhorn in 1964 [34]. The RAS method fits as an instance of the "alternate minimization" heuristic, of which this is one of the few known instances in which it converges quickly.
[6]Each iteration of the RAS method costs complexity $O(m)$, the number of non-zero entries in $A$.

| Subcase | Paper | Total Complexity |
|---|---|---|
| full positive matrix | RAS method [17, 1993] | $\widetilde{O}(n^2 n^{1/2}/\varepsilon)$ ♭ |
| | Scaling0 | $\widetilde{O}(n^2 n^{1/3}/\varepsilon^{2/3})$ ♭ |
| | Scaling1 or Scaling3 | $\widetilde{O}(n^2)$ |
| scaling factors poly bounded | RAS method [16, 2008] | $\widetilde{O}(mn^{1/2}/\varepsilon)$ ♭ |
| | Scaling0 | $\widetilde{O}(mn^{1/3}/\varepsilon^{2/3})$ ♭ |
| | Scaling1 | $\widetilde{O}(m + n^{3/2})$ ♭ |
| | Scaling2+Scaling3 | $\widetilde{O}(m + n^{4/3})$ |
| general | RAS method [16, 2008] | $\widetilde{O}(mn/\varepsilon^2)$ ♭ |
| | Scaling0 | $\widetilde{O}(mn/\varepsilon^{2/3})$ |
| | LSW method [22, 1998] | $\widetilde{O}(mn^5)$ ♭ |
| | ellipsoid [18, 1996] | $\widetilde{O}(n^4)$ ♭ |
| | interior point [5, 2004] | $\widetilde{O}(n^6)$ ♭ |
| | max flow [32, 2007] | $\geq \widetilde{\Omega}(mn^4)$ ♭ |
| | Scaling1 | $\widetilde{O}(mn + n^{5/2})$ ♭ |
| | Scaling0+Scaling3 | $\widetilde{O}(mn + n^{7/3})$ |

Table 1: $(\mathbb{1}, \mathbb{1})$-scaling.

- We use $\widetilde{O}$ to hide log factors in $n$ and $1/\varepsilon$.
- Scaling0 is a simple method just like RAS;
- Scaling1, Scaling2, and Scaling3 use Laplacian system solvers and graph sparsification.
- ♭ indicates the complexity is outperformed.

| Subcase | Paper | Total Complexity |
|---|---|---|
| full positive matrix | RAS method [16, 2008] | $\widetilde{O}(n^2 h^{1/2}/\varepsilon)$ ♭ |
| | Scaling0 | $\widetilde{O}(n^2 h^{1/3}/\varepsilon^{2/3})$ ♭ |
| | Scaling1 | $\widetilde{O}(n^2)$ |
| scaling factors poly bounded | RAS method [16, 2008] | $\widetilde{O}(mh^{1/2}/\varepsilon)$ ♭ |
| | Scaling0 | $\widetilde{O}(mh^{1/3}/\varepsilon^{2/3})$ ♭ |
| | Scaling1 | $\widetilde{O}(m + n^{3/2})$ |
| | Scaling2+Scaling3 | $\widetilde{O}(m + h^{1/2} n^{5/6})$ |
| general | RAS method [16, 2008] | $\widetilde{O}(mh^2/\varepsilon^2)$ ♭ |
| | Scaling0 | $\widetilde{O}(mn^{2/3} h^{1/3}/\varepsilon^{2/3})$ |
| | LSW method [22, 1998] | $\widetilde{O}(mn^5)$ ♭ |
| | ellipsoid [18, 1996] | $\widetilde{O}(n^4)$ ♭ |
| | interior point [5, 2004] | $\widetilde{O}(n^6)$ ♭ |
| | max flow [32, 2007] | $\geq \widetilde{\Omega}(mn^4)$ ♭ |
| | Scaling1 | $\widetilde{O}(mn + n^{5/2})$ |
| | Scaling0+Scaling3 | $\widetilde{O}(mn + n^{\frac{7}{3}} + mn^{\frac{1}{3}} h^{\frac{1}{2}})$ |

Table 2: $(r, c)$-scaling.

- Following [16], we assume $r$ and $c$ are positive integral vectors and $h = \|r\|_1 = \|c\|_1$. Obviously $h \geq n$.
- Since the complexity of maximum flow is at least $\Omega(m)$, we present a complexity lower bound to [32].

exactly $(r, c)$-scalable with polynomial scaling factors.[7] We summarize them in Table 1 and 2.

Other algorithmic approaches were also developed for matrix scaling. The results [12, 24, 25, 33] proved asymptotic convergence without giving complexity bounds. Kalantari and Khachiyan [18] used the ellipsoid method, obtaining the first poly-logarithmic dependence on the approximation parameter $\varepsilon$, with total complexity $\widetilde{O}(n^4)$.[8] Balakrishnan [5] used interior point method and obtained a total complexity $\widetilde{O}(n^6)$. Rote and Zachariasen [32] reduced the $(r, c)$-scaling problem to running $\widetilde{O}(n^4)$ instances of mincost maximum flow. Linial *et al.* [22] proposed the first strongly polynomial algorithm with a total complexity $\widetilde{O}(n^7)$.

## 1.2 Our Improvements Over Known Results

We propose four algorithms to tackle the general matrix scaling problem and also some special cases. In all cases, we have outperformed all relevant previous results, and in some cases our complexities are close to linear in terms of the input size.

To state our complexity bounds let us discuss the input conventions we use. We denote by $m$ the number of nonzero entries of $A$, and assume $m \geq n \geq d$. We assume all entries of $A$ are rational numbers with polynomial sizes (i.e., at most $\mathsf{poly}(n)$ in numerators and denominators[9]), and both $r$ and $c$ are positive integral vectors with entries at most $\mathsf{poly}(n)$. Let $h \stackrel{\text{def}}{=} \|r\|_1 = \|c\|_1 \geq n$.

---
[7]That is, $A$ can be $(r, c)$-scaled with diagonal scaling matrices $X, Y$ where each $X_{i,i}$ and $Y_{j,j}$ are in $\left[\frac{1}{\mathsf{poly}(n)}, \mathsf{poly}(n)\right]$.
[8]Throughout the paper, we use $\widetilde{O}, \widetilde{\Omega}$ and $\widetilde{\Theta}$ notions to hide polylogarithmic factors in $n$ and $1/\varepsilon$.
[9]More generally, the complexities scale linear with the bit-size of the matrix entries.

A complete listing of our results appear in Table 1 and Table 2.

Our `Scaling0` can be viewed as an accelerated version of RAS. Its total complexity is

$$\widetilde{O}(mn/\varepsilon^{2/3}) \text{ for the } (\mathbb{1},\mathbb{1})\text{-scaling, or } \widetilde{O}(mn^{2/3}h^{1/3}/\varepsilon^{2/3}) \text{ for the general } (r,c)\text{-scaling.}$$

This improves the best result of RAS by a factor $\varepsilon^{-4/3}$ for $(\mathbb{1},\mathbb{1})$-scaling, and a factor $h^{5/3}n^{-2/3}\varepsilon^{-4/3} \geq n\varepsilon^{-4/3}$ for $(r,c)$-scaling. We stress that even testing scalability requires $\varepsilon < 1/n$ (see [22]) so reducing the $\varepsilon$ dependency from $\varepsilon^{-2}$ to $\varepsilon^{-2/3}$ (and later to $\mathsf{polylog}(1/\varepsilon)$) is very meaningful.

In the $\mathsf{polylog}(1/\varepsilon)$ complexity regimes, our `Scaling1` and `Scaling3` have complexities

$$\widetilde{O}(mn + n^{7/3}) \text{ for } (\mathbb{1},\mathbb{1})\text{-scaling, or } \widetilde{O}(mn + \min\{n^{5/2}, n^{7/3} + mn^{1/3}h^{1/2}\}) \text{ for } (r,c)\text{-scaling.}$$

If $A$ is $(r,c)$-scalable with polynomially large scaling factors,[10] our complexities reduce to

$$\widetilde{O}(m + n^{4/3}) \text{ for } (\mathbb{1},\mathbb{1})\text{-scaling, or } \widetilde{O}(m + \min\{n^{3/2}, h^{1/2}n^{5/6}\}) \text{ for } (r,c)\text{-scaling.}$$

**Our Approaches.** We have four algorithms `Scaling0`, `Scaling1`, `Scaling2`, and `Scaling3`, all based on tailor-made first and second-order techniques in continuous optimization. We also combine graph sparsification, Laplacian linear system solvers, and multiplicative weight updates into the optimization process. We now elaborate more on how this is done.

Matrix scaling can be written (in several ways) as the solution to convex optimization problems. We focus on a specific convex objective in this paper, which is the log of the capacity function [13]:

$$f(x) \stackrel{\text{def}}{=} \sum_{i=1}^{d} r_i \log \left( \sum_{j \in [n]} A_{i,j} e^{x_j} \right) - c^\top x \ . \tag{1.1}$$

If $A$ is asymptotically scalable, then the (approximate) minimizer of $f(x)$ corresponds to scaling matrices $X, Y$ such that $XAY$ is an $\varepsilon$-approximate $(r,c)$-matrix (see Proposition 2.3). A similar objective was also studied by Kalantari *et al.* [16].

At a high level, `Scaling0` uses first-order optimization techniques to minimize $f(x)$, and all other methods `Scaling1`, `Scaling2`, and `Scaling3` use a *mixture* of first and second order techniques.

FIRST-ORDER FRAMEWORK. It was known that the RAS method can be viewed as a first-order method [16], but only with convergence rate $1/\varepsilon^2$. Since $f(x)$ is not Lipschitz smooth (i.e., $\nabla^2 f(x)$ does not have a bounded spectral norm), one cannot apply generic optimization methods. We propose first-order building blocks that are specific to the matrix scaling problem, and then use the linear coupling framework of [4] to combine gradient and mirror descent, in order to achieve the $1/\varepsilon^{2/3}$ convergence rate. We call this method `Scaling0` and it outperforms the RAS method in all relevant parameter regimes. Note that `Scaling0` is as simple to implement as the RAS method.

SECOND-ORDER FRAMEWORK. It turns out the Hessian $\nabla^2 f(x)$ is always Laplacian, so one can invert it efficiently using modern SDD linear system solvers and graph sparsification techniques. This gives hope for designing efficient second-order methods. Unfortunately, $f(x)$ is not self-concordant in the entire space so we cannot apply standard second-order methods (e.g. Newton method). Instead, we show $f(x)$ satisfies a special property: the second-order Taylor approximation of $f(x + \delta)$ at point $x$ is accurate for all vector $\delta$ with $\|\delta\|_\infty \leq 1/8$:

$$f(x) + \langle \nabla f(x), \delta \rangle + \tfrac{1}{6}\delta^\top \nabla^2 f(x)\delta \leq f(x + \delta) \leq f(x) + \langle \nabla f(x), \delta \rangle + \delta^\top \nabla^2 f(x)\delta \ .$$

This implies if we can repeatedly minimize

$$f(x) + \langle \nabla f(x), \delta \rangle + \tfrac{1}{6}\delta^\top \nabla^2 f(x)\delta \quad \text{over an } \ell_\infty \text{ constraint on } \delta, \tag{1.2}$$

---

[10]Namely, when $A$ can be scaled to an $(r,c)$ matrix with diagonal scaling matrices $X, Y$ that satisfying each $X_{i,i}$ and $Y_{j,j}$ are within $\left[\frac{1}{\mathsf{poly}(n)}, \mathsf{poly}(n)\right]$. This condition is satisfied at least when all entries of $A$ are within $\left[\frac{1}{\mathsf{poly}(n)}, \mathsf{poly}(n)\right]$.

and update $x \leftarrow x + \frac{1}{6}\delta$, then we can have an $\log(1/\varepsilon)$ convergence rate as opposed to $1/\mathsf{poly}(\varepsilon)$.

Our `Scaling1` algorithm uses multiplicative weight update to solve (1.2), our `Scaling2` algorithm uses accelerated gradient descent to solve (1.2), and our final and most involved algorithm `Scaling3` uses more advanced multiplicative weight update in combination with first-order techniques to solve (1.2). We remark here that `Scaling3` needs a warm-start, that is, a point $x$ where $f(x) - \inf_x\{f(x)\}$ is sufficiently small. We use `Scaling0` or `Scaling2` to find such a warm-start.

**A Parallel Work.** When preparing this paper we found out that Aleksander Madry and his students obtained similar results for the same problem. The two works are completely independent, except on a psychological level.[11]

**A Related Problem.** In the matrix balancing problem, a symmetric matrix $B \in \mathbb{R}^{n \times n}$ is $\ell_p$-balanced if the $\ell_p$-norm of its $i$-th row equals that of its $i$-th column, for every $i \in [n]$. Given any $A \in \mathbb{R}^{n \times n}$, we wish to find a diagonal matrix $D$ with positive diagonal entries, such that $B = DAD^{-1}$ is $\ell_p$-balanced. Our techniques in this paper can also be extended to matrix balancing.

## 1.3 Roadmap

In Section 2, we discuss preliminaries. In Section 3, we show diameter bounds for the scaling parameters. In Section 4, we present our first-order framework and algorithm `Scaling0`. In Section 5, we present our second-order framework. In Section 6, 7, and 8 respectively, we introduce our algorithms `Scaling1`, `Scaling2`, and `Scaling3`. Throughout this paper, we assume exact arithmetic operations for presenting the cleanest proofs; we discuss how to use logarithmic bit-length to implement our algorithms in Section 9. Most of the proofs are in the appendix.

## 2 Notations and Preliminaries

Throughout the paper, we denote by $\|v\|_p$ the $p$-norm of vector $v$ if $p \in [1, +\infty]$, and $\|v\|$ the Euclidean norm of $v$ when it is clear from the context. We denote by $\|v\|_w \stackrel{\text{def}}{=} \left(\sum_{i=1}^n w_i v_i^2\right)^{1/2}$ the $w$-normalized Euclidean norm of vector $v$ if $w$ is a positive vector. We denote by $\|v\|_\mathbf{A} = (v^\top \mathbf{A} v)^{1/2}$ the matrix-Euclidean norm. We denote by $e^v = (e^{v_i})_i$, $\log(v) = (\log v_i)_i$, and $v^{-1} = (v_i^{-1})_i$ the component-wise exponentiation, logarithm, and inversion for vector $v$. Given vectors $u, v$, we denote by $u \leq v$ the relationship that $u_i \leq v_i$ for all coordinates $i$.

Given symmetric matrices $\mathbf{M}$ and $\mathbf{N}$, we write $\mathbf{M} \preceq \mathbf{N}$ if $\mathbf{N} - \mathbf{M}$ is positive semidefinite (PSD). We say a matrix $\mathbf{M}$ is Laplacian if (1) $\mathbf{M}$ is symmetric, (2) $\mathbf{M}_{i,j} \leq 0$ for $i \neq j$, and (3) $\mathbf{M}_{i,i} = -\sum_{j \neq i} \mathbf{M}_{i,j}$. It satisfies $v^\top \mathbf{M} v = \sum_{i<j} |\mathbf{M}_{i,j}|(v_i - v_j)^2$ for every vector $v$. We say a matrix $\mathbf{M}$ is symmetric diagonally dominant (or SDD for short) if (1) $\mathbf{M}$ is symmetric and (2) $\mathbf{M}_{i,i} \geq \sum_{j \neq i} |\mathbf{M}_{i,j}|$. Obviously, a Laplacian matrix is SDD; and an SDD matrix is PSD.

Throughout this paper, $\mathbf{A} \in \mathbb{R}_{\geq 0}^{d \times n}$ is non-negative and of dimensions $d \times n$. We denote by $m$ the number of non-zero entries of $\mathbf{A}$. Without loss of generality, we assume $d \leq n \leq m$ and the maximum entry of each row of $\mathbf{A}$ is exactly 1. We denote by $\mathbf{A}_i$ the $i$-th row vector of $\mathbf{A}$. We assume all the positive entries of $\mathbf{A}$ are in the range $\left[\frac{1}{\mathsf{poly}(n)}, 1\right]$. We also assume $r \in \mathbb{R}_{>0}^d$ and $c \in \mathbb{R}_{>0}^n$ are positive integral vectors and each $r_i, c_j \in \{1, 2, \ldots, \mathsf{poly}(n)\}$.[12] We let $h \stackrel{\text{def}}{=} \|r\|_1 = \|c\|_1$.

**Definition 2.1.** *Given $r \in \mathbb{R}_{>0}^d, c \in \mathbb{R}_{>0}^n$ and $\mathbf{A} \in \mathbb{R}_{\geq 0}^{d \times n}$, and denote by $r' \in \mathbb{R}_{\geq 0}^d$ (resp. $c' \in \mathbb{R}_{\geq 0}^d$) the vector of row sums (resp. column sums) of $\mathbf{A}$. We say*

---

[11] At the time we heard of their work, our bound was $O(mn^{1/3})$. They told us that they believed they could obtain $O(m + n^{4/3})$, which undoubtedly pushed us to discover that our techniques actually yield the same bound as well.

[12] This assumption was also made for instance by Kalantari *et al.* [16].

- **A** *is an $(r,c)$-matrix if $r = r'$ and $c = c'$.*
- **A** *is an $\varepsilon$-approximate $(r,c)$-matrix if $r' = r$ and $\|c' - c\|_{c^{-1}}^2 = \sum_{j \in [n]} c_j^{-1}(c'_j - c_j)^2 \leq \varepsilon^2$.*[13]
- **A** *is $(r,c)$-scalable if there exists positive diagonal matrices $\mathbf{X}, \mathbf{Y}$ so $\mathbf{XAY}$ is an $(r,c)$-matrix.*
- **A** *is asymptotically $(r,c)$-scalable if for every $\varepsilon > 0$, there exists positive diagonal matrices $\mathbf{X}, \mathbf{Y}$ so that $\mathbf{XAY}$ is an $\varepsilon$-approximate $(r,c)$-matrix.*

It is known that the existence of $(r,c)$-scaling can be characterized by the following proposition.

**Proposition 2.2** ([33]). *A non-negative matrix $\mathbf{A} \in \mathbb{R}_{\geq 0}^{d \times n}$ is exactly $(r,c)$-scalable if and only if $\|r\|_1 = \|c\|_1$ and for every zero minor $R \times C \subseteq [d] \times [n]$ of $\mathbf{A}$,*

1. *$\sum_{i \in [d] \setminus R} r_i \geq \sum_{j \in C} c_j$ or equivalently $\sum_{i \in R} r_i \leq \sum_{j \in [n] \setminus C} c_j$.*
2. *Equality in 1 holds iff the minor $([d] \setminus R) \times ([d] \setminus C)$ is all zero as well.*

*A nonnegative matrix $\mathbf{A}$ is asymptotically $(r,c)$-scalable if condition 1 holds.*

**Proposition 2.3.** *Our objective $f(x)$ in (1.1) is convex and*

- $\nabla_j f(x) = \sum_{i=1}^d \frac{r_i \mathbf{A}_{i,j}}{\langle \mathbf{A}_i, e^x \rangle} e^{x_j} - c_j$.
- *If $\|\nabla f(x)\|_{c^{-1}}^2 \leq \varepsilon$, then $\left(\frac{r_i \mathbf{A}_{i,j} \cdot e^{x_j}}{\langle \mathbf{A}_i, e^x \rangle}\right)_{i,j}$ is an $\varepsilon$-approximate $(r,c)$-matrix.*
- *If $\mathbf{A}$ is exactly $(r,c)$-scalable, then there exists $x^*$ so that $f(x^*) = \min_x\{f(x)\}$ and $\nabla f(x^*) = 0$.*
- *If $\mathbf{A}$ is asymptotically $(r,c)$-scalable, then $\inf_x\{f(x)\} > -\infty$.*
- **A** *is not asymptotically $(r,c)$-scalable if and only if $\inf_x\{f(x)\} = -\infty$.*

## 3 New Bounds on Scaling Parameters

We first recall a few bounds for $(r,c)$-scalable matrices that are essentially from prior work.

**Lemma 3.1** (objective bound). *For every $x$ satisfying $\|x\|_\infty \leq N$, we have $f(0) - f(x) \leq 2hN$.*

*Proof of Lemma 3.1.* Denoting $x' = x + \|x\|_\infty \mathbb{1}$, we know that $f(x') = f(x)$ and $\|x'\|_\infty \leq 2\|x\|_\infty$. On the other hand, since for every $i \in [d]$, we have $\langle \mathbf{A}_i, e^{x'} \rangle \geq \langle \mathbf{A}_i, \mathbb{1} \rangle \geq 1$, it satisfies $f(0) - f(x') \leq c^\top x' \leq 2hN$. □

**Lemma 3.2** (diameter bound). *If $\mathbf{A}$ is exactly $(r,c)$ scalable, and all non-zero entries of $\mathbf{A}$ are within $[\nu, 1]$ for some $\nu > 0$. Then, the following holds:*

1. *If $\mathbf{A}$ is full then there exists a minimizer $x^*$ of $f(x)$ such that $\|x^*\|_\infty \leq \ln \frac{hn}{\nu}$.*
2. *If $\mathbf{A}$ is not full, then there exists a minimizer $x^*$ of $f(x)$ such that $\|x^*\|_\infty \leq (h + 1/2) \ln \frac{h}{\nu}$.*

In this paper, we improve (the second item of) Lemma 3.2 in two aspects. First, we allow $\mathbf{A}$ to be *asymptotically* $(r,c)$-scalable. Second, we improve the diameter bound from $\widetilde{O}(h)$ to $\widetilde{O}(n)$ for arbitrary $(r,c)$. (Recall that $r$ and $c$ are integral so $h \geq n$.)

**Lemma 3.3** (diameter bounds for the asymptotic case). *If $\mathbf{A}$ is asymptotically $(r,c)$-scalable, and all non-zero entries of $\mathbf{A}$ are within $[\nu, 1]$ for some $\nu > 0$, then, for every $\varepsilon > 0$, there exists $x^*_\varepsilon \in \mathbb{R}^n$ such that*

$$\|x^*_\varepsilon\|_\infty = O\left(n \ln \frac{nh}{\nu\varepsilon}\right) \, , \quad \|\nabla f(x^*_\varepsilon)\|_\infty \leq \varepsilon \, , \quad \text{and} \quad f(x^*_\varepsilon) - \inf_x\{f(x)\} \leq \varepsilon \, .$$

---
[13] In certain literature people have also used $\|c' - c\|_2^2 \leq \varepsilon^2$ as the definition of $\varepsilon$-approximation [16]. However, their performance loses a factor of $\|c\|_\infty$ so we used this $\|\cdot\|_{c^{-1}}$ notation to simplify our and their statements.

One can verify that Lemma 3.3 is tight (up to constant factors) for instance when $\mathbf{A}$ is a square upper-triangular matrix and the diagonal of $\mathbf{A}$ equals $r = c$.

## 4 A New First-Order Framework

In this section, we minimize $f(x)$ using a specially designed first-order optimization method, and finds an $\varepsilon$-approximate $(r, c)$-scaling with a total complexity that scales with $\varepsilon^{-2/3}$.

**High-Level Intuition.** We first illustrate why the convergence rate $\varepsilon^{-2/3}$ is reasonable from an optimization standpoint. Recall that if we are given a convex function $g(x)$ that is $O(1)$-Lipschitz smooth —meaning that its Hessian $\nabla^2 g(x)$ has a bounded spectral norm— then, using accelerated gradient descent [26, 27], one can find a point $x_1$ satisfying $g(x_1) - g(x^*) \leq O\big(\frac{\|x^*\|_2^2}{T^2}\big)$ in $T$ iterations, where $x^*$ is a minimizer of $g(x)$. At the same time, also recall that each step of gradient descent $x' = x - \nabla g(x)$ decreases the objective by at least $g(x) - g(x') \geq \frac{1}{2}\|\nabla g(x)\|_2^2$, so we can apply another $T$ steps of gradient descent on top of $x_1$, and obtain a point $x_2$ satisfying $\|\nabla g(x_2)\|_2^2 \leq O\big(\frac{\|x^*\|_2^2}{T^3}\big)$. In other words, we reach $x_2$ with $\|\nabla g(x_2)\|_2^2 \leq \varepsilon^2$ in $T \propto \varepsilon^{-2/3}$ iterations.

Unfortunately, the function $f(x)$ we are dealing in this paper is not Lipschitz smooth, so we cannot apply the above approach. This is also why previous results using first-order techniques only achieve $1/\varepsilon^2$ rate in general and $1/\varepsilon$ rate in some special cases (see Table 1 and 2).[14]

Instead, we use the linear-coupling framework of [4] to recover this $\varepsilon^{-2/3}$ convergence rate without using smoothness. To apply linear coupling, we need to design (1) a problem-specific gradient descent step, (2) a problem-specific mirror descent step, and then (3) linearly combine the analysis of the two for a faster convergence. Furthermore, in order to ensure that the output lies in an infinite-norm box, we have to ensure that gradient and mirror steps both work inside bounded convex region. This adds some extra difficulty in the analysis.

**Roadmap.** We introduce our gradient and mirror descent steps in Section 4.1 and 4.2 respectively, and present our linear coupling method LC in Algorithm 1, and analyze it in Section 4.3. In Section 4.4 we build our algorithm Scaling0 using LC as a subroutine, and present the final theorems. We introduce the following notion for convenience:[15]

**Definition 4.1** (gradient split). *At any $x \in \mathbb{R}^n$, define small and large gradients $\nabla^{\sf s}, \nabla^{\sf l} \in \mathbb{R}^n$ by*

$$\forall j \in [n]: \quad \nabla_j^{\sf s} \stackrel{\text{def}}{=} \min\{c_j, \nabla_j f(x)\} \in [-c_j, c_j] \quad \text{and} \quad \nabla_j^{\sf l} \stackrel{\text{def}}{=} \nabla_j f(x) - \nabla_j^{\sf s} \geq 0 \ .$$

*Also, define small and large coordinates $\Lambda^{\sf s}, \Lambda^{\sf l} \subseteq [n]$ by*

$$\Lambda^{\sf s} \stackrel{\text{def}}{=} \{j \in [n] \colon \nabla_j \in [-c_j, c_j]\} \quad \text{and} \quad \Lambda^{\sf l} \stackrel{\text{def}}{=} [n] \setminus \Lambda^{\sf s} = \{j \in [n] \colon \nabla_j > c_j\} \ .$$

### 4.1 A Specific Gradient Descent

We now introduce a problem-specific gradient descent. Recall that when analyzing a smooth function $g(x)$, one can show a quadratic lower bound

$$g(x) - g(x + \delta) \geq Q(x, \delta) \stackrel{\text{def}}{=} -\langle \nabla g(x), \delta \rangle - \tfrac{1}{2}\|\delta\|_2^2 \ ,$$

and thus choosing $\delta = \arg\max_\delta \{Q(\delta)\} = -\nabla g(x)$ gives a decrease $g(x) - g(x + \delta) \geq \tfrac{1}{2}\|\nabla g(x)\|_2^2$.

For our function $f(x)$, we show a similar quadratic lower bound:

---

[14]For instance, the RAS method can be viewed as performing a gradient descent step $x' = x - \nabla f(x)$ [16].

[15]Recall that each coordinate $\nabla_j f(x)$ is in the interval $[-c_j, \infty)$. This gradient splitting technique was earlier introduced to solve positive linear programming and semidefinite programming [1–3].

**Lemma 4.2.** *Given $x \in \mathbb{R}^n$, denote by $\nabla = \nabla f(x)$ and $\Lambda^{\mathsf{s}}, \Lambda^{\mathsf{l}} \subseteq [n]$ the set of small and large coordinates (see Def. 4.1). Then, for every $\delta \in \mathbb{R}^n$ where $\|\delta\|_\infty \leq 1/2$, we have*

- *if $\delta \geq 0$, then $f(x) - f(x+\delta) \geq Q^+(x,\delta) \stackrel{\text{def}}{=} \sum_{j \in \Lambda^{\mathsf{s}}} \left( -\nabla_j \cdot \delta_j - \frac{4}{3} c_j \cdot \delta_j^2 \right) + \sum_{j \in \Lambda^{\mathsf{l}}} \left( -\frac{7}{3} \nabla_j \cdot \delta_j \right)$.*
- *if $\delta \leq 0$, then $f(x) - f(x+\delta) \geq Q^-(x,\delta) \stackrel{\text{def}}{=} \sum_{j \in \Lambda^{\mathsf{s}}} \left( -\nabla_j \cdot \delta_j - \frac{4}{3} c_j \cdot \delta_j^2 \right) + \sum_{j \in \Lambda^{\mathsf{l}}} \left( -\frac{1}{2} \nabla_j \cdot \delta_j \right)$.*

*(Recall that $\delta \geq 0$ or $\delta \leq 0$ means entry-wise non-negativity or non-positivity.)*

The above quadratic lower bounds distinct from the classical one $Q(x, \delta)$ in two aspects. First, for large coordinates $j \in \Lambda^{\mathsf{l}}$, we only have a linear lower bound. Second, $Q^+$ and $Q^-$ have different forms for $\delta \geq 0$ and $\delta \leq 0$. Here is an explanation for such two distinctions. Consider even a simple univariate function $h(x) = e^x - 1$. First, we do not have $h(0) - h(\delta) \geq -h'(0)\delta - C\delta^2$ for any constant $C$, so we cannot have a quadratic lower bound. Second, we only have $h(0) - h(\delta) \geq -\frac{7}{3} h'(0)\delta$ for $\delta \geq 0$ and $h(0) - h(\delta) \geq -\frac{1}{2} h'(0)\delta$ for and $\delta \leq 0$. The two constants $\frac{7}{3}$ and $\frac{1}{2}$ must be distinct for $\delta \geq 0$ and $\delta \leq 0$. For such reasons, we only have statements for vectors $\delta \geq 0$ or $\delta \leq 0$..

Lemma 4.2 suggests us to perform gradient descent as (one of) the minimizer of $Q^+$ and $Q^-$:

**Definition 4.3** (gradient descent). *Given $x$ satisfying $\|x\|_\infty \leq N$, define the projected gradient descent step $x' \leftarrow \mathtt{Grad}^N(x)$ where $\mathtt{Grad}^N(x) \stackrel{\text{def}}{=} \arg\min_{y \in \{y_1, y_2\}} \{f(y)\}$ where*

$$y_1 = x + \arg\max_{\delta \in \Omega^+_{N,x}} \left\{Q^+(x,\delta)\right\} \quad \text{and} \quad \Omega^+_{N,x} \stackrel{\text{def}}{=} \left\{\delta \geq 0 \mid \|x + \delta\|_\infty \leq N \wedge \|\delta\|_\infty \leq 1/2\right\}$$

$$y_2 = x + \arg\max_{\delta \in \Omega^-_{N,x}} \left\{Q^-(x,\delta)\right\} \quad \text{and} \quad \Omega^-_{N,x} \stackrel{\text{def}}{=} \left\{\delta \leq 0 \mid \|x + \delta\|_\infty \leq N \wedge \|\delta\|_\infty \leq 1/2\right\}$$

*Obviously, $\mathtt{Grad}^N(x)$ can be computed in complexity $O(n + m)$.*

Note that in the definition above, we have specified a parameter $N$ which ensures that the output $x' = \mathtt{Grad}^N(x)$ is also in the box $\|x'\|_\infty \leq N$. One can also let $N = +\infty$ and this means that we put no constraint on $\|x'\|_\infty$. The next two are direct corollaries of Lemma 4.2:

**Corollary 4.4.** *If $x' = \mathtt{Grad}^N(x)$, then we have*

$$f(x) - f(x') \geq \frac{1}{2}\left(\max_{\delta \in \Omega^+_{N,x}} \{Q^+(x,\delta)\} + \max_{\delta \in \Omega^-_{N,x}} \{Q^-(x,\delta)\}\right) \geq 0 \ .$$

**Corollary 4.5.** *If $x' = \mathtt{Grad}^\infty(x)$ and $\nabla f(x) = \nabla^{\mathsf{s}} + \nabla^{\mathsf{l}}$ (see Def. 4.1), we have*

$$\|x' - x\|_\infty \leq 1/2 \quad \text{and} \quad f(x) - f(x') \geq \tfrac{3}{32} \|\nabla^{\mathsf{s}}\|^2_{c^{-1}} + \tfrac{1}{4} \|\nabla^{\mathsf{l}}\|_1 \geq \Omega\bigl(\|\nabla^{\mathsf{s}}\|^2_{c^{-1}} + \|\nabla^{\mathsf{l}}\|_1\bigr) \ .$$

*Remark* 4.6. Corollary 4.5 replaces the classical gradient descent statement on smooth functions $g(x)$ that says $g(x) - g(x') \geq \frac{1}{2} \|\nabla g(x)\|_2^2$. Corollary 4.4 is the constrained version of Corollary 4.5.

### 4.2 A Specific Mirror Descent

The mirror descent step we take is a constrained minimization with respect to the $\|\cdot\|_c^2$ norm:

**Definition 4.7** (mirror descent). *Given $z$ satisfying $\|z\|_\infty \leq N$, a feedback vector $v \in \mathbb{R}^n$, define the projected mirror descent step $z' \leftarrow \mathtt{Mirr}^N(z, v)$ as*

$$\mathtt{Mirr}^N(z, v) \stackrel{\text{def}}{=} \arg\min_{\|z'\|_\infty \leq N} \left\{\langle v, z'\rangle + \tfrac{1}{2} \|z' - z\|_c^2\right\} \ .$$

*Obviously, $\mathtt{Mirr}^N(z, v)$ can be computed in complexity $O(n)$.*

The following lemma is classical for mirror descent:

**Lemma 4.8.** *If $z' = \mathtt{Mirr}^N(z, v)$, then for every $u$ satisfying $\|u\|_\infty \leq N$, we have*

$$\langle v, z - u\rangle \leq \langle v, z - z'\rangle - \tfrac{1}{2}\|z - z'\|_c^2 + \tfrac{1}{2}\|z - u\|_c^2 - \tfrac{1}{2}\|z' - u\|_c^2 \ .$$

**Algorithm 1** $\mathtt{LC}(\mathbf{A}, N, T, y_0)$

**Input:** $\mathbf{A} \in \mathbb{R}^{d \times n}$, a non-negative matrix; $N \geq 1$, a diameter bound; $T \geq 1$, number of iterations; $y_0 \in \mathbb{R}^n$ a starting vector satisfying $\|y_0\|_\infty \leq 15N$;

1: $z_0 \leftarrow 0$ and $\tau_0 \leftarrow \frac{1}{32N}$;
2: **for** $k = 0$ **to** $T - 1$ **do**
3:     $\tau_k \leftarrow$ the unique positive root of the quadratic equation $\frac{\tau_k^2}{\tau_{k-1}^2} + \tau_k - 1 = 0$;
4:     $x_{k+1} \leftarrow \tau_k z_k + (1 - \tau_k) y_k$;      ⋄ $\tau_k \in (0, 1)$
5:     $y_{k+1} \leftarrow \mathtt{Grad}^{15N}(x_{k+1})$;      ⋄ see Def. 4.3
6:     Define $\nabla^{\mathsf{s}} \in \mathbb{R}^n$ where $\nabla^{\mathsf{s}}_j \leftarrow \min\{\nabla_j f(x_{k+1}), 1\}$;
7:     $z_{k+1} \leftarrow \mathtt{Mirr}^N(z_k, \alpha_k \nabla^{\mathsf{s}})$ where $\alpha_k = \frac{3}{64\tau_k}$;      ⋄ see Def. 4.7
8: **end for**
9: **return** $y_T$.      ⋄ $y_T$ satisfies $\|y_T\|_\infty \leq 15N$

## 4.3 Linear Coupling

We now introduce our linear-coupling algorithm $\mathtt{LC}$ (see Algorithm 1). Starting from two initial vectors $y_0$ and $z_0 = 0$, in each iteration $k = 0, 1, \ldots, T-1$, our $\mathtt{LC}$ chooses a linear combination $x_{k+1} = \tau_k z_k + (1 - \tau_k) y_k$ for some parameter $\tau_k \in (0, 1)$, and performs two updates: $y_{k+1} = \mathtt{Grad}^{15N}(x_{k+1})$ and $z_{k+1} = \mathtt{Mirr}^N(z_k, \alpha_k \nabla^{\mathsf{s}})$. Here, $\alpha_k > 0$ is the learning rate for mirror descent. The choices of $\tau_k$ and $\alpha_k$ are in Algorithm 1. From the description:

**Fact 4.9.** We always have $\|z_k\|_\infty \leq N$, $\|x_k\|_\infty \leq 15N$, and $\|y_k\|_\infty \leq 15N$.

*Proof of Fact 4.9.* $y_0$ and $z_0 = 0$ both satisfy norm bounds. $y_k$ comes from gradient descent with range $15N$ so $\|y_k\|_\infty \leq 15N$; $z_K$ comes from mirror descent with range $N$ so $\|z_k\|_\infty \leq N$; finally, $x_k$ is a convex combination of $y_{k-1}$ and $z_{k-1}$ so satisfies $\|x_k\|_\infty \leq 15N$. □

We show the following lemma which describes the one-iteration behavior of $\mathtt{LC}$:

**Lemma 4.10.** *If* $\tau_k \alpha_k \leq 3/64$, $\tau_k \in \left(0, \frac{1}{32N}\right]$, *and* $u$ *is any vector satisfying* $\|u\|_\infty \leq N$, *then*
$$0 \leq \frac{1-\tau_k}{\tau_k}\big(f(y_k) - f(u)\big) - \frac{1}{\tau_k}\big(f(y_{k+1}) - f(u)\big) + \frac{1}{2\alpha_k}\|z_k - u\|_c^2 - \frac{1}{2\alpha_k}\|z_{k+1} - u\|_c^2 \ .$$

Lemma 4.10 is the main technical contribution of this section, and relies on careful applications of Lemma 4.2 and Lemma 4.8, together with tailor-made analysis for our $f(x)$. The next theorem is a corollary of Lemma 4.10 by appropriate choices $\tau_k$ and $\alpha_k$, and telescoping $k = 0, 1, \ldots, T-1$.

**Theorem 4.11** ($\mathtt{LC}$). *If* $y_0$ *satisfies* $\|y_0\|_\infty \leq 15N$ *and* $T \geq 1$, *then the output* $y_T = \mathtt{LC}(\mathbf{A}, N, T, y_0)$ *(see Algorithm 1) satisfies that for every* $u \in \mathbb{R}^n$ *and* $\|u\|_\infty \leq N$:
$$\|y_T\|_\infty \leq 15N \quad \text{and} \quad f(y_T) - f(u) \leq O\left(\frac{N^2\big(f(y_0) - f(u) + h\big)}{(N+T)^2}\right) \ .$$

## 4.4 Complexity Statements

The $\frac{N^2}{(N+T)^2}\big(f(y_0) - f(u)\big)$ term in Theorem 4.11 can hurt the performance of $\mathtt{LC}$.[16] For this reason, as a warm start, one needs to repeatedly apply $\mathtt{LC}$ for $\log N$ times, each with $T = \Theta(N)$. We summarize this final algorithm as $\mathtt{Scaling0}$ in Algorithm 2 and present the final theorem:

---
[16] For instance, the general upper bound on $f(0) - f(x^*)$ is only $\widetilde{O}(Nh)$ (see Lemma 3.1).

**Algorithm 2** `Scaling0(A, N, T)`

**Input:** $\mathbf{A} \in \mathbb{R}^{d \times n}$, a non-negative matrix; $N \geq 1$, a diameter bound; $T \geq 1$, number of iterations;
1: $z_0 \leftarrow 0$;
2: **for** $k = 0$ **to** $\log N$ **do**
3:     $z_0 \leftarrow \text{LC}(\mathbf{A}, N, \Theta(N), z_0)$;
4: $z_1 \leftarrow \text{LC}(\mathbf{A}, N, T, z_0)$;
5: **for** $k = 1$ **to** $T$ **do**
6:     $z_{k+1} \leftarrow \text{Grad}^{\infty}(z_k)$;
7: $z \leftarrow \arg\min_{z \in \{z_1, \ldots, z_T\}} \{\|\nabla f(z)\|^2_{c^{-1}}\}$.
8: **return** $(z_1, z)$.     ⋄ $z_1$ satisfies $\|z_1\|_\infty \leq 15N$

---

**Theorem 4.12** (`Scaling0`). *If $N \geq 1$, then $(z_1, z) = \text{Scaling0}(\mathbf{A}, N, T)$ satisfies*

- *If $T \geq N$, then for every $u$ satisfying $\|u\|_\infty \leq N$, we have*

$$\|z_1\|_\infty \leq 15N \quad \text{and} \quad f(z_1) - f(u) \leq O\left(\frac{N^2 h}{T^2}\right) .$$

- *If $T \geq (N^2 h)^{1/3}$ and there exists $u$ so that $\|u\|_\infty \leq N$ and $f(u) - \inf_x\{f(x)\} \leq 1$, then*

$$\|\nabla f(z)\|^2_{c^{-1}} \leq O\left(\frac{N^2 h}{T^3}\right) .$$

*The total complexity of `Scaling0` is $O(m(N \log N + T))$.*

---

(Due to technical reasons, we do not have bound on $\|z\|_\infty$.)

Recall that to obtain an $\varepsilon$-approximate $(r, c)$-scaling, it suffices to find $z$ with $\|\nabla f(z)\|^2_{c^{-1}} \leq \varepsilon^2$ (see Proposition 2.3). Therefore, we can combine Theorem 4.12 with bounds on the scaling parameters: namely, $N \leq \widetilde{O}(n)$ for the general $(r, c)$-scaling (see Lemma 3.3), or $N \leq \widetilde{O}(1)$ if the scaling parameters are polynomially bounded (see Footnote 10). This gives us the claimed results of `Scaling0` in Table 1 and Table 2.

## 5 A New Second-Order Framework

In this section, we propose a second-order framework in order to minimize $f(x)$. Our methods `Scaling1`, `Scaling2` and `Scaling3` in subsequent sections are all be based on this framework.

We show that near any point $x$, the function value $f(x + \delta)$ is well approximated by the second-order Taylor expansion of $f(x)$, as long as $\|\delta\|_\infty \leq 1/8$:

**Lemma 5.1** (second-order approximation). *For every $x, \delta \in \mathbb{R}^n$ with $\|\delta\|_\infty \leq 1/8$, we have*

$$f(x) + \langle \nabla f(x), \delta \rangle + \frac{1}{6} \delta^\top \nabla^2 f(x) \delta \leq f(x + \delta) \leq f(x) + \langle \nabla f(x), \delta \rangle + \delta^\top \nabla^2 f(x) \delta .$$

Note that if $f(x)$ were an arbitrary convex function, such a quadratic approximation would only work for a very small region of $\delta$. It is the special property of the matrix scaling problem that allows us to prove Lemma 5.1 for all $\|\delta\|_\infty \leq 1/8$. We include the details in Appendix E.

Also, one may carefully verify that $\nabla^2 f(x)$ is a Laplacian matrix that may contain up to $n^2$ non-zero entries even if the original matrix $\mathbf{A}$ is sparse. Using classical graph sparsification techniques (see Appendix A.2), with total complexity $\widetilde{O}(m)$, one can find another Laplacian matrix $\mathbf{H} \in \mathbb{R}^{n \times n}$ satisfying $\mathbf{H} \preceq \nabla^2 f(x) \preceq 1.1\mathbf{H}$, where $\mathbf{H} only has $\widetilde{O}(n)$ non-zero entries.

**High-Level Intuition.** Using Lemma 5.1, it becomes natural to study the minimization question $\langle\nabla f(x),\delta\rangle + \frac{1}{6}\delta^\top \mathbf{H}\delta$ over all $\|\delta\|_\infty \leq 1/8$. If $\delta^*$ is such a minimizer, then one can show $f(x) - f(x+\delta^*) \geq \Omega\big(\frac{1}{\|x-x^*\|_\infty}\big)(f(x)-f(x^*))$ where $x^*$ is the minimizer of $f(x)$. This sounds like we only needed $O(N\log(1/\varepsilon))$ iterations in total if $\|x^*\|_\infty \leq N$.

Unfortunately, this approach fails because $\|x-x^*\|_\infty$ may increase by $1/8$ per iteration, so the convergence rate may drop to $1/\varepsilon$ as opposed to $\log(1/\varepsilon)$. We fix this issue by restricting our attention only to the region $\{x \in \mathbb{R}^n \mid \|x\|_\infty \leq N\}$. If this region contains $x^*$, and if we can minimize $\langle\nabla f(x),\delta\rangle + \frac{1}{6}\delta^\top \mathbf{H}\delta$ over the intersection of $\|x+\delta\|_\infty \leq N$ and $\|\delta\|_\infty \leq 1/8$, then we can always have $f(x)-f(x+\delta^*) \geq \Omega\big(\frac{1}{N}\big)(f(x)-f(x^*))$ and thus converge in $O(N\log(1/\varepsilon))$ iterations.

For the reason above, we wish to repeatedly solve the following minimization problem

$$\boxed{\min_{\delta \in \mathsf{box}^N(x)} \left\{\langle\nabla f(x),\delta\rangle + \frac{1}{6}\delta^\top \mathbf{H}\delta\right\}} \tag{5.1}$$

**Definition 5.2.** *Given any point $x \in \mathbb{R}^n$ satisfying $\|x\|_\infty \leq N$ for some $N > 1$, we define*

$$\mathsf{box}^N(x) \overset{\mathrm{def}}{=} \left\{\delta \in \mathbb{R}^n \,\Big|\, \|\delta - \alpha\|_\infty \leq \frac{1}{32}\right\} \quad \text{where} \quad \alpha_i \overset{\mathrm{def}}{=} \begin{cases} (\frac{1}{32} - N - x_i) \in (0, \frac{1}{32}], & \text{if } x_i - \frac{1}{32} < -N; \\ (N - x_i - \frac{1}{32}) \in [-\frac{1}{32}, 0), & \text{if } x_i + \frac{1}{32} > N; \\ 0, & \text{otherwise.} \end{cases}$$

**Fact 5.3.** *For all $\delta \in \mathsf{box}^N(x)$, we have $\|x+\delta\|_\infty \leq N$ and $\|\delta\|_\infty \leq \frac{1}{16}$. We also have $0 \in \mathsf{box}^N(x)$.*

Our next Lemma 5.4 says that if we can solve (5.1) up to a small additive error, then we can decrease the objective distance to $f(u)$ by a factor of $1 - \frac{1}{900N}$ up to the same small additive error.

**Lemma 5.4.** *Given $x$ with $\|x\|_\infty \leq N$ and $\mathbf{H}$ with $\mathbf{H} \preceq \nabla^2 f(x) \preceq 1.1\mathbf{H}$, the following holds:*
*(a) For any $u \in \mathbb{R}^n$ with $\|u\|_\infty \leq N$,*

$$-\min_{\delta \in \mathsf{box}^N(x)} \left\{\langle\nabla f(x),\delta\rangle + \frac{1}{6}\delta^\top \mathbf{H}\delta\right\} \geq \frac{1}{64N}\big(f(x)-f(u)\big)\ .$$

*(b) If we are given $\widehat{\delta}$ satisfying $\|\widehat{\delta}\|_\infty \leq 1/8$ and for $\varepsilon \geq 0$:*

$$\langle\nabla f(x),\widehat{\delta}\rangle + \frac{1}{6}\widehat{\delta}^\top \mathbf{H}\widehat{\delta} \leq \min_{\delta \in \mathsf{box}^N(x)}\left\{\langle\nabla f(x),\delta\rangle + \frac{1}{6}\delta^\top \mathbf{H}\delta\right\} + \varepsilon\ ,$$

*then it satisfies that for every $u \in \mathbb{R}^n$ with $\|u\|_\infty \leq N$, $f(x)-f\big(x+\frac{\widehat{\delta}}{6.6}\big) \geq \frac{1}{900N}\big(f(x)-f(u)\big)-\varepsilon$.*

## 6 Second-Order Method 1: via Multiplicative Weight Updates

In this section, we propose `Scaling1` which uses *multiplicative weight update (MWU)* and an $\ell_2$ constrained SDD system solver to tackle problem (5.1).

**High-Level Intuitions.** Denote by $h(\delta) \overset{\mathrm{def}}{=} \langle\nabla f(x),\delta\rangle + \frac{1}{6}\delta^\top \mathbf{H}\delta$ for notation simplicity.

Given any weight vector $w \in \Delta$ where $\Delta \overset{\mathrm{def}}{=} \{w \in [1/2, n]^n \mid \sum_i w_i = n\}$, instead of minimizing $h(\delta)$ over all $\delta \in \mathsf{box}^N(x) = \{\delta \in \mathbb{R}^n \mid \|\delta - \alpha\|_\infty \leq \frac{1}{32}\}$, we can minimize $h(\delta)$ over a larger set $\Omega_w = \{\delta \in \mathbb{R}^n \mid \|\delta - \alpha\|_w^2 \leq \frac{n}{1024}\}$.[17] We would like do so because $\ell_2$ constrained minimization is computationally cheap: minimizing $h(\delta)$ over $\Omega_w$ can be done using a variant of SDD linear system solvers in total complexity $\widetilde{O}(n)$ (see Appendix A.3).

Next, we wish to apply the multiplicative weight update framework. Starting from some $w_0 \in \Delta$, in each round $k = 0, 1, \ldots, T-1$, we minimize $h(\delta)$ over set $\Omega_{w_k}$ and let $\delta_k \in \Omega_{w_k}$ be an approximate minimizer. Then, we update $w_{k+1}$ from $w_k$ by penalizing the coordinates $i$ in $\delta_k$ where $|\delta_{k,i} - \alpha_i|$ is

---

[17]It is easy to verify that $\mathsf{box}^N(x) \subseteq \Omega_w$ and conversely $\|\delta - \alpha\|_\infty \leq O(\sqrt{n})$ for every $\delta \in \Omega_w$.

**Algorithm 3** `MWUbasic`$(\nabla, \mathbf{H}, \alpha, T, K, \varepsilon)$

---

**Input:** $\nabla \in \mathbb{R}^n$; $\mathbf{H} \in \mathbb{R}^{n \times n}$ a Laplacian matrix; $\alpha \in \mathbb{R}^n$ satisfying $\|\alpha\|_\infty \leq 1/32$; $T \geq 1$ number of rounds; $K \geq 1$ a parameter; $\varepsilon > 0$ an accuracy parameter.

1: $\Delta \leftarrow \{w \in [1/2, n]^n \colon \sum_i w_i = n\}$ and $w_0 \leftarrow (1, 1, \ldots, 1) \in \Delta$;
2: **for** $k = 0$ **to** $T - 1$ **do**
3:   Use Lemma A.4 to find a vector $\delta_k \in \mathbb{R}^n$ satisfying $\|\delta_k - \alpha\|_{w_k}^2 \leq \frac{n}{1024}$ and
$$\langle \nabla f(x), \delta_k \rangle + \tfrac{1}{6}\delta_k^\top \mathbf{H}\delta_k \leq \min_{\|\delta - \alpha\|_{w_k}^2 \leq n/1024} \{\langle \nabla f(x), \delta \rangle + \tfrac{1}{6}\delta^\top \mathbf{H}\delta\} + \varepsilon$$
4:   Define loss vector $\ell_k \in \mathbb{R}^n$ by $\ell_{k,i} \leftarrow -|\delta_{k,i} - \alpha_i|$.
5:   $w_{k+1} \leftarrow \arg\min_{z \in \Delta} \left\{\eta \langle \ell_k, z \rangle + \sum_{i \in [n]} \left(z_i \log \frac{z_i}{w_{k,i}} + w_{k,i} - z_i\right)\right\}$
     ⋄ *a multiplicative weight update with parameter $\eta = 1/(\sqrt{n} + K)$, see Section A.1*
6: **end for**
7: **return** $\overline{\delta} \leftarrow \frac{1}{T} \sum_{k=0}^{T-1} \delta_k$;

---

**Algorithm 4** `Scaling1`$(\mathbf{A}, N, \varepsilon)$

---

**Input:** $\mathbf{A} \in \mathbb{R}^{d \times n}$ non-negative matrix; $N \geq 1$ diameter bound; $\varepsilon \in (0, 1)$ accuracy parameter.

1: $x_0 \leftarrow 0$, $K \leftarrow \Theta(\log(1/\varepsilon))$, and $T \leftarrow \widetilde{\Theta}(\sqrt{n})$;
2: **for** $t = 0$ **to** $NK$ **do**
3:   Define $\mathsf{box}^N(x_t)$ and $\alpha \in [-1/32, 1/32]^n$ using Def. 5.2;
4:   $\mathbf{H} \leftarrow$ a matrix with $\widetilde{O}(n)$ nonzeros satisfying $\mathbf{H} \preceq \nabla^2 f(x_t) \preceq 1.1\mathbf{H}$;
5:   $\overline{\delta} \leftarrow$ `MWUbasic`$(\nabla f(x_t), \mathbf{H}, \alpha, T, K, \frac{\varepsilon}{900N})$;
6:   $x_{t+1} \leftarrow x + \frac{\overline{\delta}}{6.6}$ and $N \leftarrow N + \frac{1}{50K}$;    ⋄ *so $x_{t+1}$ satisfies $\|x_{t+1}\|_\infty \leq N$ for this new $N$*
7: **end for**
8: **return** $y \leftarrow$ the last $x_t$.

---

large. A variant of the MWU theory implies that, as long as $T = \widetilde{O}(\sqrt{n})$, the average $\overline{\delta} = \frac{1}{T}\sum_{k=0}^{T-1} \delta_k$ satisfies $\|\overline{\delta} - \alpha\|_\infty \leq O(1)$. At the same time, since objective $\delta_k$ minimizes (5.1) over a larger set $\Omega_w \supseteq \mathsf{box}^N(x)$, we also have $h(\overline{\delta}) \leq \frac{1}{T}\sum_{k=0}^{T-1} h(\delta_k) \leq \min_{\delta \in \mathsf{box}^N(x)} h(\delta)$. This gives an approximate solution to (5.1), and the total complexity is $\widetilde{O}(nT) = \widetilde{O}(n^{3/2})$ if $\mathbf{H}$ is given.

We summarize the above process as `MWUbasic` (see Algorithm 3), and show the following lemma:

**Lemma 6.1** (`MWUbasic`). *If $\mathbf{H} \in \mathbb{R}^{n \times n}$ is Laplacian, $K \geq 1$, $T \geq \Omega((n^{1/2}K + K^2)\log n)$, $\|\alpha\|_\infty \leq 1/32$, and $\varepsilon > 0$, then the output $\overline{\delta} = $ `MWUbasic`$(\mathbf{A}, \mathbf{H}, \alpha, T, K, \varepsilon)$ satisfies*

$$\|\overline{\delta} - \alpha\|_\infty \leq \frac{1}{32} + \frac{1}{8K} \quad \text{and} \quad \langle \nabla, \overline{\delta} \rangle + \frac{1}{6}\overline{\delta}^\top \mathbf{H}\overline{\delta} \leq \min_{\|\delta - \alpha\|_\infty \leq 1/32}\left\{\langle \nabla, \delta \rangle + \frac{1}{6}\delta^\top \mathbf{H}\delta\right\} + \varepsilon\ .$$

With Lemma 6.1, we can repeatedly apply `MWUbasic` to minimize (5.1) for $\widetilde{O}(N \log(1/\varepsilon))$ times. We summarize the algorithm as `Scaling1` (in Algorithm 4) and have the following final theorem:

---

**Theorem 6.2** (`Scaling1`). *If $N \geq 1$ and $\varepsilon \in (0, 1)$, the output $y = $ `Scaling1`$(\mathbf{A}, N, \varepsilon)$ satisfies*

$$\|y\|_\infty \leq 2N \quad \text{and} \quad f(y) - f(u) \leq \varepsilon \quad \text{for every $u$ with } \|u\|_\infty \leq N.$$

*Furthermore, if there exists $u$ satisfying $f(u) - \inf_x\{f(x)\} \leq \varepsilon$ and $\|u\|_\infty \leq N$, then we also have $\|\nabla f(y)\|_{c^{-1}}^2 \leq \varepsilon$. The total complexity is $\widetilde{O}(N(m + n^{3/2}))$.*

---

We can combine Theorem 6.2 with bounds on scaling parameters: namely, $N \leq \widetilde{O}(n)$ for the general $(r, c)$-scaling (see Lemma 3.3), or $N \leq \widetilde{O}(1)$ if the scaling parameters are polynomially bounded (see Footnote 10). This gives us the claimed results of `Scaling1` in Table 1 and Table 2.

# 7 Second-Order Method 2: via Accelerated Gradient Descent

In this section, we propose `Scaling2` (see Algorithm 5) which directly solves the constrained minimization problem (5.1) using a constrained version of accelerated gradient descent [26, 27]. We shall not directly use `Scaling2` to solve the matrix scaling problem; instead, we shall later use `Scaling2` as a warm-start for `Scaling3`.

---
**Algorithm 5** `Scaling2(A, N, T)`

---
**Input:** $\mathbf{A} \in \mathbb{R}^{d \times n}$, a non-negative matrix; $N \geq 1$, a diameter bound; $T \geq 1$, number of iterations;
1: $x_0 \leftarrow 0$;
2: **for** $t = 0$ **to** $\widetilde{O}(N)$ **do**
3:     $\mathbf{H} \leftarrow$ a matrix with $\widetilde{O}(n)$ nonzeros satisfying $\mathbf{H} \preceq \nabla^2 f(x_t) \preceq 1.1\mathbf{H}$;
                                                                                                 ⋄ *see Lemma A.3 for details.*
4:     $\delta \leftarrow$ approximate minimizer for $\min_{\delta \in \mathsf{box}^N(x_t)} \left\{ \langle \nabla f(x_t), \delta \rangle + \frac{1}{6} \delta^\top \mathbf{H} \delta \right\}$ .
        ⋄ *compute $\delta$ by applying $T$ steps of constrained accelerated gradient descent. See Lemma 7.1*
5:     $x_{t+1} \leftarrow x_t + \delta$;
6: **end for**
7: **return** $y \leftarrow$ the last $x_t$.                                                                                ⋄ *$y$ satisfies $\|y\|_\infty \leq N$*

---

We have the following main lemma to estimate the per-iteration performance of `Scaling2`:

**Lemma 7.1.** *In each iteration $t$ of* `Scaling2`*, if $\|x_t\|_\infty \leq N$, then we can compute $x_{t+1}$ in complexity $\widetilde{O}(m + Tn)$, and it satisfies $\|x_{t+1}\|_\infty \leq N$ and*

$$\text{either (1): } f(x_{t+1}) - f(u) \leq O\Big(\frac{Nh}{T^2}\Big) \quad \text{or} \quad (2): f(x_t) - f(x_{t+1}) \geq \Omega\Big(\frac{1}{N}\Big)(f(x_t) - f(u)) \ .$$

*Here, $u$ is any vector satisfying $\|u\|_\infty \leq N$.*

The following theorem is a direct corollary of Lemma 7.1.

**Theorem 7.2** (`Scaling2`)**.** *If $T \geq 1$, the output $y = $ `Scaling2`$(\mathbf{A}, N, T)$ satisfies*

$$\|y\|_\infty \leq N \quad \text{and} \quad f(y) - f(u) \leq O\Big(\frac{Nh}{T^2}\Big) \text{ for every } u \text{ with } \|u\|_\infty \leq N \ .$$

*The total complexity $\widetilde{O}(mN + NnT)$.*

*Proof of Theorem 7.2 from Lemma 7.1.* Whenever Line 3 is reached, either we have $f(x_{t+1}) - f(u) \leq O\big(\frac{Nh}{T^2}\big)$ so we are done, or we have $f(x_{t+1}) - f(u) \leq \big(1 - \Omega(\frac{1}{N})\big)(f(x_t) - f(u))$. The latter cannot happen more than $\widetilde{O}(N)$ times. □

# 8 Second-Order Method 3: via More Advanced MWU

In this section, we present our final (and most involved algorithm) `Scaling3` to solve the matrix $(r, c)$-scaling problem. As shown in Algorithm 6, our `Scaling3` method is almost identical to `Scaling1`, except that it calls a different subroutine `MWUfull` as opposed to `MWUbasic`.

We have the following theorem for `Scaling3`:

**Algorithm 6** Scaling3($\mathbf{A}, N, x_0, \varepsilon$)

**Input:** $\mathbf{A} \in \mathbb{R}^{d \times n}$ a non-negative matrix; $N \geq 1$ a diameter bound; $x_0 \in \mathbb{R}^d$ a starting vector with $\|x_0\|_\infty \leq N$; $\varepsilon > 0$ an accuracy parameter.
1: $t \leftarrow 0$, $K \leftarrow \Theta(\log(1/\varepsilon))$, $\rho \leftarrow \Theta(n^{1/3})$, and $T \leftarrow \widetilde{\Theta}(n^{1/3})$;
2: **repeat**
3:     Define $\mathsf{box}^N(x_t)$ and $\alpha \in [-1/32, 1/32]^n$ using Def. 5.2;
4:     $\mathbf{H} \leftarrow$ a matrix with $\widetilde{O}(n)$ nonzeros satisfying $\mathbf{H} \preceq \nabla^2 f(x_t) \preceq 1.1\mathbf{H}$;
5:     $\overline{\delta} \leftarrow \mathtt{MWUfull}\left(\nabla f(x_t), \mathbf{H}, \alpha, T, \rho, K, O(\frac{\varepsilon}{Nn^3})\right)$;
6:     $x_{t+1} \leftarrow x + \frac{\overline{\delta}}{6.6}$ and $N \leftarrow N + \frac{1}{K}$;      ⋄ *so $x_{t+1}$ satisfies $\|x_{t+1}\|_\infty \leq N$ for this new $N$*
7:     $t \leftarrow t + 1$;
8: **until** $\|\nabla f(x_t)\|^2_{c^{-1}} \leq \varepsilon$
9: **return** $y \leftarrow$ the last $x_t$.

---

**Theorem 8.1** (Scaling3). *If $x_0$ satisfies $f(x_0) - f(u) \leq Nn^{1/3}$ and $\|x_0\|_\infty \leq N$, and $\varepsilon \in (0, 1/4]$, then the output $y = \mathtt{Scaling3}(\mathbf{A}, N, x_0, \varepsilon)$ satisfies*

$$\|y\|_\infty \leq 10N \quad \text{and} \quad f(y) - f(u) \leq \varepsilon \quad \text{for every } u \text{ with } \|u\|_\infty \leq N.$$

*Furthermore, if there exists $u$ satisfying $f(u) - \inf_x\{f(x)\} \leq \varepsilon$ and $\|u\|_\infty \leq N$, then we also have $\|\nabla f(y)\|^2_{c^{-1}} \leq \varepsilon$. The total complexity is $\widetilde{O}(N(m + n^{4/3}))$.*

---

In Section 8.1, we present our $\mathtt{MWUfull}$ subroutine, illustrate its intuition, and discuss its differences to $\mathtt{MWUbasic}$.

To apply $\mathtt{Scaling3}$ to solve the $(r, c)$-scaling problem, we need $x_0$ with $f(x_0) - f(u) \leq Nn^{1/3}$. We can either use $\mathtt{Scaling0}$ or $\mathtt{Scaling2}$ to find such this warm start $x_0$. We present our total complexity statements below:

**Corollary 8.2.** *To find a point $y$ with $f(y) - f(u) \leq \varepsilon$ and $\|y\|_\infty \leq O(N)$, we can:*

1. *Either run $\mathtt{Scaling0}$ to obtain a point $x_0$, satisfying $\|x_0\|_\infty \leq 15N$ and $f(x_0) - f(u) \leq Nn^{1/3}$, and then apply $\mathtt{Scaling3}$. The total complexity is (using Theorem 4.12)*

$$\widetilde{O}\left(mN + mN^{1/2}h^{1/2}n^{-1/6} + Nn^{4/3}\right)$$

 *This gives us total complexity $\widetilde{O}\left(mn + mn^{1/3}h^{1/2} + n^{7/3}\right)$ when $N = \widetilde{O}(n)$.*

2. *Or run $\mathtt{Scaling2}$ to obtain a point $x_0$, satisfying $\|x_0\|_\infty \leq 2N$ and $f(x_0) - f(u) \leq Nn^{1/3}$, and then apply $\mathtt{Scaling3}$. The total complexity is (using Theorem 7.2)*

$$\widetilde{O}\left(mN + Nn^{5/6}h^{1/2}\right), \quad \text{or} \quad \widetilde{O}\left(m + n^{5/6}h^{1/2}\right) \text{ when } N = \widetilde{O}(1).$$

### 8.1 A More Advanced Multiplicative Weight Update Subroutine

Our main algorithmic tool for $\mathtt{Scaling3}$ is a new MWU algorithm $\mathtt{MWUfull}$ (see Algorithm 7) which can reduce $T$, the number of MWU rounds from $\widetilde{O}(n^{1/2})$ to $\widetilde{O}(n^{1/3})$, as compared to Lemma 6.1.

**Intuition Behind $\mathtt{MWUfull}$.** Similar to $\mathtt{MWUbasic}$, we run MWU for $T$ rounds, and in each round $k$, we also compute a vector $\delta_k$ which minimizes the same objective (5.1) (i.e., minimizes $\langle \nabla, \delta \rangle + \frac{1}{6}\delta^\top \mathbf{H}\delta$) and replacing the $\ell_1$ constraint with an $\ell_2$ constraint $\Omega_w = \left\{\delta \in \mathbb{R}^n \mid \|\delta - \alpha\|^2_w \leq \frac{n}{1024}\right\}$.

**Algorithm 7** `MWUfull`$(\nabla, \mathbf{H}, \alpha, T, \rho, K, \varepsilon)$

**Input:** $\nabla \in \mathbb{R}^n$; $\mathbf{H} \in \mathbb{R}^{n \times n}$ a Laplacian matrix; $\alpha \in \mathbb{R}^n$ satisfying $\|\alpha\|_\infty \leq 1/32$; $T \geq 1$ number of iterations; $\rho \geq 1$ a truncation parameter; $K \geq 1$ a parameter; $\varepsilon > 0$ an accuracy parameter.

1: $S \leftarrow \varnothing$, $v \leftarrow 0$, and $\widehat{v} \leftarrow 0$
2: $\Delta \leftarrow \{w \in [1/2, n]^n \colon \sum_i w_i = n\}$ and $w_0 \leftarrow (1, 1, \ldots, 1) \in \Delta$;
3: **for** $k = 0$ **to** $T - 1$ **do**
4:     Use Lemma A.4 to find a vector $\delta_k \in \mathbb{R}^n$ satisfying $\|\delta_k - \alpha\|_{w_k}^2 \leq \frac{n}{1024}$ and
$$\langle \nabla, \delta_k \rangle + \frac{1}{6} \delta_k^\top \mathbf{H} \delta_k \leq \min_{\|\delta - \alpha\|_{w_k}^2 \leq n/1024} \left\{ \langle \nabla, \delta \rangle + \frac{1}{6} \delta^\top \mathbf{H} \delta \right\} + \varepsilon$$
5:     Denote by $y = \delta_k$ for simplicity and assume wlog $|y_1| \geq |y_2| \geq \cdots \geq |y_n|$ in round $k$.
6:     **if** $\exists s \in [\rho] \colon |y_s| - |y_{s+1}| \geq 15$, $|y_{s+1}| \leq \frac{\rho}{2}$ and $|y_s| \geq \frac{\rho}{4}$ **then**
7:         $v_k \leftarrow (\underbrace{\text{sgn}(y_1), \ldots, \text{sgn}(y_s)}_{s}, \underbrace{0, \ldots, 0}_{n-s})$.      ⋄ *each $v_{k,j} \in \{-1, 0, 1\}$*
8:         $v_k^+ \leftarrow \left(\max\{v_{k,j}, 0\}\right)_{j=1}^n$ and $v_k^- \leftarrow \left(\min\{v_{k,j}, 0\}\right)_{j=1}^n$.
         ⋄ *therefore $v_k = v_k^+ + v_k^-$, $v_{k,j}^+ \in \{0, 1\}$ and $v_{k,j}^- \in \{-1, 0\}$*
9:         **if** $|\langle \nabla, v_k \rangle| \leq 2\varepsilon$ and $v_k^\top \mathbf{H} v_k \leq 4\varepsilon$ **then**
10:            Let $\mathbf{H}_\triangleleft \in \mathbb{R}^{s \times s}$, $\nabla_\triangleleft, \alpha_\triangleleft \in \mathbb{R}^s$ be the sub-matrix and sub-vectors of $\mathbf{H}$, $\nabla$ and $\alpha$;
         ⋄ *see Def. 8.3*
11:            $z_\triangleleft \leftarrow$ `MWUbasic`$(\nabla_\triangleleft, \mathbf{H}_\triangleleft, \alpha_\triangleleft, \widetilde{\Theta}(\sqrt{s}), 1/4, \varepsilon) \in \mathbb{R}^s$ so it satisfies $\|z_\triangleleft - \alpha_\triangleleft\|_\infty \leq \frac{1}{16}$ and
$$\langle \nabla_\triangleleft, z_\triangleleft \rangle + \frac{1}{6} (z_\triangleleft)^\top \mathbf{H}_\triangleleft z_\triangleleft \leq \min_{\|z - \alpha_\triangleleft\|_\infty \leq 1/32} \left\{ \langle \nabla_\triangleleft, z \rangle + \frac{1}{6} z^\top \mathbf{H}_\triangleleft z \right\} + \varepsilon$$
         ⋄ *this comes from Lemma 6.1*
12:            $\delta_{k,i} \leftarrow (z_\triangleleft)_i$ for all $i \in [s]$.      ⋄ *replace the first $s$ coordinates of $\delta_k$ with $z_\triangleleft$*
13:         **else**
14:            **if** $|\langle \nabla, v_k^+ \rangle| > \varepsilon$ **then**      ⋄ *necessarily $|\langle \nabla, v_k^+ \rangle| \geq 1/2$, see Lemma H.1*
15:                $v \leftarrow v + v_k^+$ and $\widehat{v} \leftarrow \widehat{v} + v_k^+$
16:            **else**      ⋄ *necessarily $|\langle \nabla, v_k^- \rangle| \geq 1/2$, see Lemma H.1*
17:                $v \leftarrow v + v_k^-$ and $\widehat{v} \leftarrow \widehat{v} - v_k^-$
18:            **end if**
19:            $S = S \cup \{k\}$.
20:         **end if**
21:     **end if**
22:     Define loss vector $\ell_k \in \mathbb{R}^n$ by $\ell_{k,i} \leftarrow -\min\{|\delta_{k,i} - \alpha_i|, \rho + 1\}$.      ⋄ *truncate it at $\rho + 1$*
23:     $w_{k+1} \leftarrow \arg\min_{z \in \Delta} \left\{ \eta \langle \ell_k, z \rangle + \sum_{i \in [n]} \left( z_i \log \frac{z_i}{w_{k,i}} + w_{k,i} - z_i \right) \right\}$
     ⋄ *a multiplicative weight update with parameter $\eta = \frac{1}{\rho + K}$, see Section A.1*
24: **end for**
25: **if** $|S| \leq \frac{T}{2K}$ **then**
26:     **return** $\overline{\delta} \leftarrow \frac{1}{T} \left( \sum_{k \in [T] \setminus S} \delta_k \right)$;
27: **else**
28:     **return** $\overline{\delta} \leftarrow \frac{v}{K \|\widehat{v}\|_\infty}$;
29: **end if**

Unlike `MWUbasic`, this time, we define the loss vector $\ell_k$ by letting $\ell_{k,i} = -\min\{|\delta_{k,i} - \alpha_i|, \rho+1\}$ (see Line 22 of `MWUfull`), so we *truncate* the vector $\delta_k - \alpha$ to $[-\rho-1, \rho+1]$, for some parameter $\rho = \Theta(n^{1/3})$. This ensures that the "width" of MWU is only $n^{1/3}$ so we only need to apply MWU for $T = \widetilde{\Theta}(n^{1/3})$ iterations.

Because of the truncation, we cannot always solve (5.1) almost optimally (like we did in `MWUbasic`). However, we observe that if truncation happens for $\ll T$ rounds, then we are still okay (see Line 26 of `MWUfull` and Lemma 8.4.a). Otherwise, we wish to find a direction $\overline{\delta}$ that at least decreases the objective by an additive amount (see Line 28 of `MWUfull` and Lemma 8.4.b).

We now discuss a bit more in details regarding how to find such a direction $\overline{\delta}$. Since $\rho = \Theta(n^{1/3})$ and $\|\delta_k\|_w^2 = O(n)$, using a simple counting argument, one can show that there are at most $s = O(n^{1/3})$ coordinates of $\delta_k$ that get truncated in each round of MWU. Without loss of generality, say these are the first $s$ coordinates of $\delta_k$.

Intuitively, we wish to replace these $s$ coordinates with zero, and apply MWU again to solve this smaller-sized problem.[18] Unfortunately, since the cross term in Hessian $\mathbf{H}$, namely,

$$\sum_{i \in [s], j \in [n]-[s]} |\mathbf{H}_{i,j}|$$

might be very large, solving this smaller-sized problem may not necessarily yield good minimizers of the original problem.

If the cross term is small (see Line 9 in `MWUfull`), then the above discussion works and we can recursively apply MWU on the smaller-sized problem (see Line 11). Otherwise, if the cross term is large (see Line 13), then we observe that moving in a direction $v_k = (\underbrace{1, \ldots, 1}_{s}, \underbrace{0, \ldots, 0}_{n-s})$ can essentially decrease the objective by constant. Since MWU guarantees that the truncated coordinates do not overlap too much across the $T$ rounds, one can show that the summation of such directions $v_k$ is a good descent direction (that we denote by $\overline{\delta}$ in Line 28), and it can decrease the objective value sufficiently by some additive amount (see Lemma 8.4.b). This concludes our high-level discussion on `MWUfull`.

**Details.** In the description of `MWUfull`, we have denoted by $v_\triangleleft$ the sub-vector of $v$ with only the first $s$ coordinates. This notion is formally introduced as follows:

**Definition 8.3.** *Let $s$ be in $[n-1]$.*

- *For any vector $x \in \mathbb{R}^n$, we write $x = (x_\triangleleft, x_\triangleright)$ where $x_\triangleleft \in \mathbb{R}^s$ and $x_\triangleright \in \mathbb{R}^{n-s}$.*
- *For any Laplacian matrix $\mathbf{H} \in \mathbb{R}^{n \times n}$, we write $\mathbf{H}_\triangleleft \in \mathbb{R}^{s \times s}$ as the Laplacian sub-matrix where $(\mathbf{H}_\triangleleft)_{i,j} = \mathbf{H}_{i,j}$ for every $i, j \in [s]$ and $i \neq j$; and we write $\mathbf{H}_\triangleright \in \mathbb{R}^{(n-s) \times (n-s)}$ as the Laplacian sub-matrix where $(\mathbf{H}_\triangleright)_{i,j} = \mathbf{H}_{s+i, s+j}$ for every $i, j \in [n-s]$ and $i \neq j$. The diagonal entries of $\mathbf{H}_\triangleleft$ and $\mathbf{H}_\triangleright$ are automatically induced by the definition of Laplacian matrices.*

We have the following lemma for `MWUfull`:

**Lemma 8.4** (`MWUfull`)**.** *If $\rho \in [10n^{1/3}, 2\sqrt{n}]$, $\|\alpha\|_\infty \leq 1/32$, $\varepsilon \in [0, 1/16]$, $K \geq 1$, and $T = \Omega((K\rho + K^2)\log n)$, letting $x$ be any vector in $\mathbb{R}^n$ and $\mathbf{H}$ be any Laplacian satisfying $\mathbf{H} \preceq \nabla^2 f(x) \preceq 1.1\mathbf{H}$. Then, the output*

$$\overline{\delta} \leftarrow \texttt{MWUfull}(\nabla f(x), \mathbf{H}, \alpha, T, \rho, K, \varepsilon)$$

*satisfies $\|\overline{\delta} - \alpha\|_\infty \leq \frac{1}{32} + \frac{2}{K}$ and either*

---

[18]More precisely, we can define $\nabla_\triangleleft$ to be the sub-vector of $\nabla$ but with only the first $s$ coordinates. We can also define $\mathbf{H}_\triangleleft$ to be essentially the sub-matrix of $\mathbf{H}$ with the upper left $s \times s$ block. Then, we can recursively use `MWUbasic` to minimize $\langle \nabla_\triangleleft, z \rangle + \frac{1}{6} z^\top \mathbf{H}_\triangleleft z$ over the $\ell_1$ constraint. Finally, we can replace the first $s$ coordinates in $\delta_k$ with $z \in \mathbb{R}^s$.

(a) $\langle \nabla f(x), \overline{\delta} \rangle + \frac{1}{6}\overline{\delta}^\top \mathbf{H}\overline{\delta} \leq \frac{1}{4} \min_{\|\delta - \alpha\|_\infty \leq 1/32} \left\{ \langle \nabla f(x), \delta \rangle + \frac{1}{6}\delta^\top \mathbf{H}\delta \right\} + 52n^3 \varepsilon$, or

(b) $\langle \nabla f(x), \overline{\delta} \rangle + \frac{1}{6}\overline{\delta}^\top \mathbf{H}\overline{\delta} \leq -\frac{1}{256} \frac{\rho}{K}$.

Above, case (a) corresponds to Line 26 of `MWUfull`, and says that we can solve the $\ell_1$ constrained problem (5.1) with a multiplicative factor; and case (b) corresponds to Line 28 of `MWUfull`, and says that the objective value can be decreased by an additive amount $\rho/K$.

We defer all the proofs to Appendix H. In particular, in Section H.1 we prove Theorem 8.1 from Lemma 8.4; and in Section H.3, we prove Lemma 8.4.

## 9 Discussion on Numerical Issues

Throughout the paper we have assumed exact computations of the gradient $\nabla f(x)$ and (the sparsified approximation $\mathbf{H}$ of) the Hessian $\nabla^2 f(x)$. Exact computations may take too much time, since in principle, we need to compute $e^{x_i}$ where $x_i$ can go up to $\widetilde{O}(n)$ (see Lemma 3.3).

We point out in this section that, in fact, we only need to compute $\xi$-additive approximations $\widetilde{\nabla}$ and $\widetilde{\nabla^2}$ satisfying $\|\widetilde{\nabla} - \nabla f(x)\|_2$ and $\|\widetilde{\nabla^2} - \nabla^2 f(x)\|_F \leq \xi$, for some $\xi \leq \frac{\varepsilon}{\mathsf{poly}(n)}$. This is because, for every $\delta$ with $\|\delta\|_\infty \leq 1$, we have:

$$\left( \langle \nabla f(x), \delta \rangle + \delta^\top \nabla^2 f(x) \delta \right) - \left( \langle \widetilde{\nabla}, \delta \rangle + \delta^\top \widetilde{\nabla^2} \delta \right) \leq 2n\xi$$

so we can still approximate the function $f(x)$ very well using approximate gradients or Hessians.

In the remainder of this section, we discuss how to obtain such approximations $\widetilde{\nabla}$ and $\widetilde{\nabla^2}$. By the definition of the gradient $\nabla f(x)$, as well as the sparsification of $\nabla^2 f(x)$ (see Appendix A.2), we only need to compute $\frac{\mathbf{A}_{i,j}e^{x_j}}{\langle \mathbf{A}_i, e^x \rangle} = \frac{\mathbf{A}_{i,j}}{\sum_{k=1}^n \mathbf{A}_{i,k}e^{x_k - x_j}}$ with an additive accuracy $\xi$. We introduce the following procedure to compute this quantity for each $i \in [d]$ and $j \in [n]$:

1. Output 0 if $\mathbf{A}_{i,j} = 0$.

2. Output 0 if there exists $k$ with $\mathbf{A}_{i,k} \neq 0$ and $x_k - x_j \geq \omega(\log \frac{n}{\xi})$.

3. For all $k$ with $\mathbf{A}_{i,k} = 0$ or $x_k - x_j = -\omega(\log \frac{n}{\xi})$, then define $z_k = 0$. Otherwise, define $z_k = e^{x_k - x_j}$ which is between $2^{-O(\log(n/\xi))}$ and $2^{O(\log(n/\xi))}$. Finally, output $\frac{\mathbf{A}_{i,j}}{\sum_{k=1}^n \mathbf{A}_{i,k}z_k}$.

The correctness follows from the following two simple observations:

1. if $\mathbf{A}_{i,k} \neq 0$ and $x_k - x_j = \omega(\log \frac{n}{\xi})$, then $\frac{\mathbf{A}_{i,j}e^{x_j}}{\langle \mathbf{A}_i, e^x \rangle} = o(\xi)$ and it is okay to output 0.

2. If $x_k - x_j = -\omega(\log \frac{n}{\xi})$, then $\mathbf{A}_{i,k}e^{x_k - x_j} = o(\xi/\mathsf{poly}(n))$ and $\sum_{s=1}^n \mathbf{A}_{i,s}e^{x_s - x_j} \geq \mathbf{A}_{i,j} \geq \frac{1}{\mathsf{poly}(n)}$, so it is also okay to set $z_k = 0$.

In sum, to obtain such $\xi$ additive accuracy, it suffices to use bit-length $O(\log \frac{n}{\xi})$ for all the arithmetic computations. Since we use the $\widetilde{O}$ notion to hide polylog factors in $n$ and $1/\varepsilon$, this does not affect our complexity statements in Table 1 and Table 2.

## Acknowledgements

We thank Isabella Lari for providing us the technical report version of their paper [16]. This material is based upon work supported by the National Science Foundation under agreement No. CCF-1412958. Any opinions, findings and conclusions or recommendations expressed in this material are those of the author(s) and do not necessarily reflect the views of the National Science Foundation.

# Appendix

## A Useful Subroutines

In this section we provide some useful subroutines that shall be used in our second-order based methods (namely, `Scaling1`, `Scaling2` and `Scaling3`).

- Section A.1 defines and analyzes a special constrained variant of the multiplicative weight update method;
- Section A.2 shows that the Hessian $\nabla^2 f(x)$ for our objective function can always be sparsified into a Laplacian matrix $\mathbf{H}$ with only $\widetilde{O}(n)$ non-zero entries;
- Section A.3 shows that we can solve an $\ell_2$ constraint variant of the SDD linear system.

### A.1 Constrained Multiplicative Weight Updates

Given vector $\beta \in \mathbb{R}_{\geq 0}^n$ satisfying $\|\beta\|_1 \leq n/2$, we study the behavior of the *multiplicative weight update (MWU)* method over a constrained set $\Delta = \{w \in \mathbb{R}^n \colon \sum_i w_i = n \bigwedge \forall i, w_i \geq \beta_i\}$.

**The MWU Process.** We start from a vector $w_0 \in \Delta$ where $w_{0,i} = \beta_i + \frac{n - \|\beta\|_1}{n} \geq \frac{1}{2}$. In each round $k = 0, 1, \ldots, T-1$, we are given a feedback vector $\ell \in [-\rho, \rho]^n$, and perform update

$$w_{k+1} = \arg\min_{z \in \Delta} \left\{ \eta \langle \ell_k, z \rangle + V_{w_k}(z) \right\} \quad \text{where} \quad V_x(y) \stackrel{\text{def}}{=} \sum_{i \in [n]} y_i \log \frac{y_i}{x_i} + x_i - y_i$$

for some positive parameter $\eta > 0$ known as the *learning rate*. It is a simple exercise to verify that $w_{k+1}$ is unique (because $V_x(y)$ is strictly convex for the positive orthant) and the update can be efficiently computed in time $O(n \log n)$ (see Section A.1.1). We have the following theorem:

**Lemma A.1.** *In MWU, if each $\ell_k \in [-\rho, \rho]^n$ and $\eta \in (0, \rho^{-1}]$, then for every $u \in \Delta$ we have*

$$\sum_{k=0}^{T-1} \langle \ell_k, w_k - u \rangle \leq \frac{n \log(2n^2)}{\eta} + 2\eta \sum_{k=0}^{T-1} \|\ell_k\|_{w_k}^2 \ .$$

*Proof of Lemma A.1.* In each round $k$, we first choose a dummy weight vector

$$\widetilde{w}_{k+1} = \arg\min_{z \geq 0} \left\{ \eta \langle \ell_k, z \rangle + V_{w_k}(z) \right\}$$

which is similar to $w_{k+1}$ but with the constraint $z \in \Delta$ replaced with $z \geq 0$. We claim that $\widetilde{w}_{k+1}$ is strictly positive in each coordinate. This follows from the fact that one can always find $\widetilde{w}_{k+1} > 0$ so that the gradient is zero:

$$0 = \nabla_i V_{w_k}(\widetilde{w}_{k+1}) + \eta \ell_{k,i} = (\log \widetilde{w}_{k+1,i} - \log w_{k,i}) + \eta \ell_{k,i} \ . \tag{A.1}$$

Next, it is easy to verify that $w_{k+1} = \arg\min_{z \in \Delta} \{V_{\widetilde{w}_{k+1}}(z)\}$ (by taking the derivative).[19] We assume for notational simplicity that $\widetilde{w}_0 \stackrel{\text{def}}{=} w_0$.

Using (A.1), we easily obtain that for every $u \in \Delta$,

$$\langle \eta \ell_k, w_k - u \rangle = \langle \nabla V_{w_k}(\widetilde{w}_{k+1}), u - w_k \rangle \stackrel{①}{=} V_{w_k}(u) - V_{\widetilde{w}_{k+1}}(u) + V_{\widetilde{w}_{k+1}}(w_k)$$

$$\stackrel{②}{\leq} V_{\widetilde{w}_k}(u) - V_{\widetilde{w}_{k+1}}(u) + V_{\widetilde{w}_{k+1}}(w_k) \ . \tag{A.2}$$

---

[19]This two-step interpretation of mirror descent is also known as the *tweaked* version by [31, 38]. Although most results can be proven from the original one-step version, this two-step interpretation often leads to cleaner proofs.

Above, equality ① is known as the "three-point equality" of Bregman divergence and can be easily checked via expanding out the definitions of $V_x(y)$; inequality ② is due to the generalized Pythagorean theorem of the Bregman divergence.[20]

On the other hand, we have

$$V_{\widetilde{w}_{k+1}}(w_k) \leq V_{w_k}(\widetilde{w}_{k+1}) + V_{\widetilde{w}_{k+1}}(w_k) \stackrel{③}{=} \langle \nabla V_{w_k}(\widetilde{w}_{k+1}), \widetilde{w}_{k+1} - w_k \rangle = \langle \eta \ell_k, w_k - \widetilde{w}_{k+1} \rangle ,$$

where the equality ③ is again due to the three-point equality and can be checked via expanding out the definitions of $V_x(y)$. Now, noticing that $\widetilde{w}_{k+1,i} = w_{k,i} \cdot e^{-\eta \ell_{k,i}}$, we have (using the fact that $\eta \ell_{k,i} \in [-1, 1]$)

$$V_{\widetilde{w}_{k+1}}(w_k) \leq \langle \eta \ell_k, w_k - \widetilde{w}_{k+1} \rangle \leq \sum_{i \in [n]} (\eta \ell_{k,i}) \cdot w_{k,i} \cdot (1 - e^{-\eta \ell_{k,i}}) \leq 2\eta^2 \cdot \sum_{i \in [n]} w_{k,i} \ell_{k,i}^2 .$$

Finally, substituting the above upper bound into (A.2) and telescoping it for $k = 1, \ldots, T$, we obtain for every $u \in \Delta$:

$$\sum_{k=0}^{T-1} \langle \ell_k, w_k - u \rangle \leq \frac{V_{\widetilde{w}_0}(u) - V_{\widetilde{w}_T}(u)}{\eta} + 2\eta \sum_{k=0}^{T-1} \|\ell_k\|_{w_k}^2 .$$

Finally, the choice $w_{0,i} \geq \frac{1}{2}$ implies a simple upper bound $V_{\widetilde{w}_0}(u) = V_{w_0}(u) = \sum_{i \in [n]} u_i \log \frac{u_i}{w_{0,i}} \leq n \log(2n^2)$. This gives the desired inequality. □

### A.1.1 Efficient Implementation

The constrained multiplicative weight update requires to compute

$$z^* = \arg\min_{z \in \Delta} \{\eta \langle \ell_k, z \rangle + V_{w_k}(z)\} \tag{A.3}$$

per round. We now show how to compute it with total complexity $O(n \log n)$. For simplicity, we only show this result when $\beta = (1/2, \ldots, 1/2)$ since we shall only use MWU for such $\beta$.

Using Lagrange multipliers, we know that the minimizer $z^*$ must be of the form:

$$\forall i \in [n]: z_i^* = w_{k,i} e^{-\eta \ell_{k,i}} e^{b_i - a}$$

where $a$ is an arbitrary real value, each $b_i > 0$ only when $z_i^* = \frac{1}{2}$, and $\sum_{i \in [n]} z_i^* = n$.

Now, let us define vector $y = (w_{k,i} e^{-\eta \ell_{k,i}})_{i=1}^n$, and assume without loss of generality that $0 < y_1 \leq y_2 \leq \cdots \leq y_n$. We have the following simple observation

**Claim A.2.** *For every $j \in [n]$, if $b_j > 0$ then for all $i \leq j$ it satisfies $b_i > 0$.*

*Proof of Claim A.2.* Assume by way of contradiction that there exists $i < j$ with $b_i = 0$ and $b_j > 0$. The optimal $z^*$ satisfies $z_i^* \geq \frac{1}{2}$ all $i \in [n]$. However, $z_i^* = y_i e^{-a} < y_i e^{b_j - a} \leq y_j e^{b_j - a} = \frac{1}{2}$ and this gives a contradiction. □

Using Claim A.2, we just need to find an index $j$ so that $z_i^* = 1/2$ for all $i \leq j$ and $z_i^* = y_i e^{-a}$ for all $i > j$. It must satisfy that

$$\sum_{i=1}^n z_i^* = \tfrac{1}{2} j + e^{-a} \sum_{j < i \leq n} y_i = n .$$

Let us denote the partial sum $Z_j \stackrel{\text{def}}{=} \sum_{j < i \leq n} y_i$.

---
[20]Namely, if $x = \arg\min_{z \in \Delta} V_{\widetilde{x}}(z)$ is the so-called Bregman projection, we have for all $u \in \Delta$: $V_{\widetilde{x}}(u) \geq V_x(u) + V_{\widetilde{x}}(x) \geq V_x(u)$. See for instance the textbook [31].

We must have $e^{-a} = \frac{n - \frac{1}{2}j}{Z_j}$. This implies that it suffices to define $j$ to be an index in $\{0, \cdots, n-1\}$ so that $\frac{n-\frac{1}{2}j}{Z_j} y_{j+1} \geq \frac{1}{2}$ and $\frac{n-\frac{1}{2}j}{Z_j} y_j < \frac{1}{2}$. Such an index $j$ always exists, and moreover, if $j$ satisfies these two conditions, then we can choose

$$z'_i = \begin{cases} \frac{1}{2} & \text{if } i \leq j; \\ \frac{n-\frac{1}{2}j}{Z_j} y_i & \text{if } i > j. \end{cases} \qquad b_i = \begin{cases} \log \frac{1}{2} - \log\left(\frac{n-\frac{1}{2}j}{Z_j} y_i\right) & \text{if } i \leq j; \\ 0 & \text{if } i > j. \end{cases}$$

and they satisfy (1) each $z'_i \geq 1/2$, (2) each $b_i \geq 0$, (3) $b_i > 0$ only when $z'_i = \frac{1}{2}$, (4) $\sum_{i \in [n]} z'_i = n$. Using theory of Lagrangian multipliers, we know that $z'$ must be a minimizer of (A.3).

The total complexity for computing (A.3) is $O(n \log n)$: one can compute $y$ in time $O(n)$, sort the coordinates of $y$ in time $O(n \log n)$, compute $Z_0, Z_1, \ldots, Z_{n-1}$ in time $O(n)$, and find $j$ and compute $z'$ in time $O(n)$.

## A.2 Laplacian Sparsification

In this subsection we show that one can use standard graph sparsification techniques to sparsify the Hessian $\nabla^2 f(x)$ of our objective function.

**Lemma A.3.** *Suppose $\mathbf{A}$ has at most $m$ non-zero entries where $m \geq n + d$. Then, for every $x \in \mathbb{R}^n$, in $\widetilde{O}(m)$ time one can find a Laplacian matrix $\mathbf{H}$ satisfying*

$$\mathbf{H} \preceq \nabla^2 f(x) \preceq 1.1 \mathbf{H} \qquad and \qquad \mathbf{H} \text{ has at most } \widetilde{O}(n+d) \text{ non-zero entries.}$$

*Proof of Lemma A.3.* Let us denote by $\mathbf{B} \in \mathbb{R}^{d \times n}$ the matrix where $\mathbf{B}_{i,j} = \frac{\mathbf{A}_{i,j} e^{x_j}}{\langle \mathbf{A}_i, e^x \rangle}$ so $\mathbf{B}$ is row normalized. Denote by $\mathbf{R} = \mathsf{diag}(r) \in \mathbb{R}^{d \times d}$ and by $\mathbf{D} = \mathsf{diag}\left((\sum_{i \in [d]} r_i \mathbf{B}_{i,j})_{j=1}^n\right) \in \mathbb{R}^{n \times n}$. Then, it is a simple exercise to verify that $\nabla^2 f(x) = \mathbf{D} - \mathbf{B}^\top \mathbf{R} \mathbf{B}$.

Note that $\nabla^2 f(x)$ is a Laplacian matrix and therefore PSD. Now, define $\mathbf{C} = \begin{pmatrix} \mathbf{0} & \mathbf{B}^\top \mathbf{R} \\ \mathbf{R} \mathbf{B} & \mathbf{0} \end{pmatrix}$, we know that $\mathbf{C}$ is symmetric, non-negative, and its row sums are the same as the diagonal matrix $\mathbf{D}' = \begin{pmatrix} \mathbf{D} & \mathbf{0} \\ \mathbf{0} & \mathbf{R} \end{pmatrix}$. In other words, we have that

$\mathbf{D}' - \mathbf{C}$ is a Laplacian matrix with at most $O(m)$ non-zero entries.

Using modern spectral graph sparsification techniques (see for instance Section 6 of Peng and Spielman [28]), one can find a Laplacian matrix $\mathbf{H}'$ with sparsity $\widetilde{O}(n+d)$, in time $\widetilde{O}(m)$, satisfying

$$\mathbf{H}' \preceq \mathbf{D}' - \mathbf{C}(\mathbf{D}')^{-1} \mathbf{C} \preceq 1.1 \mathbf{H}' .$$

Since $\mathbf{D}' - \mathbf{C}(\mathbf{D}')^{-1}\mathbf{C} = \begin{pmatrix} \mathbf{D} - \mathbf{B}^\top \mathbf{R} \mathbf{B} & \mathbf{0} \\ \mathbf{0} & * \end{pmatrix}$, we have that the top left $n \times n$ block of $\mathbf{H}' \in \mathbb{R}^{(n+d) \times (n+d)}$, if we denote it by $\mathbf{H} \in \mathbb{R}^{n \times n}$, also satisfies

$$\mathbf{H} \preceq \mathbf{D} - \mathbf{B}^\top \mathbf{R} \mathbf{B} = \nabla^2 f(x) \preceq 1.1 \mathbf{H} . \qquad \square$$

## A.3 Solving an $\ell_2$-Constrained SDD Linear System

In this subsection, we study an $\ell_2$ variant of (5.1), which is a constrained minimization problem

$$\min_{\delta \in \mathbb{R}^n} \left\{ \langle v, \delta \rangle + \delta^\top \mathbf{H} \delta \,\Big|\, \|\delta - \alpha\|_w^2 \leq n/c \right\}$$

where $w \in \mathbb{R}^n_{>0}$ is a positive weight vector, $\alpha \in \mathbb{R}^n$ is a shifting vector, $v$ is arbitrary and $\mathbf{H}$ is a Laplacian matrix. One may immediately observe that without the constraint $\|\delta - \alpha\|_w^2 \leq n/c$, this minimization can be done in nearly-linear time in the sparsity of $\mathbf{H}$, using the famous SDD linear system solvers (see Theorem A.5 below). When the $\ell_2$ constraint is present, by Lagrangian multipliers, one can solve the following min-max problem:

$$\min_{\delta \in \mathbb{R}^n} \max_{s \in \mathbb{R}_{\geq 0}} \left\{ \langle v, \delta \rangle + \delta^\top \mathbf{H} \delta + s \left( \|\delta - \alpha\|_w^2 - n/c \right) \right\} \enspace.$$

For this reason, one can perform a binary search in $s$, and for each fixed value of $s$, apply an SDD linear system solver. We summarize this result into the following lemma:

**Lemma A.4.** *For every weight vector $w \in [1/2, n]^n$, every SDD matrix $\mathbf{H} \in \mathbb{R}^{n \times n}$ with at most $m \geq n$ nonzero entries, every $v \in \mathbb{R}^n$ with $\|v\|_2^2 = O(\mathsf{poly}(n))$, every shifting vector $\alpha \in \mathbb{R}^n$ with $\|\alpha\|_w^2 \leq O(n)$, every constant $c \geq 1$, every $\varepsilon \in (0, 1]$, we can compute a vector $\delta \in \mathbb{R}^n$ satisfying*

$$\|\delta - \alpha\|_w^2 \leq n/c \quad \text{and} \quad \langle v, \delta \rangle + \delta^\top \mathbf{H} \delta \leq \min_{\|\delta - \alpha\|_w^2 \leq n/c} \left\{ \langle v, \delta \rangle + \delta^\top \mathbf{H} \delta \right\} + \varepsilon$$

*in time $\widetilde{O}(m)$, where the $\widetilde{O}$ notion hides logarithmic factors in $n, 1/\varepsilon$.*

*Proof of Lemma A.4.* First of all, we can apply a change of variable $\delta' = \delta - \alpha$ and minimize in $\|\delta'\|_w^2 \leq n/c$ for the new function $\langle v + 2\mathbf{H}\alpha, \delta' \rangle + \delta'^\top \mathbf{H} \delta'$. Therefore, we shall assume without loss of generality that $\alpha = 0$ in this proof.

Also, without loss of generality, we can assume $\frac{\varepsilon}{8n} \mathbf{I} \preceq \mathbf{H}$. This is because, instead of solving the original problem, we can minimize $\langle v, \delta \rangle + \delta^\top (\mathbf{H} + t\mathbf{I})\delta$ over $\|\delta\|_w^2 \leq n/c$ for some sufficiently small $t > 0$. To see this, consider function $h_t(\delta) = \langle v, \delta \rangle + \delta^\top (\mathbf{H} + t\mathbf{I})\delta$. Let $\delta^*$ be the minimizer of the original function in $\|\delta\|_w^2 \leq n/c$ and $\delta_t^*$ be the minimizer of $h_t(\delta)$ in $\|\delta\|_w^2 \leq n/c$. We have

$$\langle v, \delta_t^* \rangle + (\delta_t^*)^\top (\mathbf{H} + t\mathbf{I})\delta_t^* \leq \langle v, \delta^* \rangle + (\delta^*)^\top (\mathbf{H} + t\mathbf{I})\delta^*$$
$$\leq \langle v, \delta^* \rangle + (\delta^*)^\top \mathbf{H} \delta^* + t\|\delta^*\|_2^2 \leq \langle v, \delta^* \rangle + (\delta^*)^\top \mathbf{H} \delta^* + \frac{4nt}{c} \enspace.$$

Therefore, we can instead fine an $\varepsilon/2$ approximate minimizer of $h_t(\delta)$ for $t = \frac{\varepsilon}{8n}$.

We now proceed to the main proof (under the assumption that $\alpha = 0$ and $\frac{\varepsilon}{8n} \mathbf{I} \preceq \mathbf{H}$).

Consider function $g_s(\delta) = \langle v, \delta \rangle + \delta^\top \mathbf{H} \delta + s \left( \|\delta\|_w^2 - n/c \right)$. Using the standard SDD solver (see Theorem A.5), for every $s \geq 0$, we can find an $\delta_s$ such that $g_s(\delta_s) \leq \min_{\delta \in \mathbb{R}^d} g_s(\delta) + \frac{\varepsilon^3}{512 s^2 n^2}$ in time $\widetilde{O}(m)$ where $\widetilde{O}$ hides polylog factor in $n, s$ and $1/\varepsilon$.[21] Let us denote by

$$\delta^* \in \arg\min_{\|\delta\|_w^2 \leq n/c} \left\{ \langle v, \delta \rangle + \delta^\top \mathbf{H} \delta \right\} \enspace.$$

We now consider two cases.[22]

**Case 1.** In this case we assume $\|\delta^*\|_w^2 = n/c$. We always have

$$g_s(\delta_s) \leq g_s(\delta^*) + \frac{\varepsilon}{4} = \langle v, \delta^* \rangle + (\delta^*)^\top \mathbf{H} \delta^* + \frac{\varepsilon}{4}$$

and therefore, if $\|\delta_s\|_w^2 \in [n/c - \varepsilon/(4s), n/c]$, we have:

$$\langle v, \delta_s \rangle + (\delta_s)^\top \mathbf{H} \delta_s \leq g_s(\delta_s) + s \cdot \frac{\varepsilon}{4s} \leq \langle v, \delta^* \rangle + (\delta^*)^\top \mathbf{H} \delta^* + \frac{\varepsilon}{2}$$

as desired. Therefore, it remains to find a point $\delta_s$ such that $\|\delta_s\|_w^2 \in [n/c - \varepsilon/(4s), n/c]$. Our binary-search algorithm proceeds as follows:

1. Find $\delta_0$, if $\|\delta_0\|_w^2 \leq n/c$ then output $\delta_0$.

---

[21]We note that the original Theorem A.5 gives a multiplicative error bound. However, since $\mathbf{H} \succeq \frac{\varepsilon}{8n} \mathbf{I}$ and $\|v\|_2^2 \leq \mathsf{poly}(n)$, this multiplicative error bound also implies an additive error bound.

[22]Technically speaking, the algorithm does not know $\delta^*$ so does not know which of the two cases it belongs two. In any case, we can run the algorithm for both cases and outputs the one that gives a smaller objective value.

2. Let $a = 0$ and $b = t = 4c\|v\|_2$

3. Repeatedly pick $z = \frac{a+b}{2}$. If $\|\delta_z\|_w^2 \geq n/c$, then let $a = z$, otherwise, let $b = z$.

4. Until $\|\delta_z\|_w^2 \in [n/c - \varepsilon/(4z), n/c]$, and output $\delta_z$.

We first show that $\|\delta_t\|_w^2 \leq n/c - \varepsilon/(4t)$. Denote $\delta_s^* = \arg\min\{g_s(\delta)\} = -\frac{1}{2}(\mathbf{H} + s\mathbf{W})^{-1}v$, where $\mathbf{W} = \mathsf{diag}(w)$. We have

$$\|\delta_t^*\|_w^2 \leq v^\top(\mathbf{H} + t\mathbf{W})^{-1}\mathbf{W}(\mathbf{H} + t\mathbf{W})^{-1}v \leq \|v\|_2^2 \mathbf{Tr}((\mathbf{H} + t\mathbf{W})^{-1}\mathbf{W}(\mathbf{H} + t\mathbf{W})^{-1}) \leq \frac{\|v\|_2^2 n}{t^2} \leq \frac{n}{2c} \ . \tag{A.4}$$

Since $\mathbf{H} \succeq \frac{\varepsilon}{8n}\mathbf{I}$, we also have:

$$\|\delta_t - \delta_t^*\|_2^2 \leq \frac{8n}{\varepsilon}(g(\delta_t) - g(\delta_t^*)) \leq \frac{8n}{\varepsilon}\frac{\varepsilon^3}{512t^2n^2} \leq \frac{\varepsilon^2}{64t^2n}$$

and this implies $\|\delta_t - \delta_t^*\|_w \leq \frac{\varepsilon}{8t}$. Plugging this into (A.4) we have $\|\delta_t\|_w^2 \leq n/c - \varepsilon/(4t)$.

Since now we have $\|\delta_t\|_w^2 \leq n/c - \varepsilon/(4t)$, in each iteration of the binary search, we have $\|\delta_b\|_w^2 \leq n/c - \varepsilon/(4b)$ and $\|\delta_a\|_w^2 \geq n/c$. Therefore, it suffices to bound the total number of binary-search iterations. To do this, we first calculate:

$$\left|\frac{d\|\delta_s^*\|_w^2}{ds}\right| = \left|\frac{d}{ds}\left(\frac{1}{4}v^\top(\mathbf{H} + s\mathbf{W})^{-1}\mathbf{W}(\mathbf{H} + s\mathbf{W})^{-1}v\right)\right|$$

$$= \left|\left(\frac{1}{2}v^\top(\mathbf{H} + s\mathbf{W})^{-1}\mathbf{W}(\mathbf{H} + s\mathbf{W})^{-1}\mathbf{W}(\mathbf{H} + s\mathbf{W})^{-1}v\right)\right| \leq 8\|v\|_2^2\left(\frac{8n}{\varepsilon}\right)^3 n^2 \ .$$

This implies for any $s, s'$:

$$\|\delta_s^*\|_w - \|\delta_{s'}^*\|_w \leq \sqrt{8\|v\|_2^2\left(\frac{8n}{\varepsilon}\right)^3 n^2 |s - s'|}$$

Accordingly, we have

$$\|\delta_s\|_w - \|\delta_{s'}\|_w \leq \|\delta_s - \delta_s^*\|_w - \|\delta_{s'} - \delta_{s'}^*\|_w + \|\delta_s^*\|_w - \|\delta_{s'}^*\|_w \leq \frac{\varepsilon}{8s} + \frac{64\|v\|_2 n^{2.5}}{\varepsilon^{1.5}}\sqrt{|s - s'|} \ .$$

In other words, the binary search process must have terminated before $|b - a| \leq \frac{\varepsilon^5}{32768n^5\|v\|_2^2}$. This takes $O\left(\log \frac{\varepsilon^5}{32768n^5\|v\|_2^2}\right) = \widetilde{O}(1)$ iterations. Note also that throughout the binary search, $s \leq 4c\|v\|_2$, so the total complexity is $\widetilde{O}(m)$, hiding polylog factors in $n$ and $1/\varepsilon$.

**Case 2.** In the case we assume $\|\delta^*\|_w^2 < n/c$. This implies $\delta^* = \delta_0^* = -(2\mathbf{H})^{-1}v$. We know that

$$\left|\langle v, \delta^*\rangle + (\delta^*)^\top \mathbf{H}\delta^*\right| = \frac{1}{4}v^\top \mathbf{H}^{-1}v \leq \|v\|_2^2 \frac{2n}{\varepsilon} \ .$$

In this case, we find $\delta_0$ such that $g_0(\delta_0) \leq g_0(\delta^*) + \frac{\varepsilon^5}{512c^2n^4\|v\|_2^4}$. Again using $\mathbf{H} \succeq \frac{\varepsilon}{8n}\mathbf{I}$, we have:

$$\|\delta_0 - \delta_0^*\|_w \leq \sqrt{\frac{8n^2}{\varepsilon}}\sqrt{\frac{\varepsilon^5}{512c^2n^4\|v\|_2^4}} \leq \frac{\varepsilon^2}{8\|v\|_2^2 cn} \ .$$

Therefore we still have:

$$\|\delta_0\|_w^2 \leq (\|\delta^*\|_w + \|\delta^* - \delta_0\|_w)^2 \leq \frac{n}{c} + 2\sqrt{\frac{n}{c}}\frac{\varepsilon^2}{8\|v\|_2^2 cn} + \left(\frac{\varepsilon^2}{8\|v\|_2^2 cn}\right)^2 \leq \frac{n}{c} + \frac{\varepsilon^2}{4c\|v\|_2^2} \ .$$

If we just output $\delta = (1 - \lambda)\delta_0$ for $\lambda = \frac{\varepsilon^2}{4n\|v\|_2^2}$, we will have:

$$\|\delta\|_w^2 \leq (1-\lambda)^2 \|\delta_0\|_w^2 \leq \left(1 - \frac{\varepsilon^2}{2n\|v\|_2^2}\right)\left(\frac{n}{c} + \frac{\varepsilon^2}{2c\|v\|_2^2}\right) \leq \frac{n}{c}$$

and

$$\langle v, \delta \rangle + (\delta)^\top \mathbf{H}\delta \leq (1-\lambda)\left(\langle v, \delta_0 \rangle + (\delta_0)^\top \mathbf{H}\delta_0\right)$$
$$\leq \langle v, \delta^* \rangle + (\delta^*)^\top \mathbf{H}\delta^* + \frac{\varepsilon}{2} + \lambda \left|\langle v, \delta_0 \rangle + (\delta_0)^\top \mathbf{H}\delta_0\right|$$
$$\leq \langle v, \delta^* \rangle + (\delta^*)^\top \mathbf{H}\delta^* + \frac{\varepsilon}{2} + \lambda\|v\|_2^2 \frac{2n}{\varepsilon} \leq \langle v, \delta^* \rangle + (\delta^*)^\top \mathbf{H}\delta^* + \varepsilon$$

as desired. □

**Theorem A.5.** *Given a SDD matrix $\mathbf{M} \in \mathbb{R}^{n \times n}$ with $m \geq n$ nonzero entries, for every vector $v \in \mathbb{R}^n$ in the (column) span of $\mathbf{M}$ with 2-norm bounded by $\mathsf{poly}(n)$, every $\varepsilon' > 0$, we can find a vector $z \in \mathbb{R}^n$ satisfying*

$$\|z - z^*\|_{\mathbf{M}}^2 \leq \varepsilon' \|z^*\|_{\mathbf{M}}^2 \quad \text{and} \quad v^\top z + \frac{1}{2} z^\top \mathbf{M} z \leq (1 - \varepsilon') \min_{z \in \mathbb{R}^n}\{v^\top z + \frac{1}{2} z^\top \mathbf{M} z\}$$

*in time $\widetilde{O}(m \log(1/\varepsilon'))$, where the $\widetilde{O}$ notation hides logarithmic factors in $n$.*

*Proof of Theorem A.5.* Let $z^* \in \mathbb{R}^n$ be the exact minimizer for the right hand side, and denote by $\mathsf{OPT} = -\min_{z \in \mathbb{R}^n}\{v^\top z + \frac{1}{2} z^\top \mathbf{M} z\}$. Using standard SDD linear system solvers (originally proposed by Spielman and Teng [35], and later simplified by for instance [19, 21]) we can find a vector $z$ satisfying

$$(z - z^*)^\top \mathbf{M}(z - z^*) \leq \varepsilon' (z^*)^\top \mathbf{M} z^* \quad \text{in time } \widetilde{O}((n + m)\log(1/\varepsilon')) \ . \tag{A.5}$$

(See the appendix of [19] for this exact statement.) Using the fact that $\mathbf{M}z^* = -v$ and $\mathsf{OPT} = \frac{1}{2}(z^*)^\top \mathbf{M} z^* = -\frac{1}{2} v^\top z^*$, we can simplify (A.5) and obtain $v^\top z + \frac{1}{2} z^\top \mathbf{M} v \leq -(1 - \varepsilon')\mathsf{OPT}$. □

## B  Missing Details for Section 2

Recall that

$$f(x) \stackrel{\text{def}}{=} \sum_{i=1}^d r_i \log \langle \mathbf{A}_i, e^x \rangle - c^\top x \ . \tag{1.1}$$

**Proposition 2.3.** *Our objective $f(x)$ in (1.1) is convex and*

- $\nabla_j f(x) = \sum_{i=1}^d \frac{r_i \mathbf{A}_{i,j}}{\langle \mathbf{A}_i, e^x \rangle} e^{x_j} - c_j$.
- *If $\|\nabla f(x)\|_{c^{-1}}^2 \leq \varepsilon$, then $\left(\frac{r_i \mathbf{A}_{i,j} \cdot e^{x_j}}{\langle \mathbf{A}_i, e^x \rangle}\right)_{i,j}$ is an $\varepsilon$-approximate $(r,c)$-matrix.*
- *If $\mathbf{A}$ is exactly $(r,c)$-scalable, then there exists $x^*$ so that $f(x^*) = \min_x\{f(x)\}$ and $\nabla f(x^*) = 0$.*
- *If $\mathbf{A}$ is asymptotically $(r,c)$-scalable, then $\inf_x\{f(x)\} > -\infty$.*
- *$\mathbf{A}$ is not asymptotically $(r,c)$-scalable if and only if $\inf_x\{f(x)\} = -\infty$.*

*Proof of Proposition 2.3.* The first item is obtained after a simple calculation. The second item follows as the $i^{th}$ row sum is given by $\sum_{j=1}^n \frac{r_i \mathbf{A}_{i,j}}{\langle \mathbf{A}_i, e^x \rangle} e^{x_j} = r_i \cdot \frac{\langle \mathbf{A}_i, e^x \rangle}{\langle \mathbf{A}_i, e^x \rangle} = r_i$ and the $j^{th}$ column sum is given by $c'_j = \sum_{i=1}^d \frac{r_i \mathbf{A}_{i,j}}{\langle \mathbf{A}_i, e^x \rangle} e^{x_j} = \nabla_j f(x) + c_j$. Thus, $\|c' - c\|_{c^{-1}}^2 = \|\nabla f(x)\|_{c^{-1}}^2 \leq \varepsilon$ implies that the above matrix is an $\varepsilon$-approximate $(r,c)$-matrix. For the third item, let $x,y$ be the scaling vectors so that $\mathsf{diag}(y)\mathbf{A}\mathsf{diag}(x)$ is $(r,c)$-scaled. Then setting $x^* = \log(x)$ we have $\nabla f(x^*) = 0$. The fourth item follows from Lemma 3.3 established later. To prove the fifth item, it suffices to show if $\mathbf{A}$ is not scalable then $f(x)$ can go to $-\infty$. Using the characterization Proposition 2.2, one can find a direction $x \in \mathbb{R}^n$ to move so that $f(xt)$ tends to $-\infty$ as $t \to \infty$. □

## C  Missing Details for Section 3

### C.1  Proof of Lemma 3.2

**Lemma 3.2** (norm bound). *If $\mathbf{A}$ is exactly $(r,c)$ scalable, and all non-zero entries of $\mathbf{A}$ are within $[\nu, 1]$ for some $\nu > 0$. Then, the following holds:*

1. *If $\mathbf{A}$ is full then there exists a minimizer $x^*$ of $f(x)$ such that $\|x^*\|_\infty \leq \ln \frac{hn}{\nu}$.*

2. *If $\mathbf{A}$ is not full, then there exists a minimizer $x^*$ of $f(x)$ such that $\|x^*\|_\infty \leq (h+1/2)\ln \frac{h}{\nu}$.*

*Proof of Lemma 3.2.* Let $x^*$ be any minimizer of $f(x)$ and since $\nabla f(x^*) = 0$, there must exist $y \in \mathbb{R}^d$ such that the matrix $\mathsf{diag}(y)\mathbf{A}\mathsf{diag}(e^{x^*})$ is an $(r,c)$-matrix.

1. Without loss of generality, we can assume $x_1^* \geq x_2^* \geq \cdots \geq x_n^* = 0$. It remains to bound $x_1^*$. Since $\mathbf{A}_{i,n} \leq 1$ for every $i \in [d]$ and $\sum_{i \in [d]} y_i \mathbf{A}_{i,n} e^{x_n^*} = c_n \geq 1$, there must exist $i \in [d]$ such that $y_i \geq \frac{1}{n}$. Therefore, since $y_i \mathbf{A}_{i,1} e^{x_1} \leq r_i \leq h$ and $\mathbf{A}_{i,1} \geq \nu$, we must have $x_1 \leq \ln \frac{h}{\mathbf{A}_{i,1} y_i} \leq \ln \frac{hn}{\nu}$.

2. It was shown in [16, Theorem 5.1 and Lemma 5.1] that one can assume $|x_j^*| \leq (h+1/2)\ln\frac{h}{\nu}$.[23]

□

### C.2  Proof of Lemma 3.3

In this section we prove the following lemma:

**Lemma 3.3.** *If $\mathbf{A}$ is asymptotically $(r,c)$-scalable, and all non-zero entries of $\mathbf{A}$ are within $[\nu, 1]$ for some $\nu > 0$, then, for every $\varepsilon > 0$, there exists $x_\varepsilon^* \in \mathbb{R}^n$ such that*

$$\|x_\varepsilon^*\|_\infty = O\left(n \ln \frac{nh}{\nu \varepsilon}\right) \;, \quad \|\nabla f(x_\varepsilon^*)\|_\infty \leq \varepsilon \;, \quad \text{and} \quad f(x_\varepsilon^*) - \inf_x \{f(x)\} \leq \varepsilon \;.$$

In order to prove this Lemma 3.3, we need three structural lemmas.

#### C.2.1  Structural Lemmas

The first structural lemma states that if $\|\nabla f(x)\|_\infty$ is very small for some $x \in \mathbb{R}^n$, then one can modify $x$ to decrease its $\ell_\infty$ norm, without increasing $\|\nabla f(x)\|_\infty$ too much:

**Lemma C.1** (structural lemma 1). *Suppose we are given $x \in \mathbb{R}^n$ satisfying $\|\nabla f(x)\|_\infty \leq \varepsilon$. Define $\rho = \ln \frac{n^2 h}{\nu^2 \varepsilon}$, and assume without loss of generality that $x_1 \leq x_2 \leq \cdots \leq x_n$. Now, suppose there exists $s \in [n-1]$ such that $x_{s+1} - x_s = \rho_s \geq \rho$, then we choose $y \in \mathbb{R}^n$ as:*

$$y_i = \begin{cases} x_i & \text{if } i \leq s; \\ x_i - \rho_s + \rho & \text{if } i > s. \end{cases}$$

*and it satisfies $\|\nabla f(y)\|_\infty \leq \left(1 + \frac{1}{n}\right)\varepsilon$.*

---
[23]They only stated the result for $\max_{j \in [n]} \{x_j^*\}$, however, their same proof also implies the upper bound on $-x_j^*$.

*Proof of Lemma C.1.* Recall that for each $i \in [n]$, $\nabla_i f(y) = \sum_{j=1}^{d} r_j \frac{\mathbf{A}_{j,i}}{\sum_{k=1}^{n} \mathbf{A}_{j,k} e^{y_k - y_i}} - c_i$. It suffices to show that, for all $i \in [n]$ and $j \in [d]$ such that $\mathbf{A}_{j,i} \neq 0$, we have

$$\left| \frac{1}{\sum_{k=1}^{n} \mathbf{A}_{j,k} e^{y_k - y_i}} - \frac{1}{\sum_{k=1}^{n} \mathbf{A}_{j,k} e^{x_k - x_i}} \right| \leq \frac{\varepsilon}{nh}.$$

We now divide the proof into the following two cases:

- Suppose $i \leq s$. In this case, $\sum_{k=1}^{n} \mathbf{A}_{j,k} e^{y_k - y_i} = \sum_{k=1}^{s} \mathbf{A}_{j,k} e^{x_k - x_i} + \sum_{k=s+1}^{n} \mathbf{A}_{j,k} e^{y_k - y_i}$. Now, if there exists $k \geq s+1$ such that $\mathbf{A}_{j,k} \neq 0$, then we have:
  - $\mathbf{A}_{j,k} e^{x_k - x_i} \geq \nu e^{\rho}$, and this implies $\frac{1}{\sum_{k=1}^{n} \mathbf{A}_{j,k} e^{x_k - x_i}} \leq \frac{1}{\nu e^{\rho}} \leq \frac{\varepsilon}{2nh}$; and
  - $\mathbf{A}_{j,k} e^{y_k - y_i} \geq \nu e^{\rho}$, and this implies $\frac{1}{\sum_{k=1}^{n} \mathbf{A}_{j,k} e^{y_k - y_i}} \leq \frac{1}{\nu e^{\rho}} \leq \frac{\varepsilon}{2nh}$.

  Together, we have $\left| \frac{1}{\sum_{k=1}^{n} \mathbf{A}_{j,k} e^{y_k - y_i}} - \frac{1}{\sum_{k=1}^{n} \mathbf{A}_{j,k} e^{x_k - x_i}} \right| \leq \frac{\varepsilon}{nh}$.

  If there is no such $k$, then $\sum_{k=1}^{n} \mathbf{A}_{j,k} e^{y_k - y_i} = \sum_{k=1}^{n} \mathbf{A}_{j,k} e^{x_k - x_i}$.

- Suppose $i > s$. In this case, we have $x_i - x_k \geq \rho_s$ for $k \leq s$ and $y_k - y_i = x_k - x_i$ for $k \geq s+1$. Hence, if $\mathbf{A}_{j,k} = 0$ for all $k \leq s$, we are done. Otherwise, we must have some $k \leq s$ such that $\nu \leq \mathbf{A}_{j,k} \leq 1$ and therefore, we have the following inequalities:

$$\sum_{k=1}^{s} \mathbf{A}_{jk} e^{x_k - x_i} \leq \sum_{k=1}^{s} e^{-\rho_s} \leq s e^{-\rho_s}$$

and as a consequence, we have:

$$\sum_{k=1}^{s} \mathbf{A}_{jk} e^{y_k - y_i} = \sum_{k=1}^{s} \mathbf{A}_{jk} e^{x_k - x_i + (\rho_s - \rho)} = e^{\rho_s - \rho} \cdot \sum_{k=1}^{s} \mathbf{A}_{jk} e^{x_k - x_i} \leq s e^{-\rho}.$$

Since we are assuming that $\mathbf{A}_{j,i} \neq 0$, and therefore $\mathbf{A}_{j,i} \geq \nu$, we also have

$$\nu \leq \mathbf{A}_{j,i} \leq \sum_{k=1}^{n} \mathbf{A}_{j,k} e^{x_k - x_i} \text{ as well as } \nu \leq \mathbf{A}_{j,i} \leq \sum_{k=1}^{n} \mathbf{A}_{j,k} e^{y_k - y_i}$$

Together, we have

$$\left| \frac{1}{\sum_{k=1}^{n} \mathbf{A}_{j,k} e^{y_k - y_i}} - \frac{1}{\sum_{k=1}^{n} \mathbf{A}_{j,k} e^{x_k - x_i}} \right| \leq \frac{s e^{-\rho}}{\nu^2} \leq \frac{\varepsilon}{nh}. \qquad \square$$

The second structural lemma says that if a matrix can be scaled sufficiently close to an $(r,c)$-matrix, then it must be (asymptotically) $(r,c)$-scalable.

**Lemma C.2** (structural lemma 2). *Suppose $\mathbf{D}$ is a non-negative matrix, with $r = \mathbf{D}\mathbb{1}$ and $\|\mathbf{D}^\top \mathbb{1} - c\|_\infty < \frac{1}{n}$, then $\mathbf{D}$ is asymptotically $(r,c)$-scalable.*

*Proof of Lemma C.2.* Suppose by way of contradiction that $\mathbf{D}$ is not asymptotically $(r,c)$-scalable. Then, by the characterization Proposition 2.2, there must be a zero minor $R \times C \subseteq [d] \times [n]$ of $\mathbf{D}$ such that $\sum_{i \in [d] \setminus R} r_i < \sum_{j \in C} c_j$.

Since $r_i, c_j$ are integers, we know that $\sum_{i \in [d] \setminus R} r_i \leq \sum_{j \in C} c_j - 1$. However, we also know that

$$\sum_{i \in [d] \setminus R} r_i = \sum_{i \in [d] \setminus R} \sum_{j \in [n]} \mathbf{D}_{i,j} = \sum_{j \in [n]} \sum_{i \in [d] \setminus R} \mathbf{D}_{i,j} \geq \sum_{j \in C} \sum_{i \in [d] \setminus R} \mathbf{D}_{i,j}$$

$$= \sum_{j \in C} \sum_{i \in [d]} \mathbf{D}_{i,j} \geq \sum_{j \in C} \left( c_j - \frac{1}{n} \right) > \sum_{j \in C} c_j - 1.$$

This contradiction completes the proof. $\qquad \square$

The third structural lemma gives an alternative characterization to the scalability of matrices:

**Lemma C.3** (structural lemma 3). *If $\mathbf{A}$ is asymptotically $(r,c)$-scalable, then up to row and column permutations, $\mathbf{A}$ can be written as a block upper-triangular matrix $\mathbf{A} = (\mathbf{B}_{u,v})_{u \in [p], v \in [p]}$, where*

- *each $\mathbf{B}_{u,v}$ is a submatrix $(\mathbf{A}_{i,j})_{i \in L_u, j \in R_v}$, where $L_u \cap L_{u'} = R_v \cap R_{v'} = \varnothing$ for $u \neq u'$ and $v \neq v'$.*
- *each $\mathbf{B}_{u,u}$ is exactly $(r|_{L_u}, c|_{R_u})$-scalable.*
- *each $\mathbf{B}_{u,v} = \mathbf{0}$ if $u > v$.*

*Proof of Lemma C.3.* We prove this by induction. If $\mathbf{A}$ is already exactly $(r, c)$ scalable then we are done. Otherwise, we claim that there must be a zero minor $L \times R$ of $\mathbf{A}$ so that $\sum_{i \in \overline{L}} r_i = \sum_{j \in R} c_j$. This is because if for all zero minors it satisfies $\sum_{i \in \overline{L}} > \sum_{j \in R} c_j$, then $\mathbf{A}$ must be exactly scalable owing to Proposition 2.2.

Now, if we permute the rows and columns, we can put this $L \times R$ minor in the lower bottom of $\mathbf{A}$, and write $\mathbf{A} = \begin{pmatrix} \mathbf{B} & \mathbf{C} \\ \mathbf{0} & \mathbf{D} \end{pmatrix}$. In addition, since $\sum_{i \in \overline{L}} r_i = \sum_{j \in R} c_j$ and $\sum_{i \in L} r_i = \sum_{j \in \overline{R}} c_j$, we claim that $\mathbf{B}$ is asymptotically $(r|_{\overline{L}}, c|_R)$-scalable, and $\mathbf{D}$ is asymptotically $(r|_L, c|_{\overline{R}})$-scalable. This is because for every $L' \times R'$ zero minor of $\mathbf{B}$, we have $L' \cup L \times R'$ is also a zero minor of $\mathbf{A}$, so it satisfies

$$\sum_{i \in \overline{L} \setminus L'} r_i = \sum_{i \in [d] \setminus (L' \cup L)} r_i \geq \sum_{j \in R'} c_j \ .$$

And similarly for $\mathbf{D}$. In sum, we can recurse on $\mathbf{B}$ and $\mathbf{D}$ until we find the desired block upper-triangular form. □

### C.2.2 Proof of Lemma 3.3

We are now ready to prove Lemma 3.3. Basically, we can start from a vector $x$ with $\|\nabla f(x)\|_\infty$ being sufficiently small, and then apply Lemma C.1 to make $\|x\|_\infty \leq \widetilde{O}(n)$. Finally, we use Lemma C.2 and Lemma C.3 to also bound the objective value.

*Proof of Lemma 3.3.* As $\nabla f(x) = \nabla f(x + t \cdot \mathbb{1})$ for any $t \in \mathbb{R}$, we can always work with vectors $x \in \mathbb{R}^n$ such that $\|x\|_1 = n$. Therefore, repeatedly applying Lemma C.1, we know that if $\mathbf{A}$ is asymptotically $(r, c)$-scalable, then for every $\varepsilon > 0$, we can obtain an $x_\varepsilon \in \mathbb{R}^n$ such that $\|x_\varepsilon\|_\infty = O\left(n \ln \frac{nh}{\nu \varepsilon}\right)$ and $\|\nabla f(x_\varepsilon)\|_\infty \leq \varepsilon$.

Now, we define matrix $\mathbf{A}'$ as $\mathbf{A}'_{i,j} = \frac{\mathbf{A}_{i,j} e^{(x_\varepsilon)_j}}{\langle \mathbf{A}_i, e^{x_\varepsilon} \rangle}$. We know that $\mathbf{A}'' = (r_i \mathbf{A}'_{i,j})_{i,j}$ is an $\varepsilon$-approximate $(r, c)$-matrix. Therefore, we claim that we can always write $\mathbf{A}''$ as $\mathbf{A}'' = (1-s)\mathbf{B} + s\mathbf{D}$ for some $s \in [0,1]$, where $\mathbf{B}$ is an exact $(r, c)$-matrix and $\mathbf{D}$ is not asymptotically $(r, c)$-scalable.

This follows from an argument similar to [22]: since $\mathbf{A}''$ is asymptotically $(r, c)$-scalable, we can write $\mathbf{A}''$ in the upper triangular form according to Lemma C.3. This implies one can find an exact $(r, c)$-matrix $\mathbf{C}$ where the non-zero entries of $\mathbf{C}$ is a subset of the non-zero entries of $\mathbf{A}''$. Now, one can subtract a $t\mathbf{C}$ from $\mathbf{A}''$ where $t > 0$ is the largest possible real value such that the entries of $\mathbf{A}'' - t\mathbf{C}$ are still non-negative. If the remaining matrix $\mathbf{A}'' - t\mathbf{C}$ is still asymptotically $(r, c)$-scalable, we can repeat this process until we reach a non-scalable matrix $\mathbf{D}$. Finally, since the row sums of $\mathbf{A}''$ already equal to $r$, we must have $s \in [0, 1]$.

Next, $\mathbf{D} = \frac{\mathbf{A}'' - (1-s)\mathbf{B}}{s}$ is not asymptotically $(r, s)$-scalable, so by Lemma C.2, we have $\frac{1}{n} \leq \|\mathbf{D}^\top \mathbb{1} - c\|_\infty = \frac{\|(\mathbf{A}'')^\top \mathbb{1} - c\|_\infty}{s} \leq \frac{\varepsilon}{s}$, which implies $s \leq \varepsilon n$.

Finally, defining $\mathbf{B}' = \mathsf{diag}(r)^{-1}\mathbf{B}$, we have for every $x \in \mathbb{R}^d$:

$$f(x_\varepsilon + x) - f(x_\varepsilon) = \sum_{i=1}^d r_i \log \frac{\langle \mathbf{A}_i, e^{x_\varepsilon + x}\rangle}{\langle \mathbf{A}_i, e^{x_\varepsilon}\rangle} - \langle c, x\rangle = \sum_{i=1}^d r_i \log \langle \mathbf{A}'_i, e^x\rangle - \langle c, x\rangle$$

$$\geq \sum_{i=1}^d r_i \log\big((1-s)\langle \mathbf{B}'_i, e^x\rangle\big) - \langle c, x\rangle$$

$$= \sum_{i=1}^d r_i \log(1-s) + \left(\sum_{i=1}^d r_i \log \langle \mathbf{B}'_i, e^x\rangle - \langle c, x\rangle\right)$$

$$\overset{\text{①}}{\geq} -2hs \geq -4hn\varepsilon \ .$$

Above, inequality ① uses the fact that $\mathbf{B}$ is already an $(r,c)$-matrix so $\sum_{i=1}^d r_i \log\langle \mathbf{B}'_i, e^x\rangle - \langle c, x\rangle \geq 0$ for every $x \in \mathbb{R}^n$ (see Proposition 2.3).

Dividing $\varepsilon$ by $4hn$ gives the desired result. $\square$

## D  Missing Details for Section 4

### D.1  Proof of Lemma 4.2

**Lemma 4.2.** *Given $x \in \mathbb{R}^n$, denote by $\nabla = \nabla f(x)$ and $\Lambda^{\mathsf{s}}, \Lambda^{\mathsf{l}} \subseteq [n]$ the set of small and large coordinates (see Def. 4.1). Then, for every $\delta \in \mathbb{R}^n$ where $\|\delta\|_\infty \leq 1/2$, we have*

- *if $\delta \geq 0$, then $f(x) - f(x+\delta) \geq Q^+(x,\delta) \overset{\text{def}}{=} \sum_{j \in \Lambda^{\mathsf{s}}}\big(-\nabla_j \cdot \delta_j - \frac{4}{3}c_j \cdot \delta_j^2\big) + \sum_{j \in \Lambda^{\mathsf{l}}}\big(-\frac{7}{3}\nabla_j \cdot \delta_j\big)$.*
- *if $\delta \leq 0$, then $f(x) - f(x+\delta) \geq Q^-(x,\delta) \overset{\text{def}}{=} \sum_{j \in \Lambda^{\mathsf{s}}}\big(-\nabla_j \cdot \delta_j - \frac{4}{3}c_j \cdot \delta_j^2\big) + \sum_{j \in \Lambda^{\mathsf{l}}}\big(-\frac{1}{2}\nabla_j \cdot \delta_j\big)$.*

*(Recall that $\delta \geq 0$ or $\delta \leq 0$ means entry-wise non-negativity or non-positivity.)*

*Proof of Lemma 4.2.* Note that

$$f(x) - f(x+\delta) = \int_{\tau=0}^1 \langle \nabla f(x+\tau\delta), -\delta\rangle \, \mathrm{d}\tau = \sum_{j=1}^n \int_{\tau=0}^1 -(\nabla_j f(x+\tau\delta) \cdot \delta_j) \, \mathrm{d}\tau$$

We now consider the two cases separately:

- If $\delta \geq 0$, we have for all $j \in [n]$,

$$-(\nabla_j f(x+\tau\delta) \cdot \delta_j) = -\Big(\sum_{i=1}^d \frac{r_i \mathbf{A}_{i,j}}{\langle \mathbf{A}_i, e^{x+\tau\delta}\rangle} e^{x_j + \tau\delta_j} - c_j\Big) \cdot \delta_j \geq -\Big(\sum_{i=1}^d \frac{r_i \mathbf{A}_{i,j}}{\langle \mathbf{A}_i, e^x\rangle} e^{x_j + \tau\delta_j} - c_j\Big) \cdot \delta_j$$

$$\geq -\Big(\sum_{i=1}^d \frac{r_i \mathbf{A}_{i,j}}{\langle \mathbf{A}_i, e^x\rangle} e^{x_j}(1 + \frac{4}{3}\tau\delta_j) - c_j\Big) \cdot \delta_j = -\nabla_j \cdot \delta_j - \frac{4}{3}\tau\delta_j^2 \cdot \sum_{i=1}^d \frac{r_i \mathbf{A}_{i,j}}{\langle \mathbf{A}_i, e^x\rangle} e^{x_j}$$

If $j \in \Lambda^{\mathsf{s}}$, then $\sum_{i=1}^d \frac{r_i \mathbf{A}_{i,j}}{\langle \mathbf{A}_i, e^x\rangle} e^{x_j} = \nabla_j + c_j \leq 2c_j$ so

$$\int_{\tau=0}^1 -(\nabla_j f(x+\tau\delta) \cdot \delta_j) \, \mathrm{d}\tau \geq -\nabla_j \cdot \delta_j - \frac{4}{3}\delta_j^2 c_j \ .$$

If $j \in \Lambda^{\mathsf{l}}$, then $\sum_{i=1}^d \frac{r_i \mathbf{A}_{i,j}}{\langle \mathbf{A}_i, e^x\rangle} e^{x_j} = \nabla_j + c_j \leq 2\nabla_j f(x)$ so

$$\int_{\tau=0}^1 -(\nabla_j f(x+\tau\delta) \cdot \delta_j) \, \mathrm{d}\tau \geq -\nabla_j \cdot \delta_j - \frac{4}{3}\delta_j^2 \nabla_j \geq -\frac{7}{3}\nabla_j \cdot \delta_j$$

- If $\delta \leq 0$, we have for all $j \in [n]$,

$$-(\nabla_j f(x+\tau\delta) \cdot \delta_j) = -\Big(\sum_{i=1}^{d} \frac{r_i \mathbf{A}_{i,j}}{\langle \mathbf{A}_i, e^{x+\tau\delta}\rangle} e^{x_j+\tau\delta_j} - c_j\Big) \cdot \delta_j \geq -\Big(\sum_{i=1}^{d} \frac{r_i \mathbf{A}_{i,j}}{\langle \mathbf{A}_i, e^{x}\rangle} e^{x_j+\tau\delta_j} - c_j\Big) \cdot \delta_j$$

$$\geq -\Big(\sum_{i=1}^{d} \frac{r_i \mathbf{A}_{i,j}}{\langle \mathbf{A}_i, e^{x}\rangle} e^{x_j}(1+\tfrac{3}{4}\tau\delta_j) - c_j\Big) \cdot \delta_j = -\nabla_j \cdot \delta_j - \tfrac{3}{4}\tau\delta_j^2 \cdot \sum_{i=1}^{d} \frac{r_i \mathbf{A}_{i,j}}{\langle \mathbf{A}_i, e^{x}\rangle} e^{x_j}$$

If $j \in \Lambda^{\mathsf{s}}$, then $\sum_{i=1}^{d} \frac{r_i \mathbf{A}_{i,j}}{\langle \mathbf{A}_i, e^{x}\rangle} e^{x_j} = \nabla_j + c_j \leq 2c_j$ so

$$\int_{\tau=0}^{1} -(\nabla_j f(x+\tau\delta) \cdot \delta_j)\,\mathrm{d}\tau \geq -\nabla_j \cdot \delta_j - \tfrac{3}{4}\delta_j^2 c_j \geq -\nabla_j \cdot \delta_j - \tfrac{4}{3}\delta_j^2 c_j \ .$$

If $j \in \Lambda^{\mathsf{l}}$, then $\sum_{i=1}^{d} \frac{r_i \mathbf{A}_{i,j}}{\langle \mathbf{A}_i, e^{x}\rangle} e^{x_j} = \nabla_j + c_j \leq 2\nabla_j f(x)$ so

$$\int_{\tau=0}^{1} -(\nabla_j f(x+\tau\delta) \cdot \delta_j)\,\mathrm{d}\tau \geq -\nabla_j \cdot \delta_j - \tfrac{3}{4}\delta_j^2 \nabla_j \geq -\tfrac{1}{2}\nabla_j \cdot \delta_j \ . \qquad \square$$

## D.2 Proof of Lemma 4.8

**Lemma 4.8.** *If $z' = \mathtt{Mirr}^N(z, v)$, for every $u$ satisfying $\|u\|_\infty \leq N$, we have*

$$\langle v, z-u\rangle \leq \langle v, z-z'\rangle - \tfrac{1}{2}\|z-z'\|_c^2 + \tfrac{1}{2}\|z-u\|_c^2 - \tfrac{1}{2}\|z'-u\|_c^2 \ .$$

*Proof of Lemma 4.8.* [24] Denoting by $w = \big((z'_i - z_i)c_i\big)_{i=1}^{n} \in \mathbb{R}^n$, we compute that

$$\langle v, z-u\rangle = \langle v, z-z'\rangle + \langle v, z'-u\rangle$$
$$\overset{①}{\leq} \langle v, z-z'\rangle + \langle -w, z'-u\rangle$$
$$\overset{②}{=} \langle v, z-z'\rangle + \tfrac{1}{2}\|z-u\|_c^2 - \tfrac{1}{2}\|z'-u\|_c^2 - \tfrac{1}{2}\|z'-z\|_c^2 \ .$$

Here, ① is due to the minimality of $z' = \arg\min_{\|x\|_\infty \leq N}\{\|x-z\|_c^2 + \langle v, x\rangle\}$, which implies that $\langle \tfrac{\mathrm{d}}{\mathrm{d}x}(\|x-z\|_c^2 + \langle v, x\rangle)\big|_{x=z'}, u-z'\rangle = \langle w+v, u-z'\rangle \geq 0$ for all $\|u\|_\infty \leq N$. Equality ② can be verified by directly expanding out the three Euclidean norm squares, and is known as the "three-point equality" for Bregman divergence. $\qquad \square$

## D.3 Proof of Lemma 4.10

**Lemma 4.10.** *If $\tau_k \alpha_k \leq 3/64$, $\tau_k \in \big(0, \tfrac{1}{32N}\big]$, and $u$ is any vector satisfying $\|u\|_\infty \leq N$, then*

$$0 \leq \tfrac{1-\tau_k}{\tau_k}\big(f(y_k) - f(u)\big) - \tfrac{1}{\tau_k}\big(f(y_{k+1}) - f(u)\big) + \tfrac{1}{2\alpha_k}\|z_k - u\|_c^2 - \tfrac{1}{2\alpha_k}\|z_{k+1} - u\|_c^2 \ .$$

*Proof of Lemma 4.10.* Denoting by $\nabla = \nabla f(x) = \nabla^{\mathsf{s}} + \nabla^{\mathsf{l}}$ according to Def. 4.1, we use convexity to derive that

$$f(x_{k+1}) - f(u) \leq \langle \nabla f(x_{k+1}), x_{k+1}-u\rangle = \langle \nabla f(x_{k+1}), x_{k+1}-z_k\rangle + \langle \nabla^{\mathsf{l}}, z_k-u\rangle + \langle \nabla^{\mathsf{s}}, z_k-u\rangle \tag{D.1}$$

We bound the three terms on the right hand side of (D.1) separately.

---

[24] This proof can be found for instance in the textbook [6].

**The first term in (D.1).** We have $x_{k+1} = \tau_k z_k + (1-\tau_k) y_k$ and therefore

$$\langle \nabla f(x_{k+1}), x_{k+1} - z_k \rangle = \frac{1-\tau_k}{\tau_k} \langle \nabla f(x_{k+1}), y_k - x_{k+1} \rangle \leq \frac{1-\tau_k}{\tau_k} \left( f(y_k) - f(x_{k+1}) \right) . \tag{D.2}$$

**The second term in (D.1).** Define $u' \in \mathbb{R}^n$ where $u'_i = \min\{z_{k,i}, u_i\}$. Then, we have

$$\langle \nabla^{\mathsf{l}}, z_k - u \rangle \leq \langle \nabla^{\mathsf{l}}, z_k - u' \rangle \tag{D.3}$$

because $\nabla^{\mathsf{l}}$ is a non-negative vector. Now, define $v = 8\tau_k u' - 7\tau_k z_k + (1-\tau_k) y_k$ and compare this to $x_{k+1} = \tau_k z_k + (1-\tau_k) y_k$, we have that $x_{k+1} - v = 8\tau_k(z_k - u') \geq 0$. It is easy to verify that $\|v\|_\infty \leq 15N$ because $\|z_k\|_\infty \leq N$, $\|u'\|_\infty \leq N$, $\|z_k - u'\|_\infty \leq 2N$, and $\|y_k\|_\infty \leq 15N$.

Now, consider two types of indices, let

- $A \subseteq [n]$ be the set of indices $i$ where $\nabla^{\mathsf{l}}_i > 0$ and $x_{k+1,i} > v_i + 1/2$, and
- $B \subseteq [n]$ be the set of indices $i$ where $\nabla^{\mathsf{l}}_i > 0$ and $x_{k+1,i} \leq v_i + 1/2$.

We apply Corollary 4.4 by choosing the following $\delta \in [-1/2, 0]^n$:

$$\delta_j = \begin{cases} -1/2, & \text{if } j \in A; \\ v_i - x_{k+1,i} \in [-1/2, 0], & \text{if } j \in B; \\ 0, & \text{if } j \notin A \cup B. \end{cases}$$

It must satisfy that

$$f(x_{k+1}) - f(y_{k+1}) \overset{\text{①}}{\geq} \frac{1}{2} Q^-(x_{k+1}, \delta) \overset{\text{②}}{\geq} -\frac{1}{4} \sum_{j \in \Lambda^{\mathsf{l}}} \nabla_j \cdot \delta_j \overset{\text{③}}{\geq} -\frac{1}{4} \sum_{j \in A} \nabla^{\mathsf{l}}_j \cdot \delta_j - \frac{1}{4} \sum_{j \in B} \nabla^{\mathsf{l}}_j \cdot \delta_j$$

$$= \frac{1}{8} \sum_{j \in A} \nabla^{\mathsf{l}}_j + \frac{1}{4} \sum_{j \in B} \nabla^{\mathsf{l}}_j \cdot (x_{k+1,j} - v_j) \overset{\text{④}}{\geq} \frac{1}{16N} \sum_{j \in A} \nabla^{\mathsf{l}}_j \cdot (z_{k,j} - u'_j) + 2\tau_k \sum_{j \in B} \nabla^{\mathsf{l}}_j \cdot (z_{k,j} - u'_j)$$

$$\overset{\text{⑤}}{\geq} 2\tau_k \sum_{j \in A \cup B} \nabla^{\mathsf{l}}_j \cdot (z_{k,j} - u'_j) = 2\tau_k \langle \nabla^{\mathsf{l}}, z_k - u' \rangle \overset{\text{⑥}}{\geq} 2\tau_k \langle \nabla^{\mathsf{l}}, z_k - u \rangle \tag{D.4}$$

Above,

- Inequality ① comes from Corollary 4.4.

  (Note that we can apply Corollary 4.4 because $\|x_{k+1} + \delta\|_\infty \leq 15N$, which comes from the definition of $\delta$ and the facts $\|v\|_\infty \leq 15N$ and $\|x_{k+1}\|_\infty \leq 15N$.)

- Inequality ② follows from the definition of $Q^-$ and the fact $\delta \leq 0$, see Lemma 4.2.
- Inequality ③ uses $\nabla_j \geq \nabla^{\mathsf{l}}_j$ for every $j \in \Lambda^{\mathsf{l}}$ and the fact $\delta \leq 0$.
- Inequality ④ uses $\|z_k\|_\infty \leq N$ and $\|u'\|_\infty \leq N$ so $z_{k,j} - u'_j \in [0, 2N]$, as well as the fact that $x_{k+1} - v = 8\tau_k(z_k - u') \geq 0$.
- Inequality ⑤ uses our assumption on $\tau_k$.
- Inequality ⑥ comes from (D.3).

**The third term in (D.1).** We first apply Lemma 4.8 to obtain

$$\langle \nabla^{\mathsf{s}}, z_k - u \rangle \leq \langle \nabla^{\mathsf{s}}, z_k - z_{k+1} \rangle - \frac{1}{2\alpha_k} \|z_k - z_{k+1}\|_c^2 + \frac{1}{2\alpha_k} \|z_k - u\|_c^2 - \frac{1}{2\alpha_k} \|z_{k+1} - u\|_c^2 . \tag{D.5}$$

Define $v = 8\tau_k z_{k+1} - 7\tau_k z_k + (1-\tau_k) y_k$ and compare this to $x_{k+1} = \tau_k z_k + (1-\tau_k) y_k$, we have that $x_{k+1} - v = 8\tau_k(z_k - z_{k+1})$. It is easy to verify that $\|v\|_\infty \leq 15N$ because $\|z_k\|_\infty \leq N$, $\|z_{k+1}\|_\infty \leq N$, $\|z_k - z_{k+1}\|_\infty \leq 2N$, and $\|y_k\|_\infty \leq 15N$.

We rewrite the first two terms on the right hand side of (D.5) as:

$$\langle \nabla^{\mathsf{s}}, z_k - z_{k+1} \rangle - \frac{1}{2\alpha_k} \| z_k - z_{k+1} \|_c^2 = \frac{1}{8\tau_k} \langle \nabla^{\mathsf{s}}, x_{k+1} - v \rangle - \frac{1}{2 \cdot 8^2 \tau_k^2 \alpha_k} \| x_{k+1} - v \|_c^2$$
$$\leq \frac{1}{8\tau_k} \big( \langle \nabla^{\mathsf{s}}, x_{k+1} - v \rangle - \frac{4}{3} \| x_{k+1} - v \|_c^2 \big) \ . \quad (\text{D.6})$$

Above, the inequality uses our provided upper bound to $\alpha_k \tau_k$. Next, we define $v'$ by

$$v'_j = \begin{cases} v_j, & \text{if } -\nabla^{\mathsf{s}}_j \cdot (v_j - x_{k+1,j}) - \frac{4}{3} c_j \cdot (v_j - x_{k+1,j})^2 \geq 0; \\ x_{k+1,j}, & \text{if } -\nabla^{\mathsf{s}}_j \cdot (v_j - x_{k+1,j}) - \frac{4}{3} c_j \cdot (v_j - x_{k+1,j})^2 \leq 0. \end{cases}$$

Therefore, (D.6) also implies

$$\langle \nabla^{\mathsf{s}}, z_k - z_{k+1} \rangle - \frac{1}{2\alpha_k} \| z_k - z_{k+1} \|_c^2 \leq \frac{1}{8\tau_k} \big( \langle \nabla^{\mathsf{s}}, x_{k+1} - v' \rangle - \frac{4}{3} \| x_{k+1} - v' \|_c^2 \big) \ .$$

Define $A^{--}, A^-, A^+, A^{++} \subseteq [n]$ to respectively be the set of indices $i$ where $v'_i - x_{k+1,i}$ is in the range $(-\infty, -1/2], (-1/2, 0), (0, 1/2),$ and $[1/2, \infty)$. Define $\delta^+, \delta^- \in \mathbb{R}^n$ as the vector where

$$\delta^+_j = \begin{cases} -\frac{3}{8} \nabla^{\mathsf{s}}_j / c_j, & \text{if } j \in A^{++}; \\ v'_j - x_{k+1,j}, & \text{if } j \in A^+; \\ 0, & \text{otherwise.} \end{cases} \quad \text{and} \quad \delta^-_j = \begin{cases} -\frac{3}{8} \nabla^{\mathsf{s}}_j / c_j, & \text{if } j \in A^{--}; \\ v'_j - x_{k+1,j}, & \text{if } j \in A^-; \\ 0, & \text{otherwise.} \end{cases}$$

One can carefully verify that $\delta^+ \in [0, 1/2]^n$ and $\| x_{k+1} + \delta^+ \|_\infty \leq 15N$.[25] Similarly, it also satisfies $\delta^- \in [-1/2, 0]^n$ and $\| x_{k+1} + \delta^- \|_\infty \leq 15N$. Therefore, applying Corollary 4.4 we have

$$f(x_{k+1}) - f(y_{k+1}) \geq \frac{1}{2} Q^+(x_{k+1}, \delta^+) + \frac{1}{2} Q^-(x_{k+1}, \delta^-) \ . \quad (\text{D.7})$$

---

[25] We verify them coordinate by coordinate. For each coordinate $j \in [n]$,

- If $j \notin A^+ \cup A^{++}$ then $\delta^+_j = 0$ so $|x_{k+1,j} + \delta^+_j| = |x_{k+1,j}| \leq 15N$.
- If $j \in A^+$, we have $\delta^+_j \in [0, 1/2]$ by the definition of $A^+$, as well as $|x_{k+1,j} + \delta^+_j| = |v'_j| \leq 15N$.
- If $j \in A^{++}$ then we have $v'_j - x_{k+1,j} > 0$, so according to the definition of $v'$, we must have

$$v'_j = v_j, \quad v_j - x_{k+1,j} > 0, \quad \text{and} \quad \nabla^{\mathsf{s}}_j \leq -\frac{4}{3} c_j (v_j - x_{k+1,j}) < 0$$

However, $\nabla^{\mathsf{s}}_j \in [-c_j, c_j]$ so this means $\delta^+_j = -\frac{3}{8} \nabla^{\mathsf{s}}_j / c_j \in (0, 3/8]$. At the same time, $x_{k+1,j} + \delta^+_j < x_{k+1,j} + \frac{1}{2} \leq v'_j \leq 15N$.

Denote by $\delta = \delta^+ + \delta^-$, and noticing that $\delta_j^+$ and $\delta_j^-$ cannot be both nonzero, we have

$$Q^+(x_{k+1}, \delta^+) + Q^-(x_{k+1}, \delta^-)$$
$$\stackrel{\text{①}}{=} \sum_{j \in \Lambda^s} \left(-\nabla_j \cdot \delta_j - \frac{4}{3} c_j \cdot \delta_j^2\right) + \sum_{j \in \Lambda^l, \delta_j > 0} \left(-\frac{7}{3}\nabla_j \cdot \delta_j\right) + \sum_{j \in \Lambda^l, \delta_j < 0} \left(-\frac{1}{2}\nabla_j \cdot \delta_j\right)$$
$$\stackrel{\text{②}}{=} \sum_{j \in \Lambda^s \cap (A^+ \cup A^-)} \left(-\nabla_j^s \cdot \delta_j - \frac{4}{3} c_j \cdot (\delta_j)^2\right) + \sum_{j \in \Lambda^s \cap (A^{++} \cup A^{--})} \left(-\nabla_j^s \cdot \delta_j - \frac{4}{3} c_j \cdot (\delta_j)^2\right)$$
$$+ \sum_{j \in \Lambda^l \cap A^-} \left(-\frac{1}{2}\nabla_j \cdot \delta_j\right) + \sum_{j \in \Lambda^l \cap A^{--}} \left(-\frac{1}{2}\nabla_j \cdot \delta_j\right)$$
$$\stackrel{\text{③}}{\geq} \sum_{j \in \Lambda^s} \left(-\nabla_j^s \cdot (v_j' - x_{k+1,j}) - \frac{4}{3} c_j \cdot (v_j' - x_{k+1,j})^2\right) + \sum_{j \in \Lambda^l \cap A^-} \left(-\frac{1}{2}\nabla_j \cdot \delta_j\right) + \sum_{j \in \Lambda^l \cap A^{--}} \left(-\frac{1}{2}\nabla_j \cdot \delta_j\right)$$
$$\stackrel{\text{④}}{\geq} \sum_{j \in \Lambda^s} \left(-\nabla_j^s \cdot (v_j' - x_{k+1,j}) - \frac{4}{3} c_j \cdot (v_j' - x_{k+1,j})^2\right) + \frac{1}{2}\sum_{j \in \Lambda^l} \left(-\nabla_j^s \cdot (v_j' - x_{k+1,j}) - \frac{4}{3} c_j \cdot (v_j' - x_{k+1,j})^2\right)$$
$$\geq \frac{1}{2} \sum_{j \in [n]} \left(-\nabla_j^s \cdot (v_j' - x_{k+1,j}) - \frac{4}{3} c_j \cdot (v_j' - x_{k+1,j})^2\right) = \frac{1}{2}\left(\langle \nabla^s, x_{k+1} - v\rangle - \frac{4}{3}\|x_{k+1} - v\|_c^2\right) .$$
(D.8)

Above,

- Equality ① uses the definitions of $Q^-, Q^+$, see Lemma 4.2.
- Equality ② uses the fact that whenever $j \in \Lambda^l$ it must satisfy $\delta_j \leq 0$.

  Indeed, it satisfies $\nabla_j^s = c_j$ so either $v_j' = x_{k+1,j}$ (in such a case $\delta_j = 0$) or $v_j' = v_j$ (in such a case $v_j - x_{k+1,j} < 0$ so $\delta_j < 0$).

- Inequality ③ is because (1) if $j \in A^+ \cup A^-$ then $\delta_j = v_j' - x_{k+1,j}$ or (2) if $j \in A^{++} \cup A^{--}$ then $\arg\max_{t \in \mathbb{R}} \{-\nabla_j^s \cdot t - \frac{4}{3} c_j \cdot t^2\} = -\frac{3}{8}\nabla_j^s / c_j = \delta_j$.

- Inequality ④ is because

  - For every $j \in \Lambda^l \cap A^-$, we have $-\frac{1}{2}\nabla_j \cdot \delta_j \geq -\frac{1}{2}\nabla_j^s \cdot \delta_j \geq \frac{1}{2}\left(-\nabla_j^s \cdot \delta_j - \frac{4}{3} c_j (\delta_j)^2\right) = \frac{1}{2}\left(-\nabla_j^s \cdot (v_j' - x_{k+1,j}) - \frac{4}{3} c_j \cdot (v_j' - x_{k+1,j})^2\right)$.
  - For every $j \in \Lambda^l \cap A^{--}$, we have $-\frac{1}{2}\nabla_j \cdot \delta_j = \frac{3}{16} c_j \geq \frac{1}{2}\left(-\nabla_j^s \cdot (v_j' - x_{k+1,j}) - \frac{4}{3} c_j \cdot (v_j' - x_{k+1,j})^2\right)$, where the last inequality is because when $\nabla_j^s = c_j$ —which holds since $j \in \Lambda^l$— it satisfies $\max_{t \in \mathbb{R}}\{-\nabla_j^s \cdot t - \frac{4}{3} c_j \cdot t^2\} = \frac{3}{16} c_j$.

Combining (D.5), (D.6), (D.7), and (D.8), we have

$$\langle \nabla^s, z_k - u\rangle \leq \frac{1}{2\tau_k}(f(x_{k+1}) - f(y_{k+1})) + \frac{1}{2\alpha_k}\|z_k - u\|_c^2 - \frac{1}{2\alpha_k}\|z_{k+1} - u\|_c^2 . \quad (D.9)$$

**Finally.** Combining the three cases above, namely (D.2), (D.4), and (D.9), and plugging them back to (D.1), we have

$$f(x_{k+1}) - f(u) \leq \frac{1 - \tau_k}{\tau_k}\left(f(y_k) - f(x_{k+1})\right) + \frac{1}{\tau_k}\left(f(x_{k+1}) - f(y_{k+1})\right) + \frac{1}{2\alpha_k}\|z_k - u\|_c^2 - \frac{1}{2\alpha_k}\|z_{k+1} - u\|_c^2 .$$

After rearranging, we finish the proof of the lemma. □

## D.4 Proof of Theorem 4.11

**Theorem 4.11.** *If $y_0$ satisfies $\|y_0\|_\infty \leq 15N$ and $T \geq 1$, then the output $y_T = \text{LC}(\mathbf{A}, N, T, y_0)$ (see Algorithm 1) satisfies that for every $u \in \mathbb{R}^n$ and $\|u\|_\infty \leq N$:*

$$\|y_T\|_\infty \leq 15N \quad \text{and} \quad f(y_T) - f(u) \leq O\Big(\frac{N^2\big(f(y_0) - f(u) + h\big)}{(N+T)^2}\Big) .$$

*Proof of Theorem 4.11.* First of all, we show inductively that $\tau_k \leq \frac{2}{64N+k}$. Recall that $\tau_0 = \frac{1}{32N}$ and $\tau_k$ is the unique positive root of the quadratic equation $\frac{\tau_k^2}{\tau_{k-1}^2} + \tau_k - 1 = 0$. Suppose this upper bound holds for $\tau_{k-1}$ and we wish to prove that for $\tau_k$. If $\tau_k \geq \frac{2}{64N+k}$ then

$$\frac{\tau_k^2}{\tau_{k-1}^2} + \tau_k - 1 \geq \Big(\frac{64N + (k-1)}{64N + k}\Big)^2 + \frac{2}{64N+k} - 1 = \frac{1}{(64N+k)^2} > 0$$

which is a contradiction to the fact that $\tau_k$ is a root. Thus, we must have $\tau_k < \frac{2}{64N+k}$.

Now, since we have chosen $\tau_k \alpha_k = 3/64$, we rewrite Lemma 4.10 as

$$0 \leq \frac{1-\tau_k}{\tau_k^2}\big(f(y_k) - f(u)\big) - \frac{1}{\tau_k^2}\big(f(y_{k+1}) - f(u)\big) + \frac{32}{3}\|z_k - u\|_c^2 - \frac{32}{3}\|z_{k+1} - u\|_c^2 .$$

Since $\frac{1-\tau_k}{\tau_k^2} = \frac{1}{\tau_{k-1}^2}$, we can telescope the above inequality for $k = 0, 1, \ldots, T-1$ and obtain

$$\frac{1}{\tau_{T-1}^2}\big(f(y_T) - f(u)\big) \leq \frac{1-\tau_0}{\tau_0^2}\big(f(y_0) - f(u)\big) + \frac{32}{3}\|z_0 - u\|_c^2 - \frac{32}{3}\|z_T - u\|_c^2 . \tag{D.10}$$

Plugging our bound $\tau_k \leq \frac{2}{64N+k}$ and $\tau_0 = \frac{1}{32N}$, and noticing that $z_0 = 0$, we immediately have that

$$f(y_T) - f(u) \leq O\Big(\frac{N^2\big(f(y_0) - f(u)\big) + \|u\|_c^2}{(N+T)^2}\Big) . \qquad \square$$

## D.5 Proof of Theorem 4.12

We first establish a relationship between $\|\nabla^{\mathsf{s}}\|_{c^{-1}}^2 + \|\nabla^{\mathsf{l}}\|_1$ and $\|\nabla f(x)\|_{c^{-1}}^2$:

**Claim D.1.** *If $\nabla f(x) = \nabla^{\mathsf{s}} + \nabla^{\mathsf{l}}$ using Def. 4.1, and if $\|\nabla^{\mathsf{s}}\|_{c^{-1}}^2 + \|\nabla^{\mathsf{l}}\|_1 \leq t \in [0, 1]$, then $\|\nabla f(x)\|_{c^{-1}}^2 \leq 3t$ and thus $x$ gives a $3t$-approximate $(r, c)$-scaling for $\mathbf{A}$.*

*Proof of Claim D.1.* It satisfies $\|\nabla f(z_k')\|_{c^{-1}}^2 = \|\nabla^{\mathsf{s}}\|_{c^{-1}}^2 + \sum_{j:\, \nabla_j f(z_k') > c_j} \big(\nabla_j f(z_k')\big)^2/c_j \leq 3\|\nabla^{\mathsf{s}}\|_{c^{-1}}^2 + 2\sum_{j:\, \nabla_j f(z_k') > c_j} \big(\nabla_j f(z_k') - c_j\big)^2/c_j \leq 3\|\nabla^{\mathsf{s}}\|_{c^{-1}}^2 + 2\|\nabla^{\mathsf{l}}\|_1^2 \leq 3t$. $\square$

---

**Theorem 4.12** (Scaling0). *If $N \geq 1$, then $(z_1, z) = \text{Scaling0}(\mathbf{A}, N, T)$ satisfies*

- *If $T \geq N$, then for every $u$ satisfying $\|u\|_\infty \leq N$, we have*

$$\|z_1\|_\infty \leq 15N \quad \text{and} \quad f(z_1) - f(u) \leq O\big(\tfrac{N^2 h}{T^2}\big) .$$

- *If $T \geq (N^2 h)^{1/3}$ and there exists $u$ so that $\|u\|_\infty \leq N$ and $f(u) - \inf_x\{f(x)\} \leq 1$, then*

$$\|\nabla f(z)\|_{c^{-1}}^2 \leq O\big(\tfrac{N^2 h}{T^3}\big) .$$

*The total complexity of Scaling0 is $O(m(N \log N + T))$.*

---

*Proof of Theorem 4.12.* Recall that `Scaling0` repeatedly applies `LC`. In each of the first $\log(N)$ applications of `LC`, it satisfies according to Theorem 4.11 that

$$\|y_T\|_\infty \leq 15N \quad \text{and} \quad f(y_T) - f(u) \leq O\Big(\frac{1}{2}\big(f(y_0) - f(u)\big) + h\Big) \ .$$

Since it satisfies $f(0) - f(u) \leq 2h\|u\|_\infty \leq O(hN)$ according to Lemma 3.1, after $\log(N)$ applications of `LC` we must have $\|z_0\|_\infty \leq 15N$ and $f(z_0) - f(u) \leq O(h)$. Then, applying `LC` for $T$ iterations, we have

$$\|z_1\|_\infty \leq 15N \text{ and } f(z_1) - f(u) \leq O(N^2 h/T^2) \text{ as desired.}$$

Finally, applying Corollary 4.5 for $T$ steps, we have that there exists some $k \in [T]$ so that $z_k$ satisfies $\|\nabla^{\mathsf{s}}\|^2_{c^{-1}} + \|\nabla^{\mathsf{l}}\|_1 \leq O\big(\frac{f(z_1) - \inf_x\{f(x)\}}{T}\big) \leq O\big(\frac{f(z_1) - f(u) + 1}{T}\big) \leq O\big(\frac{N^2 h}{T^3}\big)$, where $\nabla f(z_k) = \nabla^{\mathsf{s}} + \nabla^{\mathsf{l}}$ is the gradient splitting using Def. 4.1. Therefore, as long as $N^2 h/T^3 \leq 1$, it must satisfy $\|\nabla f(z_k)\|^2_{c^{-1}} \leq O(\|\nabla^{\mathsf{s}}\|^2_{c^{-1}} + \|\nabla^{\mathsf{l}}\|_1) \leq O\big(\frac{N^2 h}{T^3}\big)$ owing to Claim D.1. □

# E  Missing Details for Section 5

## E.1  Proof of Lemma 5.1

**Lemma 5.1** (second-order approximation). *For every $x, \delta \in \mathbb{R}^n$ with $\|\delta\|_\infty \leq 1/8$, we have*

$$f(x) + \langle \delta, \nabla f(x)\rangle + \frac{1}{6}\delta^\top \nabla^2 f(x)\delta \leq f(x+\delta) \leq f(x) + \langle \delta, \nabla f(x)\rangle + \delta^\top \nabla^2 f(x)\delta \ .$$

First of all, we claim that it suffices to prove Lemma 5.1 for $x = 0$. This is because, if we define a new matrix $\mathbf{A}'$ by setting $\mathbf{A}'_{i,j} = \mathbf{A}_{i,j} \cdot e^{x_j}$, and a corresponding function $g(\cdot)$ defined in the same way as $f(x)$ but with this new matrix $\mathbf{A}'$, then, it satisfies $f(y) = g(y-x) + c^\top x$, $\nabla f(y) = \nabla g(y-x)$ and $\nabla^2 f(y) = \nabla^2 g(y-x)$ for any $y \in \mathbb{R}^n$

Now, for a fixed direction $(z_1, \cdots, z_d)$, slightly abusing notation, we also define the univariate function $f(t) \stackrel{\text{def}}{=} f(tz_1, \cdots, tz_n)$. We make the following claim which directly implies Lemma 5.1.

**Claim E.1.** *If $\|z\|_\infty \leq 1$ and $|t| \leq \frac{1}{8}$, we have:*

$$\left| f(t) - \left(f(0) + \frac{df}{dt}\Big|_{t=0} t + \frac{1}{2}\frac{d^2 f}{dt^2}\Big|_{t=0} t^2\right) \right| \leq \frac{1}{3}\frac{d^2 f}{dt^2}\Big|_{t=0} t^2$$

*Proof of Claim E.1.* The proof is just a direct calculation. We have

$$f'(t) = \sum_{i=1}^d \left( r_i \left( \frac{\sum_{j=1}^n \mathbf{A}_{i,j} z_j e^{tz_j}}{\sum_{j=1}^n \mathbf{A}_{i,j} e^{tz_j}} \right) \right) - \langle c, z \rangle \ . \tag{E.1}$$

We consider a fixed $i \in [n]$ and focus on the term $\frac{\sum_{j=1}^n \mathbf{A}_{i,j} z_j e^{tz_j}}{\sum_{j=1}^n \mathbf{A}_{i,j} e^{tz_j}}$. Let us denote

$$h_k(t) \stackrel{\text{def}}{=} \sum_{j=1}^n \mathbf{A}_{i,j} z_j^k e^{tz_j} \quad \text{and} \quad g(t) \stackrel{\text{def}}{=} \frac{h_1(t)}{h_0(t)} = \frac{\sum_{j=1}^n \mathbf{A}_{i,j} z_j e^{tz_j}}{\sum_{j=1}^n \mathbf{A}_{i,j} e^{tz_j}} \ . \tag{E.2}$$

We wish to bound the higher-order derivatives of $f(t)$ by bounding higher-order derivatives of $g$.

Since $h_0(t) g(t) = h_1(t)$, by the chain rule of derivative, we have for all $s \in \mathbb{N}$

$$\frac{d^s h_1}{dt^s} = \sum_{k=0}^s \binom{s}{k} \frac{d^k g}{dt^k} \frac{d^{s-k} h_0}{dt^{s-k}} = \sum_{k=0}^s \binom{s}{k} h_{s-k} \frac{d^k g}{dt^k}$$

where the last equality is due to $\frac{\mathrm{d}h_k}{\mathrm{d}t} = h_{k+1}$. Now,
$$\frac{\mathrm{d}^s g}{\mathrm{d}t^s} = \frac{1}{h_0}\left(h_{s+1} - h_s g_0 - \sum_{k=1}^{s-1}\binom{s}{k}h_{s-k}\frac{\mathrm{d}^k g}{\mathrm{d}t^k}\right) \ .$$

Again, for notation simplicity, let us denote $g_k \stackrel{\text{def}}{=} \frac{\mathrm{d}^k g}{\mathrm{d}t^k}$, we then have:
$$|g_s| \leq \frac{1}{|h_0|}\left(|h_{s+1} - h_s g_0| + \sum_{k=1}^{s-1}\binom{s}{k}|h_{s-k}||g_k|\right) \leq \frac{1}{|h_0|}(|h_{s+1} - h_s g_0|) + \sum_{k=1}^{s-1}\binom{s}{k}|g_k| \ ,$$

where the second inequality is because when $|z_j| \leq 1$, we always have $|h_s| \leq |h_0| = h_0$. This implies
$$|g_s(0)| \leq \frac{1}{|h_0(0)|}(|h_{s+1}(0) - h_s(0)g_0(0)|) + \sum_{k=1}^{s-1}\binom{s}{k}|g_k(0)|$$

Let us first check the term $h_0(0) \cdot (h_{s+1}(0) - h_s(0)g_0(0)) = h_0 h_{s+1}(0) - h_s(0)h_1(0)$. We know that
$$h_0(0)h_{s+1}(0) - h_s(0)h_1(0) = \sum_{1 \leq j_1 < j_2 \leq d} \mathbf{A}_{i,j_1}\mathbf{A}_{i,j_2}\left(z_{j_1}^{s+1} + z_{j_2}^{s+1} - z_{j_1}^s z_{j_2} - z_{j_2}^s z_{j_1}\right)$$
$$= \sum_{1 \leq j_1 < j_2 \leq d} \mathbf{A}_{i,j_1}\mathbf{A}_{i,j_2}(z_{j_1} - z_{j_2})^2\left(\sum_{k=0}^{s-1} z_{j_1}^k z_{j_2}^{s-1-k}\right)$$
$$\leq s \sum_{1 \leq j_1 < j_2 \leq d} \mathbf{A}_{i,j_1}\mathbf{A}_{i,j_2}(z_{j_1} - z_{j_2})^2$$
$$= s(h_0(0)h_2(0) - h_1(0)^2) = sh_0(0)^2 g_1(0) \ .$$

In the same manner, we can also show $h_0(0)h_{s+1}(0) - h_s(0)h_1(0) \geq -sh_0(0)^2 g_1(0)$. Therefore,
$$|g_s(0)| \leq s|g_1(0)| + \sum_{k=1}^{s-1}\binom{s}{k}|g_k(0)| \ .$$

At this point, we can inductively prove that $|g_s(0)| \leq 2^s s!|g_1(0)|$ for $s \geq 1$. This is true for $s = 1$. For $s \geq 2$, we have:
$$|g_s(0)| \leq |g_1(0)| \cdot \left(s + \sum_{k=1}^{s-1}\binom{s}{k}2^k k!\right) = |g_1(0)| \cdot \left(s + s!\sum_{k=1}^{s-1}\frac{2^k}{(s-k)!}\right) \leq |g_1(0)|(s + s!(2^s - 1)) \leq 2^s s!|g_1(0)| \ .$$

Plugging this upper bound $|g_s(0)| \leq 2^s s!|g_1(0)|$ back to (E.1) and (E.2), and summing up over all indices $i \in [n]$, we immediately have $\forall s \geq 2$,
$$\left|\frac{\mathrm{d}^s f}{\mathrm{d}t^s}\Big|_{t=0}\right| \leq 2^s s!\frac{\mathrm{d}^2 f}{\mathrm{d}t^2}\Big|_{t=0} \ .$$

Using Taylor expansion, we can conclude that for all $t \leq \frac{1}{8}$,
$$\left|f(t) - \left(f(0) + \frac{\mathrm{d}f}{\mathrm{d}t}\Big|_{t=0}t + \frac{1}{2}\frac{\mathrm{d}^2 f}{\mathrm{d}t^2}\Big|_{t=0}t^2\right)\right| \leq \frac{1}{3}\frac{\mathrm{d}^2 f}{\mathrm{d}t^2}\Big|_{t=0}t^2 \ . \qquad \square$$

### E.2 Proof of Lemma 5.4

**Lemma 5.4.** *Given $x$ with $\|x\|_\infty \leq N$ and Laplacian matrix $\mathbf{H}$ with $\mathbf{H} \preceq \nabla^2 f(x) \preceq 1.1\mathbf{H}$, we have*
*(a) For every $u \in \mathbb{R}^n$ with $\|u\|_\infty \leq N$,*
$$-\min_{\delta \in \mathsf{box}^N(x)}\left\{\langle \nabla f(x), \delta\rangle + \tfrac{1}{6}\delta^\top \mathbf{H}\delta\right\} \geq \tfrac{1}{64N}(f(x) - f(u)) \ .$$

(b) For $\varepsilon \geq 0$, If we are given $\widehat{\delta}$ satisfying $\|\widehat{\delta}\|_\infty \leq 1/8$ and

$$\langle \nabla f(x), \widehat{\delta}\rangle + \tfrac{1}{6}\widehat{\delta}^\top \mathbf{H}\widehat{\delta} \leq \min_{\delta \in \mathsf{box}^N(x)} \left\{ \langle \nabla f(x), \delta\rangle + \tfrac{1}{6}\delta^\top \mathbf{H}\delta \right\} + \varepsilon \ ,$$

then it satisfies that for every $u \in \mathbb{R}^n$ with $\|u\|_\infty \leq N$, $f(x)-f\big(x+\tfrac{\widehat{\delta}}{6.6}\big) \geq \tfrac{1}{900N}\big(f(x)-f(u)\big)-\varepsilon$.

*Proof of Lemma 5.4.* Let $y = \tfrac{64N-1}{64N}x + \tfrac{1}{64N}u$ then one can carefully verify that $y - x \in \mathsf{box}^N(x)$. By the convexity of $f(\cdot)$, we have $f(y) \leq (1 - \tfrac{1}{64N})f(x) + \tfrac{1}{64N}f(u)$ and therefore $f(x) - f(y) \geq \tfrac{1}{64N}\big(f(x) - f(u)\big)$. This implies

$$f(x) - \min_{\delta \in \mathsf{box}^N(x)} \{f(x+\delta)\} \geq \frac{1}{64N}\big(f(x) - f(u)\big) \ .$$

Now, using Lemma 5.1 and $\nabla^2 f(x) \succeq \mathbf{H}$, we further have

$$-\min_{\delta \in \mathsf{box}^N(x)} \{\langle \nabla f(x), \delta\rangle + \tfrac{1}{6}\delta^\top \mathbf{H}\delta\} \geq \frac{1}{64N}\big(f(x) - f(u)\big) \ .$$

The conditions in item (b) imply

$$-\Big(\langle \nabla f(x), \frac{\widehat{\delta}}{6.6}\rangle + 1.1\frac{\widehat{\delta}^\top}{6.6}\mathbf{H}\frac{\widehat{\delta}}{6.6}\Big) = -\frac{1}{6.6}\Big(\langle \nabla f(x), \widehat{\delta}\rangle + \tfrac{1}{6}\widehat{\delta}^\top \mathbf{H}\widehat{\delta}\Big) \geq \frac{1}{900N}\big(f(x) - f(u)\big) - \varepsilon \ .$$

At the same time, applying Lemma 5.1 again together with $\nabla^2 f(x) \preceq 1.1\mathbf{H}$, we have

$$f(x) - f(x+\widehat{\delta}/6.6) \geq -\Big(\langle \nabla f(x), \frac{\widehat{\delta}}{6.6}\rangle + \frac{\widehat{\delta}^\top}{6.6}\nabla^2 f(x)\frac{\widehat{\delta}}{6.6}\Big) \geq \frac{1}{900N}\big(f(x) - f(u)\big) - \varepsilon \ . \quad \square$$

# F  Missing Details for Section 6

## F.1  Proof of Lemma 6.1

**Lemma 6.1** (MWUbasic). *If $\mathbf{H} \in \mathbb{R}^{n \times n}$ is Laplacian, $K \geq 1$, $T \geq \Omega((n^{1/2}K + K^2)\log n)$, $\|\alpha\|_\infty \leq 1/32$, and $\varepsilon > 0$, then the output $\overline{\delta} = \mathsf{MWUbasic}(\mathbf{A}, \mathbf{H}, \alpha, T, K, \varepsilon)$ satisfies*

$$\|\overline{\delta} - \alpha\|_\infty \leq \frac{1}{32} + \frac{1}{8K} \quad \text{and} \quad \langle \nabla, \overline{\delta}\rangle + \tfrac{1}{6}\overline{\delta}^\top \mathbf{H}\overline{\delta} \leq \min_{\|\delta-\alpha\|_\infty \leq 1/32} \left\{\langle \nabla, \delta\rangle + \tfrac{1}{6}\delta^\top \mathbf{H}\delta\right\} + \varepsilon \ .$$

*Proof of Lemma 6.1.* Our MWUbasic runs runs the constrained multiplicative weight update method introduced in Section A.1 with $T$ iterations and $\beta_i = \tfrac{1}{2}$ for each $i \in [n]$.

Define $\mathsf{OPT} = -\min_{\|\delta-\alpha\|_\infty \leq 1/32}\{\langle \nabla, \delta\rangle + \tfrac{1}{6}\delta^\top \mathbf{H}\delta\} \geq 0$ .

In each iteration $k$, we use Lemma A.4 to find a vector $\delta_k \in \mathbb{R}^n$ satisfying

$$\|\delta_k - \alpha\|_{w_k}^2 \leq \frac{n}{1024} \quad \text{and} \quad \langle \nabla f(x), \delta_k\rangle + \tfrac{1}{6}\delta_k^\top \mathbf{H}\delta_k \leq \min_{\|\delta-\alpha\|_{w_k}^2 \leq n/1024} \left\{\langle \nabla f(x), \delta\rangle + \tfrac{1}{6}\delta^\top \mathbf{H}\delta\right\} + \varepsilon$$

Since any vector $\delta$ with $\|\delta - \alpha\|_\infty \leq 1/32$ also satisfies $\|\delta - \alpha\|_{w_k}^2 = \sum_{i \in [n]}(\delta_i - \alpha_i)^2 w_{k,i} \leq \tfrac{1}{1024}\sum_{i \in [n]} w_{k,i} = n/1024$, we must have

$$\langle \nabla f(x), \delta_k\rangle + \tfrac{1}{6}\delta_k^\top \mathbf{H}\delta_k \leq -\mathsf{OPT} + \varepsilon \ .$$

Define the loss vector $\ell_k$ by setting $\ell_{k,i} = -|\delta_{k,i} - \alpha_i|$. Since each $(\delta_{k,i} - \alpha_i)^2 \leq \tfrac{n}{1024w_{k,i}} \leq n$, we can choose $\rho = \sqrt{n}$ and $\eta = 1/(\sqrt{n}+K)$ in order to apply Lemma A.1. We have that for every

$u \in \Delta$:

$$\frac{1}{T}\sum_{k=0}^{T-1}\langle \ell_k, -u\rangle \leq \frac{1}{T}\sum_{k=0}^{T-1}\left(\langle -\ell_k, w_k\rangle + 2\eta\|\ell_k\|_{w_k}^2\right) + \frac{n\log(2n^2)}{\eta T}$$

$$\leq \frac{1}{T}\sum_{k=0}^{T-1}\left(\sqrt{\|\ell_k\|_{w_k}^2 \cdot \|w_k\|_1} + 2\eta\|\ell_k\|_{w_k}^2\right) + \frac{n\log(2n^2)}{\eta T}$$

$$\leq \frac{1}{T}\sum_{k=0}^{T-1}\left(\frac{n}{\sqrt{2048}} + \frac{2}{\sqrt{n}+K}\frac{n}{2048}\right) + \frac{n(\sqrt{n}+K)\log(2n^2)}{T} \ .$$

Therefore, as long as $T \geq \Omega((n^{1/2}K + K^2)\log n)$, we must have $\frac{1}{T}\sum_{k=0}^{T-1}\langle \ell_k, -u\rangle \leq \frac{n}{32} + \frac{n}{8K}$. This implies that $\frac{1}{T}\sum_{k=0}^{T-1}|\delta_{k,i} - \alpha_i| \leq \frac{1}{32} + \frac{1}{8K}$ for every $i \in [n]$ (because we can choose $u = n \times \mathbf{e}_i$).

In sum, if we define $\overline{\delta} = \frac{1}{T}\sum_{k=0}^{T-1}\delta_k$, we have

$$\|\overline{\delta} - \alpha\|_\infty \leq \frac{1}{32} + \frac{1}{8K} \quad \text{and} \quad \langle \nabla f(x), \overline{\delta}\rangle + \frac{1}{6}\overline{\delta}^\top \mathbf{H}\overline{\delta} \leq -\mathsf{OPT} + \varepsilon \ . \qquad \square$$

**Theorem 6.2** (`Scaling1`). *If $N \geq 1$ and $\varepsilon \in (0,1)$, the output $y = \mathtt{Scaling1}(\mathbf{A}, N, \varepsilon)$ satisfies*

$$\|y\|_\infty \leq 2N \quad \text{and} \quad f(y) - f(u) \leq \varepsilon \quad \text{for every } u \text{ with } \|u\|_\infty \leq N.$$

*Furthermore, if there exists $u$ satisfying $f(u) - \inf_x\{f(x)\} \leq \varepsilon$ and $\|u\|_\infty \leq N$, then we also have $\|\nabla f(y)\|_{c^{-1}}^2 \leq \varepsilon$. The total complexity is $\widetilde{O}(N(m + n^{3/2}))$.*

*Proof of Theorem 6.2.* In each iteration $t$, we have $\|x_t\|_\infty \leq N$ and Lemma 6.1 finds us a vector $\overline{\delta}$ with

$$\|\overline{\delta} - \alpha\|_\infty \leq \frac{1}{32} + \frac{1}{8K} \quad \text{and} \quad \langle \nabla, \overline{\delta}\rangle + \frac{1}{6}\overline{\delta}^\top \mathbf{H}\overline{\delta} \leq \min_{\|\delta - \alpha\|_\infty \leq 1/32}\left\{\langle \nabla, \delta\rangle + \frac{1}{6}\delta^\top \mathbf{H}\delta\right\} + \frac{\varepsilon}{900N} \ .$$

We have $\|\overline{\delta}\|_\infty \leq \|\overline{\delta} - \alpha\|_\infty + \|\alpha\|_\infty < 1/16$ and

$$\left\|\frac{\overline{\delta}}{6.6} - \alpha\right\|_\infty \leq \frac{1}{6.6}\left\|\overline{\delta} - \alpha\right\|_\infty + \left(1 - \frac{1}{6.6}\right)\|\alpha\|_\infty \leq \frac{1}{32} + \frac{1}{50K} \ .$$

Since any $\delta$ satisfying $\|\delta - \alpha\|_\infty \leq \frac{1}{32}$ also satisfies $\|x_t + \delta\|_\infty \leq N$ (see Fact 5.3), we immediately have $\|x_t + \frac{\overline{\delta}}{6.6}\|_\infty \leq N + \frac{1}{50K}$.

In sum, owing to Lemma 5.4, if we let $x_{t+1} = x_t + \overline{\delta}/6.6$ it satisfies

$$\|x_{t+1}\|_\infty \leq N + \frac{1}{50K} \quad \text{and} \quad f(x_t) - f(x_{t+1}) \geq \max\left\{0, \frac{1}{900N}(f(x_t) - f(u) - \varepsilon)\right\} \ .$$

This means we can find a point $y$ satisfying $\|y\|_\infty \leq 2N$ and $f(y) - f(u) \leq \varepsilon$ if we choose $K = \Theta(\log(1/\varepsilon))$.

If $f(u) - \inf_x\{f(x)\} \leq \varepsilon$, then using Claim D.1 we also have $\|\nabla f(y)\|_{c^{-1}}^2 \leq \varepsilon$.

The per iteration complexity of `Scaling1` is $\widetilde{O}(m + n^{3/2})$, in which the complexity $\widetilde{O}(m)$ is for obtaining the sparsification matrix $\mathbf{H}$ of the Hessian $\nabla^2 f(x_t)$, and the complexity $\widetilde{O}(n^{3/2})$ is for the multiplicative weight update subroutine `MWUbasic`. $\square$

# G  Missing Details for Section 7

## G.1  Proof of Lemma 7.1

**Lemma 7.1.** *In each iteration $t$ of* Scaling2, *if $\|x_t\|_\infty \leq N$, then we can compute $x_{t+1}$ in complexity $\widetilde{O}(m+Tn)$, and it satisfies $\|x_{t+1}\|_\infty \leq N$ and*

$$\text{either (1): } f(x_{t+1}) - f(u) \leq O\Big(\frac{Nh}{T^2}\Big) \quad \text{or} \quad \text{(2): } f(x_t) - f(x_{t+1}) \geq \Omega\Big(\frac{1}{N}\Big)(f(x_t) - f(u)) \ .$$

*Here, $u$ is any vector satisfying $\|u\|_\infty \leq N$.*

*Proof of Lemma 7.1.* We use accelerated gradient descent to directly solve the following minimization problem:

$$0 \geq \mathsf{OPT} \stackrel{\text{def}}{=} \min_{\delta \in \mathsf{box}^N(x)} \Big\{ \langle \nabla f(x), \delta \rangle + \frac{1}{6}\delta^\top \mathbf{H}\delta \Big\} \tag{G.1}$$

We re-scale the problem by letting $\widehat{\delta}_i = \sqrt{\mathbf{H}_{i,i}}\delta_i$ and $\widehat{\mathbf{H}}_{i,j} = \frac{\mathbf{H}_{i,j}}{\sqrt{\mathbf{H}_{i,i}\mathbf{H}_{j,j}}}$. This re-scaling translates the problem into a new one $\min_{\widehat{\delta} \in \widehat{\mathsf{box}}} \big\{ \langle g, \widehat{\delta} \rangle + \frac{1}{6}\widehat{\delta}^\top \widehat{\mathbf{H}}\widehat{\delta} \big\}$, where the new matrix $\widehat{\mathbf{H}}$ is still Laplacian but has spectral norm at most 1. This means that the quadratic function $\frac{1}{6}\widehat{\delta}^\top \widehat{\mathbf{H}}\widehat{\delta}$ is 1-smooth (i.e., its Hessian has constant spectral norm) so we can apply the constrained version of accelerated gradient descent (see for instance [4, Theorem 4.1]), and obtain a vector $\delta \in \mathsf{box}^N(x_t)$ satisfying

$$\langle \nabla f(x_t), \delta \rangle + \frac{1}{6}\delta^\top \mathbf{H}\delta - \mathsf{OPT} \leq O\Big(\frac{\|\widehat{\delta}^*\|_2^2}{T^2}\Big) \ .$$

Here, $\delta^*$ is any minimizer of (G.1) and $\widehat{\delta}^*$ is its re-scaled version (namely, $\widehat{\delta}^*_i = \sqrt{\mathbf{H}_{i,i}}\delta^*_i$). Since $\sum_{j \in [n]} (\nabla_j f(x_t) + c_j) = \sum_{i \in [d], j \in [n]} r_i \frac{\mathbf{A}_{i,j} e^{(x_t)_j}}{\langle \mathbf{A}_i, e^{x_t} \rangle} = \sum_{i \in [d]} r_i = h$, we have

$$\|\widehat{\delta}^*\|_2^2 = \sum_{j \in [n]} (\delta^*)_j^2 \mathbf{H}_{j,j} \leq \sum_{j \in [n]} (\delta^*)_j^2 \nabla_{j,j} f(x_t) \leq \sum_{j \in [n]} (\delta^*)_j^2 (\nabla_j f(x_t) + c_j) \leq \sum_{j \in [n]} (\nabla_j f(x_t) + c_j) = h \ .$$

Therefore, we can get a point $\delta \in \mathsf{box}^N(x_t)$ satisfying

$$\langle \nabla f(x_t), \delta \rangle + \frac{1}{6}\delta^\top \mathbf{H}\delta - \mathsf{OPT} \leq O\Big(\frac{h}{T^2}\Big) \ . \tag{G.2}$$

At this point, if we have $f(x_t) - f(u) \leq O(\frac{Nh}{T^2})$ then we are done, because we can always assume that $f(x_{t+1}) \leq f(x_t)$. (This is because we can always choose $x_{t+1} = x_t$.) Otherwise, we must have $-\mathsf{OPT} \geq \Omega(\frac{h}{T^2})$ according to Lemma 5.4.a, and this together with (G.2) implies we have $\langle \nabla f(x_t), \delta \rangle + \frac{1}{6}\delta^\top \mathbf{H}\delta \leq \mathsf{OPT}$. According to Lemma 5.4.b, this means we can move to $x_{t+1} = x_t + \delta$ and it satisfies $f(x_t) - f(x_{t+1}) \geq \Omega(\frac{1}{N})(f(x_t) - f(u))$. Note that we also have $\|x_{t+1}\|_\infty \leq N$ because $\delta \in \mathsf{box}^N(x_t)$ (see Fact 5.3).

Finally, since each iteration of the accelerated gradient runs in time $\widetilde{O}(n)$ because $\mathbf{H}$ has sparsity at most $\widetilde{O}(n)$ (see Lemma A.3), we have the desired complexity bound. □

# H  Missing Details for Section 8

## H.1  Proof of Theorem 8.1

We first show that Theorem 8.1 is a simple corollary of Lemma 8.4.

> **Theorem 8.1** (`Scaling3`). *If $x_0$ satisfies $f(x_0) - f(u) \leq Nn^{1/3}$ and $\|x_0\|_\infty \leq N$, and $\varepsilon \in (0, 1/4]$, then the output $y = \mathtt{Scaling3}(\mathbf{A}, N, x_0, \varepsilon)$ satisfies*
> 
> $$\|y\|_\infty \leq 10N \quad \text{and} \quad f(y) - f(u) \leq \varepsilon \quad \text{for every } u \text{ with } \|u\|_\infty \leq N.$$
> 
> *Furthermore, if there exists $u$ satisfying $f(u) - \inf_x \{f(x)\} \leq \varepsilon$ and $\|u\|_\infty \leq N$, then we also have $\|\nabla f(y)\|_{c^{-1}}^2 \leq \varepsilon$. The total complexity is $\widetilde{O}(N(m + n^{4/3}))$.*

*Proof of Theorem 8.1.* At the beginning of each iteration $t$, we have $\|x_t\|_\infty \leq N$. Lemma 8.4 finds us a vector $\overline{\delta}$ with $\|\overline{\delta} - \alpha\|_\infty \leq \frac{1}{32} + \frac{2}{K}$, and therefore

$$\left\|\frac{\overline{\delta}}{6.6} - \alpha\right\|_\infty \leq \frac{1}{6.6}\left\|\overline{\delta} - \alpha\right\|_\infty + \left(1 - \frac{1}{6.6}\right)\|\alpha\|_\infty \leq \frac{1}{32} + \frac{1}{K}$$

Recall that any $\delta$ satisfying $\|\delta - \alpha\|_\infty \leq \frac{1}{32}$ also satisfies $\|x_t + \delta\|_\infty \leq N$ (see Fact 5.3), we immediately have that $\|x_t + \frac{\overline{\delta}}{6.6}\|_\infty \leq N + \frac{1}{K}$.

Lemma 8.4 also tells us that $\overline{\delta}$ satisfies one of the following:

1. Either $\langle \nabla, \overline{\delta}\rangle + \frac{1}{6}\overline{\delta}^\top \mathbf{H}\overline{\delta} \leq 0.25 \min_{\|\delta - \alpha\|_\infty \leq 1/32} \left\{\langle \nabla, \delta\rangle + \frac{1}{6}\delta^\top \mathbf{H}\delta\right\} + \frac{\varepsilon}{20000N}$.

   In this case, owing to Lemma 5.4, if we let $x_{t+1} = x_t + \frac{\overline{\delta}}{6.6}$ it satisfies

   $$\|x_{t+1}\|_\infty \leq N + \frac{1}{K} \quad \text{and} \quad f(x_t) - f(x_{t+1}) \geq \max\left\{0, \frac{1}{10000N}(f(x_t) - f(u))\right\} - \frac{\varepsilon}{20000N}.$$

   This also means that if $f(x_t) - f(u) \geq \varepsilon$, then $f(x_{t+1}) \leq f(x_t)$.

2. Or $\langle \nabla f(x), \overline{\delta}\rangle + \frac{1}{6}\overline{\delta}^\top \mathbf{H}\overline{\delta} \leq -\frac{1}{256}\frac{\rho}{K}$, this means that

   $$\|x_{t+1}\|_\infty \leq N + \frac{1}{K} \quad \text{and} \quad f(x_t) - f(x_{t+1}) \geq \frac{1}{256}\frac{\rho}{K}.$$

Obviously, the second case can only happen no more than $O\left(\frac{(f(x_0) - \inf_x\{f(x)\})K}{\rho}\right) = \widetilde{O}(N)$ times and thus diameter $N$ can not be increased by more than $O\left(\frac{(f(x_0) - \inf_x\{f(x)\})}{\rho}\right) = O(N)$. For the first case, if we choose $K = \Theta(\log(1/\varepsilon))$ and allow the first case to run for $NK$ iterations, then we must have a point $y$ satisfying $\|y\|_\infty \leq 10N$ and $f(y) - f(u) \leq \varepsilon$.

Finally, if $f(u) - \inf_x\{f(x)\} \leq \varepsilon$, then using Claim D.1 we also have $\|\nabla f(y)\|_{c^{-1}}^2 \leq \varepsilon$.

We argue that the per-iteration complexity of `Scaling3` is $\widetilde{O}(m + n^{4/3})$. First of all, it takes complexity $\widetilde{O}(m)$ to obtain the sparsification matrix $\mathbf{H}$ of the Hessian $\nabla^2 f(x_t)$. Next, each round of subroutine `MWUfull` runs in complexity $\widetilde{O}(n)$,[26] and there are at most $T = \widetilde{O}(n^{1/3})$ rounds. □

### H.2 Three Structural Lemmas

In order to prove Lemma 8.4, we need to establish three structural lemmas.

**Notations.** Throughout this subsection, we use $w$ to denote $w_k$ which is the weight vector at round $k$ of `MWUfull`. We also denote by $y = \delta_k$ the approximate minimizer of $\langle \nabla, \delta\rangle + \frac{1}{6}\delta^\top \mathbf{H}\delta$ over the $\ell_2$ constraint $\|\delta - \alpha\|_{w_k}^2 \leq n/1024$ at iteration $k$ of `MWUfull` (see Line 5). We also use $\nabla$ and $\nabla^2$ to represent the gradient and Hessian at the current point, and it satisfies $\mathbf{H} \preceq \nabla^2 \preceq 1.1\mathbf{H}$.

Suppose $|y_1| \geq \cdots |y_n|$. Then,

---

[26]This includes also Line 11 of `MWUfull`. Since $s \leq O(n^{1/3})$ when Line 11 is reached, the recursive call to `MWUbasic` on the smaller-sized problem (with dimension $s$) costs running time at most $\widetilde{O}(s^{3/2}) = \widetilde{O}(n^{1/2})$.

- In Section H.2.1, we show that that whenever there is a large gap between $|y_s|$ and $|y_{s+1}|$, then either (a) $|\sum_{i=1}^{s} \nabla_i| \geq \frac{1}{2}$ or (b) the Hessian $\nabla^2$ is almost disconnected. See Lemma H.1.
- In Section H.2.2, we show that if Case (b) happens, we can reduce the problem to a smaller-sized one with respect to the first $s$ coordinates. See Lemma H.3.
- In Section H.2.3, we show that if Case (a) happens, we can sufficiently decrease the objective by moving in the direction $(1, \cdots 1, 0, \cdots 0)$. See Lemma H.4.

### H.2.1 Structural Lemma 1

We show that that whenever there is a large gap between $|y_s|$ and $|y_{s+1}|$, then either $|\sum_{i=1}^{s} \nabla_i| \geq \frac{1}{2}$ or the Hessian $\nabla^2$ is almost disconnected.

**Lemma H.1.** *Let $w \in \mathbb{R}^n$ be a weight vector satisfying $w_i \in [1/2, n]$, let $\alpha \in \mathbb{R}^n$ be a vector satisfying $\|\alpha\|_\infty \leq 1$ and $\|\alpha\|_w^2 \leq n/1024$, and let $\varepsilon$ be in $\left(0, \frac{1}{4}\right]$. If $y$ satisfies*

$$\|y - \alpha\|_w^2 \leq \frac{n}{1024} \quad \text{and} \quad \langle \nabla, y \rangle + \frac{1}{6} y^\top \mathbf{H} y \leq \min_{\delta \in \mathbb{R}^n, \|\delta - \alpha\|_w^2 \leq n/1024} \left\{ \langle \nabla, \delta \rangle + \frac{1}{6} \delta^\top \mathbf{H} \delta \right\} + \varepsilon$$

*and without loss of generality $y_1 \geq y_2 \geq \cdots \geq y_n$, then, for every $s \in [n-1]$, the following holds*

1. *If $y_s \geq 15$ and $y_s - y_{s+1} \geq 15$, then letting $v = (\underbrace{1, \ldots, 1}_{s}, \underbrace{0, \ldots, 0}_{n-s})$, we have*

$$\text{either} \quad \left( \sum_{i \in [s]} \nabla_i \leq -\frac{1}{2}, \quad \left| \sum_{i \in [s]} \nabla_i \right| \geq v^\top \nabla^2 v \right) \quad \text{or} \quad \left( \left| \sum_{i \in [s]} \nabla_i \right| \leq \varepsilon, \quad v^\top \nabla^2 v \leq \varepsilon \right)$$

2. *If $y_{s+1} \leq -15$ and $y_s - y_{s+1} \geq 15$, then letting $v = (\underbrace{0, \ldots, 0}_{s}, \underbrace{-1, \ldots, -1}_{n-s})$, we have*

$$\text{either} \quad \left( \sum_{i \in [s]} \nabla_i \geq \frac{1}{2}, \quad \left| \sum_{i \in [s]} \nabla_i \right| \geq v^\top \nabla^2 v \right) \quad \text{or} \quad \left( \left| \sum_{i \in [s]} \nabla_i \right| \leq \varepsilon, \quad v^\top \nabla^2 v \leq \varepsilon \right)$$

*Proof of Lemma H.1.* We only prove the first case and the second is symmetric. The proof follows by "trying" to find a point $y'$ satisfying $\|y' - \alpha\|_w^2 \leq n/1024$ and

$$\langle \nabla, y \rangle + \frac{1}{6} y^\top \mathbf{H} y > \langle \nabla, y' \rangle + \frac{1}{6} (y')^\top \mathbf{H} y' + \varepsilon$$

which gives a contradiction. In fact, we simply choose $y' = y - v$. Since $\|\alpha\|_\infty \leq 1$ and $y_s \geq 15$, we know that for every $i \in [s]$, it satisfies $(y'_i - [\alpha]_i)^2 \leq (y_i - [\alpha]_i)^2$. This implies $\|y' - \alpha\|_w^2 \leq \|y - \alpha\|_w^2 \leq n/1024$.

We now calculate the change of the quadratic function when we move from $y$ to $y'$. It must be upper bounded by $\varepsilon$ because of the optimality of $y$:

$$\varepsilon \geq \langle \nabla, y \rangle + \frac{1}{6} y^\top \mathbf{H} y - \langle \nabla, y' \rangle - \frac{1}{6} (y')^\top \mathbf{H} y' = \sum_{i \in [s]} \nabla_i + \frac{1}{6} \sum_{i \in [s], j \in [n] \setminus [s]} |\mathbf{H}_{i,j}| [(y_i - y_j)^2 - (y_i - 1 - y_j)^2]$$

$$\geq \sum_{i \in [s]} \nabla_i + \frac{1}{6} \sum_{i \in [s], j \in [n] \setminus [s]} |\mathbf{H}_{i,j}| [2(y_i - y_j) - 1] \geq \sum_{i \in [s]} \nabla_i + \frac{29}{6} \sum_{i \in [s], j \in [n] \setminus [s]} |\mathbf{H}_{i,j}|$$

$$\geq \sum_{i \in [s]} \nabla_i + \frac{29}{6} v^\top \mathbf{H} v \geq \sum_{i \in [s]} \nabla_i + 4 \cdot v^\top \nabla^2 v \ .$$

The above the calculation uses the fact that $\mathbf{H}$ is Laplacian so $\delta^\top \mathbf{H} \delta = \sum_{i<j} |\mathbf{H}_{i,j}| (\delta_i - \delta_j)^2$. We next consider two cases:

1. $|\sum_{i\in[s]} \nabla_i| \geq \frac{1}{2}$. We must have $\sum_{i\in[s]} \nabla_i \leq -\frac{1}{2}$. Therefore, using $\varepsilon \leq 1/4$, we have
$$\sum_{i\in[s]} \nabla_i + 4v^\top \nabla^2 v \leq \varepsilon \implies v^\top \nabla^2 v \leq \left|\sum_{i\in[s]} \nabla_i\right| .$$

2. $|\sum_{i\in[s]} \nabla_i| < \frac{1}{2}$. In this case we denote by $a_i = \sum_{j=1}^s \mathbf{A}_{i,j}$ for each $i \in [d]$, and have (which follows from the definitions of $\nabla$ and $\nabla^2$ for our function $f(\cdot)$)
$$\sum_{j\in[s]} \nabla_j = \sum_{i\in[d]} a_i - \sum_{j\in[s]} c_j \quad \text{and} \quad v^\top \nabla^2 v = \sum_{i\in[d]} \frac{a_i(r_i - a_i)}{r_i} .$$
Now, let us split $a_i$ into $a_i = [a_i] + \{a_i\}$ where $[a_i]$ is an integer and $\{a_i\} \in [-1/2, 1/2]$. Since each $r_i$ is an integer and $a_i \in [0, r_i]$, we know
$$\frac{a_i(r_i - a_i)}{r_i} \geq \frac{1}{2}|\{a_i\}| \quad \text{and} \quad v^\top \nabla^2 v \geq \frac{1}{2}\sum_{i\in[d]} |\{a_i\}| .$$
On the other hand, since $\sum_{j=1}^s c_j$ is an integer, apply Fact H.2 on $a = \sum_{i\in[d]} \{a_i\}$ and $b = \sum_{i\in[d]} [a_i] - \sum_{j=1}^s c_j$, we know that
$$v^\top \nabla^2 v \geq \frac{1}{2}\sum_{i\in[d]} |\{a_i\}| \geq \frac{1}{2}|a| \geq \frac{1}{2}|a+b| = \frac{1}{2}\left|\sum_{i\in[s]} \nabla_i\right|$$
which implies that
$$\varepsilon \geq \sum_{i\in[s]} \nabla_i + 4v^\top \nabla^2 v \geq \left|\sum_{i\in[s]} \nabla_i\right|$$
Therefore, we must have $\left|\sum_{i\in[s]} \nabla_i\right| \leq \varepsilon$ and $v^\top \nabla^2 v \leq \varepsilon$. □

**Fact H.2.** *Let $a \in \mathbb{R}$ and $b \in \mathbb{N}$. Suppose $0 \leq |a+b| < \frac{1}{2}$, then $|a| \geq |a+b|$.*

*Proof.* Can be verified by case analysis. □

### H.2.2 Structural Lemma 2

Our next lemma states that, the Hessian is almost disconnected —precisely, if the case "$\left|\sum_{i\in[s]} \nabla_i\right| \leq \varepsilon$ and $v^\top \nabla^2 v \leq \varepsilon$" takes place in Lemma H.1— then we can reduce the problem to a smaller one corresponding only to the first $s$ coordinates (see (H.1) below), and then append the solution of this smaller problem (denoted by $z_\triangleleft$) with the last $n-s$ coordinates of $y$ (denoted by $y_\triangleright$). The final solution $z = (z_\triangleleft, y_\triangleright)$ satisfies $\|z - \alpha\|_w^2 \leq n/1024$ and has a small quadratic value $\langle \nabla, z\rangle + \frac{1}{6}z^\top \mathbf{H} z$ (see (H.2) below).

**Lemma H.3.** *Let $w \in \mathbb{R}^n$ be a weight vector satisfying $w_i \in [1/2, n]$, let $\alpha \in \mathbb{R}^n$ be a vector satisfying $\|\alpha\|_\infty \leq 1$ and $\|\alpha\|_w^2 \leq n/1024$, and let $\varepsilon$ be in $\left(0, \frac{1}{4}\right]$. Suppose $y$ satisfies*
$$\|y - \alpha\|_w^2 \leq \frac{n}{1024} \quad \text{and} \quad \langle \nabla, y\rangle + \frac{1}{6}y^\top \mathbf{H} y \leq \min_{\delta \in \mathbb{R}^n, \|\delta - \alpha\|_w^2 \leq n/1024} \left\{\langle \nabla, \delta\rangle + \frac{1}{6}\delta^\top \mathbf{H}\delta\right\} + \varepsilon .$$

*Now, suppose there exists $s \in [n-1]$ satisfying*

1. *$|y_i| \geq 15$ for all $i = 1, 2, \ldots, s$, and*
2. *$|\langle \nabla, v\rangle| \leq \varepsilon$ and $v^\top \nabla^2 v \leq \varepsilon$ where $v = (\underbrace{\mathrm{sgn}(y_1), \ldots, \mathrm{sgn}(y_s)}_{s}, \underbrace{0, \ldots, 0}_{n-s})$*

and suppose we are given $z_\triangleleft \in \mathbb{R}^s$ satisfying

$$\|z_\triangleleft - \alpha\|_\infty \leq \frac{1}{16} \quad \text{and} \quad \langle \nabla_\triangleleft, z_\triangleleft \rangle + \frac{1}{6} z_\triangleleft^\top \mathbf{H}_\triangleleft z_\triangleleft \leq \frac{1}{2} \min_{\delta \in \mathbb{R}^s, \|\delta - \alpha_\triangleleft\|_\infty \leq 1/32} \left\{ \langle \nabla_\triangleleft, \delta \rangle + \frac{1}{6} \delta^\top \mathbf{H}_\triangleleft \delta \right\} + \varepsilon \quad \text{(H.1)}$$

Then, letting $z = (z_\triangleleft, y_\triangleright)$, we have

$$\|z - \alpha\|_w^2 \leq \frac{n}{1024} \quad \text{and} \quad \langle \nabla, z \rangle + \frac{1}{6} z^\top \mathbf{H} z \leq \frac{1}{2} \min_{\delta \in \mathbb{R}^n, \|\delta - \alpha\|_\infty \leq 1/32} \left\{ \langle \nabla, \delta \rangle + \frac{1}{6} \delta^\top \mathbf{H} \delta \right\} + 14 n^3 \varepsilon \quad . \quad \text{(H.2)}$$

*Proof of Lemma H.3.* Define $\delta^* = (\delta_\triangleleft^*, \delta_\triangleright^*)$ where

$$\delta_\triangleleft^* \in \arg\min_{\delta \in \mathbb{R}^s, \|\delta - \alpha_\triangleleft\|_\infty \leq 1/32} \left\{ \langle \nabla_\triangleleft, \delta \rangle + \frac{1}{6} \delta^\top \mathbf{H}_\triangleleft \delta \right\} \quad , \quad \text{and}$$

$$\delta_\triangleright^* \in \arg\min_{\delta \in \mathbb{R}^{n-s}, \|\delta - \alpha_\triangleright\|_\infty \leq 1/32} \left\{ \langle \nabla_\triangleright, \delta \rangle + \frac{1}{6} \delta^\top \mathbf{H}_\triangleright \delta \right\}$$

Let $y' = y - v$. Define $y_\triangleright'' \stackrel{\text{def}}{=} (1 - \lambda) y_\triangleright + \lambda \delta_\triangleright^*$ for parameter $\lambda = \frac{1}{2n^2}$, and $y'' \stackrel{\text{def}}{=} (y_\triangleleft', y_\triangleright'')$. Since each $w_i \in \left[\frac{1}{2}, n\right]$, we have

$$\|y'' - \alpha\|_w^2 = \sum_{i \in [n]} w_i (y_i'' - \alpha_i)^2$$

$$\leq \sum_{i \in [s]} w_i (y_i' - \alpha_i)^2 + \sum_{i \in \{s+1, \cdots, n\}} \left( (1 - \lambda) w_i (y_i - \alpha_i)^2 + \lambda w_i ((\delta^*)_i - \alpha_i)^2 \right)$$

$$\leq \sum_{i \in [s]} w_i \left( (y_i - \alpha_i)^2 - 1 \right) + \sum_{i \in \{s+1, \cdots, n\}} \left( w_i (y_i - \alpha_i)^2 + \lambda w_i ((\delta^*)_i - \alpha_i)^2 \right)$$

$$\leq \sum_{i \in [n]} w_i (y_i - \alpha_i)^2 - \frac{1}{2} s + \sum_{i \in \{s+1, \cdots, n\}} \lambda w_i ((\delta^*)_i - \alpha_i)^2$$

$$\leq \frac{n}{1024} - \frac{1}{2} s + \sum_{i \in \{s+1, \cdots, n\}} \lambda w_i ((\delta^*)_i - \alpha_i)^2 \leq \frac{n}{1024} - \frac{1}{2} + n^2 \lambda \leq \frac{n}{1024} \quad .$$

We now calculate the value of the quadratic function on $y''$ comparing to $y$:

$$\varepsilon \stackrel{①}{\geq} \langle \nabla, y - y'' \rangle + \frac{1}{6} y^\top \mathbf{H} y - \frac{1}{6} (y'')^\top \mathbf{H} y''$$

$$\stackrel{②}{\geq} \langle \nabla_\triangleleft, y_\triangleleft - y_\triangleleft'' \rangle + \frac{1}{6} y_\triangleleft^\top \mathbf{H}_\triangleleft y_\triangleleft - \frac{1}{6} (y_\triangleleft'')^\top \mathbf{H}_\triangleleft y_\triangleleft'' + \left( \langle \nabla_\triangleright, y_\triangleright - y_\triangleright'' \rangle + \frac{1}{6} y_\triangleright^\top \mathbf{H}_\triangleright y_\triangleright - \frac{1}{6} (y_\triangleright'')^\top \mathbf{H}_\triangleright y_\triangleright'' \right) - \frac{2}{3} \|y_s''\|_\infty^2 v^\top \mathbf{H} v$$

$$= \langle \nabla_\triangleleft, y_\triangleleft - y_\triangleleft' \rangle + \frac{1}{6} y_\triangleleft^\top \mathbf{H}_\triangleleft y_\triangleleft - \frac{1}{6} (y_\triangleleft')^\top \mathbf{H}_\triangleleft y_\triangleleft' + \left( \langle \nabla_\triangleright, y_\triangleright - y_\triangleright'' \rangle + \frac{1}{6} y_\triangleright^\top \mathbf{H}_\triangleright y_\triangleright - \frac{1}{6} (y_\triangleright'')^\top \mathbf{H}_\triangleright y_\triangleright'' \right) - \frac{2}{3} \|y_s''\|_\infty^2 v^\top \mathbf{H} v$$

$$\stackrel{③}{\geq} \langle \nabla_\triangleleft, v_\triangleleft \rangle + \left( \langle \nabla_\triangleright, y_\triangleright - y_\triangleright'' \rangle + \frac{1}{6} y_\triangleright^\top \mathbf{H}_\triangleright y_\triangleright - \frac{1}{6} (y_\triangleright'')^\top \mathbf{H}_\triangleright y_\triangleright'' \right) - \frac{2}{3} \|y_s''\|_\infty^2 v^\top \mathbf{H} v - \frac{2}{3} \sqrt{n} \varepsilon \quad .$$

$$\stackrel{④}{\geq} -4 n \varepsilon + \langle \nabla_\triangleright, y_\triangleright - y_\triangleright'' \rangle + \frac{1}{6} y_\triangleright^\top \mathbf{H} y_\triangleright - \frac{1}{6} (y_\triangleright'')^\top \mathbf{H} y_\triangleright'' \quad .$$

Above, inequality ① is by the approximate optimality of $y$; inequality ② uses the fact that $v = (\text{sgn}(y_1), \ldots, \text{sgn}(y_s), 0, \ldots, 0)$ and thus for any vector $p$:

$$4 \|p\|_\infty v^\top \mathbf{H} v \geq \sum_{i \in [s], j \in [n-s]} |\mathbf{H}_{i,j}| (p_i - p_j)^2 = (p^\top \mathbf{H} p) - (p^\top \mathbf{H}_\triangleleft p + p^\top \mathbf{H}_\triangleright p) \geq 0 \quad ; \quad \text{(H.3)}$$

In equality ③ uses

$$\left| y_\triangleleft^\top \mathbf{H}_\triangleleft y_\triangleleft - (y_\triangleleft')^\top \mathbf{H}_\triangleleft y_\triangleleft' \right| \leq \|y_\triangleleft\|_\infty \sum_{i, j \in [s]} |\mathbf{H}_{i,j}| (\text{sgn}(y_i) - \text{sgn}(y_j))^2 \leq 2 \|y_\triangleleft\|_\infty v^\top \mathbf{H} v \leq 4 \sqrt{n} \varepsilon$$

Inequality ④ uses (1) $\sum_{i\in[s]} \text{sgn}(y_i)\nabla_i \geq -\varepsilon$, (2) $v^\top \mathbf{H} v \leq \varepsilon$, and (3) $\|y''_s\|_\infty \leq \max\{\|y\|_\infty, \|\delta^*\|_\infty\} \leq 2\sqrt{n}$, where (3) comes from the fact that $\|y-\alpha\|_w^2 \leq n/1024$ implies $|y_i - \alpha_i| \leq \sqrt{n/(1024 w_i)} \leq \sqrt{n}$.

Next, define convex function $g(\delta) = \langle \nabla_\triangleright, \delta\rangle + \frac{1}{6}\delta^\top \mathbf{H}_\triangleright \delta$ for $\delta \in \mathbb{R}^{n-s}$, then we have just proved

$$\varepsilon \geq \langle \nabla, y - y''\rangle + \frac{1}{6}y^\top \mathbf{H} y - \frac{1}{6}(y'')^\top \mathbf{H} y'' \geq g(y_\triangleright) - g(y''_\triangleright) - 4n\varepsilon = g(y_\triangleright) - g((1-\lambda)y_\triangleright + \lambda \delta^*_\triangleright) - 4n\varepsilon$$
$$\geq \lambda(g(y_\triangleright) - g(\delta^*_\triangleright)) - 4n\varepsilon \ .$$

This implies

$$g(y_\triangleright) \leq g(\delta^*_\triangleright) + \frac{5n\varepsilon}{\lambda} \leq g(\delta^*_\triangleright) + 10n^3 \varepsilon \ . \tag{H.4}$$

Finally, we are ready to analyze our choice $z = (z_\triangleleft, y_\triangleright)$. We first compute that

$$\|z - \alpha\|_w^2 = \sum_{i\in[n]} w_i(z_i - \alpha_i)^2 = \sum_{i\in[n]} w_i(y_i - \alpha_i)^2 + \sum_{i\in[s]} w_i\big(((z_\triangleleft)_i - \alpha_i)^2 - (y_i - \alpha_i)^2\big)$$
$$\leq \frac{n}{1024} + \sum_{i\in[s]} w_i\big((\frac{1}{16^2} - (15-1)^2)\big) \leq \frac{n}{1024} \ .$$

Therefore, the first assertion in (H.2) is satisfied. Next, denote by

$$\overline{\delta} \in \arg\min_{\delta \in \mathbb{R}^n, \|\delta - \alpha\|_\infty \leq 1/32} \left\{\langle \nabla, \delta\rangle + \frac{1}{6}\delta^\top \mathbf{H} \delta\right\} \ ,$$

we also have

$$\langle \nabla, z\rangle + \frac{1}{6}z^\top \mathbf{H} z \overset{①}{\leq} \left(\langle \nabla_\triangleleft, z_\triangleleft\rangle + \frac{1}{6}(z_\triangleleft)^\top \mathbf{H}_\triangleleft z_\triangleleft\right) + \left(\langle \nabla_\triangleright, z_\triangleright\rangle + \frac{1}{6}(z_\triangleright)^\top \mathbf{H}_\triangleright z_\triangleright\right) + \frac{2}{3}\|z\|_\infty^2 v^\top \mathbf{H} v$$
$$\overset{②}{\leq} \frac{1}{2}\left(\langle \nabla_\triangleleft, \delta^*_\triangleleft\rangle + \frac{1}{6}(\delta^*_\triangleleft)^\top \mathbf{H}_\triangleleft \delta^*_\triangleleft\right) + \left(\langle \nabla_\triangleright, \delta^*_\triangleright\rangle + \frac{1}{6}(\delta^*_\triangleright)^\top \mathbf{H}_\triangleright \delta^*_\triangleright\right) + 10n^3 \varepsilon + 3n\varepsilon + \varepsilon$$
$$\overset{③}{\leq} \frac{1}{2}\left(\langle \nabla_\triangleleft, \delta^*_\triangleleft\rangle + \frac{1}{6}(\delta^*_\triangleleft)^\top \mathbf{H}_\triangleleft \delta^*_\triangleleft\right) + \frac{1}{2}\left(\langle \nabla_\triangleright, \delta^*_\triangleright\rangle + \frac{1}{6}(\delta^*_\triangleright)^\top \mathbf{H}_\triangleright \delta^*_\triangleright\right) + 14n^3\varepsilon$$
$$\overset{④}{\leq} \frac{1}{2}\left(\langle \nabla_\triangleleft, \overline{\delta}_\triangleleft\rangle + \frac{1}{6}(\overline{\delta}_\triangleleft)^\top \mathbf{H}_\triangleleft \overline{\delta}_\triangleleft\right) + \frac{1}{2}\left(\langle \nabla_\triangleright, \overline{\delta}_\triangleright\rangle + \frac{1}{6}(\overline{\delta}_\triangleright)^\top \mathbf{H}_\triangleright \overline{\delta}_\triangleright\right) + 14n^3\varepsilon$$
$$\leq \frac{1}{2}\left(\langle \nabla, \overline{\delta}\rangle + \frac{1}{6}(\overline{\delta})^\top \mathbf{H} \overline{\delta}\right) + 14n^3\varepsilon \ .$$

Above, inequality ① uses (H.3) again; inequality ② uses the assumption on $z_\triangleleft$ and (H.4); inequality ③ uses the fact that $\|\alpha\|_\infty \leq 1/32$ and thus $\langle \nabla_\triangleright, \delta^*_\triangleright\rangle + \frac{1}{6}(\delta^*_\triangleright)^\top \mathbf{H}_\triangleright \delta^*_\triangleright \leq 0$ from the definition of $\delta^*_\triangleright$; inequality ④ uses the optimality of $\delta^*_\triangleleft$ and $\delta^*_\triangleright$, and the fact that $\overline{\delta} = (\overline{\delta}_\triangleleft, \overline{\delta}_\triangleright)$; inequality ⑤ uses the fact that $\mathbf{H} \succeq \begin{pmatrix} \mathbf{H}_\triangleleft & 0 \\ 0 & \mathbf{H}_\triangleright \end{pmatrix}$.

This finishes the proof of the second assertion in (H.2). □

### H.2.3 Structural Lemma 3

The next lemma focuses on the case when the condition $|\sum_{i=1}^s \nabla_i| \geq \frac{1}{2}$ holds for many rounds, say, $T_0$ rounds of `MWUfull`. When we apply Lemma H.4 later we shall choose $T_0 = \Theta(T)$.

More specifically, whenever the condition $|\sum_{i=1}^s \nabla_i| \geq \frac{1}{2}$, we define a subset $S$ to consist of these first $s$ coordinates of the vector $y = \delta_k$ before reordering.[27] If there are $T_0$ iterations that this condition holds, then we denote by $S_1, S_2, \ldots, S_{T_0}$ the corresponding subsets.

---
[27] Recall that for notational convenience, we have re-ordered the coordinates of $y$ in Line 5 of `MWUfull`.

Lemma H.4 below shows that if one aggregate properly these subsets, we can find a direction $\delta$ so that the quadratic function $\langle \nabla, \delta \rangle + \frac{1}{6}\delta^\top \nabla^2 \delta$ is sufficiently small:

**Lemma H.4.** *Let $S_1, \cdots S_{T_0}$ be subsets of $[n]$, and satisfy that each $i \in [n]$ is contained in at most $P$ such subsets. Suppose it satisfies*

$$\forall t \in [T_0]: \quad \left|\sum_{i \in S_t} \nabla_i\right| \geq \frac{1}{2} \quad \text{and} \quad \left|\sum_{i \in S_t} \nabla_i\right| \geq v_{S_t}^\top \nabla^2 v_{S_t}$$

*where $v_{S_t}$ is a vector with all coordinates in $S_t$ being $1$ and all coordinates in $[n] \setminus S_t$ being $0$. Then, the following holds: letting $\delta = -\frac{1}{8P} \sum_{t \in [T_0]} \operatorname{sgn}(\langle \nabla, v_{S_t} \rangle) \cdot v_{S_t}$, we have:*

$$\|\delta\|_\infty \leq \frac{1}{8} \quad \text{and} \quad \langle \nabla, \delta \rangle + \frac{1}{6}\delta^\top \nabla^2 \delta \leq -\frac{T}{32P}.$$

*Proof of Lemma H.4.* The proof is by direct calculation. First, since each $i$ is contained in at most $P$ subsets from $S_1, \ldots, S_{T_0}$, we have $\|\delta\|_\infty \leq 1/8$.

We now focus on the linear term $\langle \nabla, \delta \rangle$. By assumption, for every $t \in [T_0]$, we have $\left|\sum_{i \in S_t} \nabla_i\right| \geq \frac{1}{2}$. Therefore, we have

$$\langle \nabla, \delta \rangle = -\frac{1}{8P} \sum_{t \in [T_0]} \operatorname{sgn}(\langle \nabla, v_{S_t}\rangle) \cdot \langle \nabla, v_{S_t} \rangle \leq -\frac{T}{16P}.$$

We then move to the quadric term $\delta^\top \nabla^2 \delta$. Denoting by $\tau_t = -\operatorname{sgn}\left(\sum_{i \in S_t} \nabla_i\right)$, we have

$$\delta^\top \nabla^2 \delta \overset{①}{=} \frac{1}{2(8P)^2} \sum_{i,j \in [n]} |\nabla_{i,j}^2| \left(\sum_{t \in [T_0]} \tau_t \mathbf{1}_{i \in S_t} - \sum_{t \in [T_0]} \tau_t \mathbf{1}_{j \in S_t}\right)^2$$

$$\overset{②}{\leq} \frac{1}{2(8P)^2} \sum_{i,j \in [n]} |\nabla_{i,j}^2| \left(\sum_{t \in [T_0]} \mathbf{1}_{i \in S_t} \mathbf{1}_{j \notin S_t} + \sum_{t \in [T_0]} \mathbf{1}_{i \notin S_t} \mathbf{1}_{j \in S_t}\right)^2$$

$$\overset{③}{\leq} \frac{1}{(8P)^2} \sum_{i,j \in [n]} |\nabla_{i,j}^2| \left(\left(\sum_{t \in [T_0]} \mathbf{1}_{i \in S_t} \mathbf{1}_{j \notin S_t}\right)^2 + \left(\sum_{t \in [T_0]} \mathbf{1}_{i \notin S_t} \mathbf{1}_{j \in S_t}\right)^2\right)$$

$$\overset{④}{\leq} \frac{1}{64P} \sum_{i,j \in [n]} |\nabla_{i,j}^2| \left(\left(\sum_{t \in [T_0]} \mathbf{1}_{i \in S_t} \mathbf{1}_{j \notin S_t}\right) + \left(\sum_{t \in [T_0]} \mathbf{1}_{i \notin S_t} \mathbf{1}_{j \in S_t}\right)\right)$$

$$= \frac{1}{32P} \sum_{t \in [T_0]} \sum_{i,j \in [n]} |\nabla_{i,j}^2| \mathbf{1}_{i \in S_t} \mathbf{1}_{j \notin S_t} = \frac{1}{16P} \sum_{t \in [T_0]} v_{S_t}^\top \nabla^2 v_{S_t}^\top$$

$$\leq \frac{1}{16P} \sum_{t \in [T_0]} \left|\sum_{i \in S_t} \nabla_i\right| = \frac{1}{16P} \sum_{t \in [T_0]} |\langle \nabla, v_{S_t}\rangle| = -\frac{1}{2}\langle \nabla, \delta \rangle$$

Above, equality ① uses the fact that $\nabla^2$ is a Laplacian matrix; inequality ② uses the fact that $|\tau_t(\mathbf{1}_{i \in S_t} - \mathbf{1}_{j \in S_t})| \leq \mathbf{1}_{i \in S_t}\mathbf{1}_{j \notin S_t} + \mathbf{1}_{i \notin S_t}\mathbf{1}_{j \in S_t}$; inequality ③ uses $(a+b)^2 \leq 2a^2 + 2b^2$; inequality ④ uses $\sum_{t \in [T_0]} \mathbf{1}_{i \notin S_t}\mathbf{1}_{j \in S_t} \leq \sum_{t \in [T_0]} \mathbf{1}_{i \notin S_t} \leq P$.

Putting the two bounds together, we have $\langle \nabla, \delta \rangle + \frac{1}{6}\delta^\top \nabla^2 \delta \leq -\frac{T}{32P}$. $\square$

### H.3 Proof of Lemma 8.4

In this subsection we prove Lemma 8.4. Let us first recall its statement:

**Lemma 8.4** (MWUfull). *If $\rho \in [10n^{1/3}, 2\sqrt{n}]$, $\|\alpha\|_\infty \leq 1/32$, $\varepsilon \in [0, 1/16]$, $K \geq 1$, and $T = \Omega((K\rho + K^2)\log n)$, letting $x$ be any vector in $\mathbb{R}^n$ and $\mathbf{H}$ be any Laplacian satisfying $\mathbf{H} \preceq \nabla^2 f(x) \preceq 1.1\mathbf{H}$. Then, the output*

$$\overline{\delta} \leftarrow \mathtt{MWUfull}(\nabla f(x), \mathbf{H}, \alpha, T, \rho, K, \varepsilon)$$

*satisfies $\|\overline{\delta} - \alpha\|_\infty \leq \frac{1}{32} + \frac{2}{K}$ and either*
*(a) $\langle \nabla f(x), \overline{\delta}\rangle + \frac{1}{6}\overline{\delta}^\top \mathbf{H}\overline{\delta} \leq \frac{1}{4}\min_{\|\delta - \alpha\|_\infty \leq 1/32}\{\langle \nabla f(x), \delta\rangle + \frac{1}{6}\delta^\top \mathbf{H}\delta\} + 52n^3\varepsilon$, or*
*(b) $\langle \nabla f(x), \overline{\delta}\rangle + \frac{1}{6}\overline{\delta}^\top \mathbf{H}\overline{\delta} \leq -\frac{1}{256}\frac{\rho}{K}$.*

Before proving Lemma 8.4, we make the following observations to help the readers understand our MWUfull method. In each round $k$ of MWUfull, denoting by $y = \delta_k$ and assume that $|y_1| \geq \cdots \geq |y_n|$ (see Line 5 of MWUfull), we have

- If $\|y\|_\infty \leq \rho$, then we do not perform truncation in Line 22 of MWUfull.
- If $\|y\|_\infty > \rho$, by a simple counting argument (see Claim H.5 below), MWUfull must be able to find $s \in [\rho]$ such that $|y_s| - |y_{s+1}| \geq 15$ and $|y_{s+1}| \leq \rho/2, |y_s| \geq \rho/4$, thus reach Line 7.

**Claim H.5.** *If $\|y\|_\infty \geq \rho$, then there must be an $s \in [\rho]$ such that $|y_s| - |y_{s+1}| \geq 15$ and $|y_{s+1}| \leq \rho/2, |y_s| \geq \rho/4$.*

*Proof of Claim H.5.* Since $\|y\|_\infty \geq \rho$ and

$$\sum_{i \in [\rho/2]} w_i \left(\frac{\rho}{2} - |\alpha_i|\right)^2 \geq \frac{1}{2} \cdot \frac{\rho}{2} \left(\frac{\rho}{2} - 1\right)^2 \geq 2n \;,$$

we know that there must be an $i \in [\rho/2]$ such that $|y_i| \geq \rho/2$ and $|y_{i+1}| \leq \rho/2$. If there exists $j \in \{i, i+1, \cdots, i + \rho/60\}$ such that $|y_j| - |y_{j+1}| \geq 15$, then we can already finish the proof by picking the smallest such $j$. Otherwise, we have:

$$\forall r \in [\rho/60]: \quad |y_{i+r}| \geq \frac{\rho}{2} - 15r$$

Therefore,

$$\sum_{i \in [n]} w_i(y_i - \alpha_i)^2 \geq \frac{1}{2} \sum_{r \in [(\rho-4)/60]} \left(\frac{\rho}{2} - 15r - 1\right)^2 \geq \frac{\rho - 4}{60} \times \frac{1}{16}\rho^2 = \frac{1}{960}\rho^2(\rho - 4) > n \;.$$

This contradicts $\|y - \alpha\|_w^2 \leq n/1024$ and completes the proof. $\square$

We are now ready to prove Lemma 8.4.

*Proof of Lemma 8.4.* Recall that in MWUfull, the subset $S$ records the set of rounds $k$ in which Line 13 has been reached. There are two cases, $|S| \leq \frac{T}{2K}$ and $|S| > \frac{T}{2K}$. In the former case, MWUfull returns on Line 26, and and in the latter case, MWUfull returns on Line 28.

1. Suppose $|S| \leq \frac{T}{2K}$. We observe that at the each of each round $k \notin [S]$, the vector $\delta_k$ satisfies

   (a) Observation 1: $\|\delta_k\|_\infty \leq \rho$.
   
   This is by Claim H.5, which states that if $\|\delta_k\|_\infty \leq \rho$ then we must have reached Line 7 of MWUfull; however, if Line 9 is true, then we must have modified $\delta_k$ so that its large coordinates are all replaced with $z_\triangleleft$ so cannot be greater than $\rho$. In other words, if we have not reached Line 19 and put $k$ into set $S$, then it must satisfy $\|\delta_k\|_\infty \leq \rho$.

(b) Observation 2: $\langle \nabla, \delta_k \rangle + \frac{1}{6}(\delta_k)^\top \mathbf{H} \delta_k \leq 0.5 \min_{\|z-\alpha\|_\infty \leq 1/32} \left\{ \langle \nabla, \delta \rangle + \frac{1}{6} \delta^\top \mathbf{H} \delta \right\} + 52n^3 \varepsilon$.

This is because, either we have not reached Line 7 in which case Line 4 of the algorithm already puts such guarantee on $\delta_k$; or if we have reached Line 7, then we must have reached Line 11, but in this case, $z_\triangleleft$ must satisfy (using Lemma 6.1)

$$\langle \nabla_\triangleleft, z_\triangleleft \rangle + \frac{1}{6}(z_\triangleleft)^\top \mathbf{H}_\triangleleft z_\triangleleft \leq \min_{\|z-\alpha_\triangleleft\|_\infty \leq 1/32} \left\{ \langle \nabla_\triangleleft, z \rangle + \frac{1}{6} z^\top \mathbf{H}_\triangleleft z \right\} + \varepsilon \ .$$

Applying our structural lemma Lemma H.3 with $4\varepsilon$, we have that $\delta_k$ satisfies the desire inequality, after updating $\delta_{k,i} = (z_\triangleleft)_i$ for $i \in [s]$ on Line 12.

Using Observation 1, following the same calculation as in the proof Lemma 6.1 (for the `MWUbasic` method), but instead this time we can pick $\eta = \frac{1}{\rho+K}$ since $\|\ell_k\|_\infty \leq \rho + 1$, we have that for every $u \in \Delta$.

$$\frac{1}{T} \sum_{k=0}^{T-1} \langle \ell_k, -u \rangle \leq \frac{1}{T} \sum_{k=0}^{T-1} \left( \langle -\ell_k, w_k \rangle + 2\eta \|\ell_k\|_{w_k}^2 \right) + \frac{n \log(2n^2)}{\eta T}$$

$$\leq \frac{1}{T} \sum_{k=0}^{T-1} \left( \sqrt{\|\ell_k\|_{w_k}^2 \cdot \|w_k\|_1} + 2\eta \|\ell_k\|_{w_k}^2 \right) + \frac{n \log(2n^2)}{\eta T}$$

$$\leq \frac{1}{T} \sum_{k=0}^{T-1} \left( \frac{n}{\sqrt{1024}} + \frac{2}{\rho+K} \frac{n}{2048} \right) + \frac{n(\rho+K) \log(2n^2)}{T} \ .$$

Therefore, as long as $T \geq \Omega((\rho K + K^2) \log n)$, we must have

$$\frac{1}{T} \sum_{k=0}^{T-1} \langle \ell_k, -u \rangle \leq \frac{n}{32} + \frac{n}{8K} \ . \tag{H.5}$$

However, this implies that $\frac{1}{T} \sum_{k \in [T] \setminus S} |\delta_{k,i} - \alpha_i| \leq \frac{1}{32} + \frac{1}{8K}$ for every $i \in [n]$ (because we can choose $u = n \times \mathbf{e}_i$). Since we have chose $\overline{\delta} \leftarrow \frac{1}{T} \left( \sum_{k \in [T] \setminus S} \delta_k \right)$, we therefore have for all $i \in [n]$

$$|\overline{\delta}_i - \alpha_i| \leq \frac{1}{T} \sum_{k \in [T]-S} |\delta_{k,i} - \alpha_i| + \frac{1}{T}|S||\alpha_i| \leq \frac{1}{32} + \frac{2}{K} \ .$$

Also, by Observation 2 and convexity, we have:

$$\langle \nabla, \overline{\delta} \rangle + \frac{1}{6}\overline{\delta}^\top \mathbf{H} \overline{\delta} \leq \frac{1}{T} \left( \sum_{k \in [T]-S} \left( \langle \nabla, \delta_k \rangle + \frac{1}{6}(\delta_k)^\top \mathbf{H} \delta_k \right) + \sum_{k \in [S]} 0 \right)$$

$$\leq \frac{1}{4} \min_{\|z-\alpha\|_\infty \leq 1/32} \left\{ \langle \nabla, \delta \rangle + \frac{1}{6} \delta^\top \mathbf{H} \delta \right\} + 52n^3 \varepsilon \ .$$

2. Suppose $|S| > \frac{T}{2K}$.

We first bound the infinite norm of $\widehat{v}$. Recall that in each round of `MWUfull`, $\widehat{v}_i$ for each coordinate $i$ either stays the same or increases by one. In particular, if $\widehat{v}_i$ increases by one, then we must have: $|y_i| \geq \rho/4$ (because Line 19 must have been reached), and this implies $|\ell_{k,i}| \geq \frac{\rho}{4} - 1$. Using (H.5) again, and choosing $u = n\mathbf{e}_i$, we conclude that this cannot happen for more than $T/(\rho/4 - 1)$ times. In other words, we have

$$\left( \frac{\rho}{4} - 1 \right) \|\widehat{v}\|_\infty \leq T \ . \tag{H.6}$$

We next show that the output $\overline{\delta}$ must satisfy

$$\langle \nabla, \overline{\delta}\rangle + \frac{1}{6}\overline{\delta}^\top \nabla^2 \overline{\delta} \leq -\frac{T}{32\|\widehat{v}\|_\infty K} \quad. \tag{H.7}$$

For each round $k \in S$, because Line 19 is reached, there must either $v'_k = v^-_k$ or $v'_k = v^+_k$ that satisfies $|\langle \nabla, v'_k\rangle| > \varepsilon$ or $(v'_k)^\top \mathbf{H} v'_k > \varepsilon$.[28] Applying Lemma H.1, we know that it must satisfies

$$|\langle \nabla, v'_k\rangle| \geq \frac{1}{2} \quad \text{and} \quad |\langle \nabla, v'_k\rangle| \geq v'^\top_k \nabla^2 v'_k \quad.$$

Let us denotes the non-zero coordinates of $v'_k$ as $S_k$. Applying Lemma H.4 on $\{S_k \mid k \in [S]\}$ along with $P = \|\widehat{v}\|_\infty$, we immediately have (H.7).

Finally, combining (H.6) and (H.7), we have

$$\langle \nabla, \overline{\delta}\rangle + \frac{1}{6}\overline{\delta}^\top \nabla^2 \overline{\delta} \leq -\frac{1}{256}\frac{\rho}{K}$$

Moreover, we know that $\|\overline{\delta} - \alpha\|_\infty \leq \|\alpha\|_\infty + \|\overline{\delta}\|_\infty \leq \frac{1}{32} + \frac{1}{K}$. □

---

[28]Otherwise, we would have satisfied $|\langle \nabla, v_k\rangle| \leq |\langle \nabla, v^+_k\rangle| + |\langle \nabla, v^-_k\rangle| \leq 2\varepsilon$ and $v^\top_k \mathbf{H} v_k \leq 2\left((v^+_k)^\top \mathbf{H} v^+_k + (v^-_k)^\top \mathbf{H} v^-_k\right) \leq 4\varepsilon$ and thus cannot reach Line 19 of `MWUfull`.